\documentclass[twocolumn]{aastex62}

\newcommand{\be}{\begin{equation}}
\newcommand{\ee}{\end{equation}}
\newcommand{\bex}{\begin{equation}\notag}
\newcommand{\eex}{\end{equation}\notag}
\newcommand{\bea}{\begin{eqnarray}}
\newcommand{\eea}{\end{eqnarray}}
\newcommand{\beax}{\begin{eqnarray*}}
\newcommand{\eeax}{\end{eqnarray*}}
\newcommand{\ba}{\begin{array}}
\newcommand{\ea}{\end{array}}
\newcommand{\half}{\frac{1}{2}}
\newcommand{\oneptfive}{\frac{3}{2}}
\newcommand{\vecB}{{\mathbf B}}
\newcommand{\vecE}{{\mathbf E}}
\newcommand{\vecU}{{\mathbf U}}

\newcommand{\vecA}{{\mathbf A}}
\newcommand{\vecJ}{{\mathbf J}}
\newcommand{\vecV}{{\mathbf V}}
\newcommand{\vecSigma}{{\mathbf \Sigma}}
\newcommand{\vecVlos}{{\mathbf V_{LOS}}}

\newcommand{\veczhat}{{\mathbf {\hat z}}}
\newcommand{\vecrhat}{{\mathbf {\hat r}}}
\newcommand{\vecnhat}{{\mathbf {\hat n}}}
\newcommand{\unitv}[1]{\mbox{\boldmath$\hat{#1}$}}

\newcommand{\scrJ}{{\mathcal{J}}}
\newcommand{\scrB}{{\mathcal{B}}}
\newcommand{\scrP}{{\mathcal{P}}}

\newcommand{\Pdot}{\dot P}
\newcommand{\Tdot}{\dot T}

\newcommand{\grad}{\mbox{\boldmath$\nabla$}}

\received{December 18, 2019}
\accepted{March 23, 2020}
\submitjournal{ApJ Supplement Series}

\shorttitle{PDFI\_SS Electric Fields}
\shortauthors{Fisher et al.}

\begin{document}

\title{The PDFI\_SS Electric Field Inversion Software}

\correspondingauthor{George Fisher}
\email{fisher@ssl.berkeley.edu}

\author[0000-0002-6912-5704]{George H. Fisher}
\affiliation{Space Sciences Laboratory \\
University of California \\
7 Gauss Way \\
Berkeley, CA 94720-7450, USA}

\author[0000-0001-8975-7605]{Maria D. Kazachenko}
\affiliation{Astrophysical and Planetary Sciences \\
University of Colorado \\
2000 Colorado Avenue \\
Boulder, CO 80309, USA}
\affiliation{Space Sciences Laboratory \\
University of California \\
7 Gauss Way \\
Berkeley, CA 94720-7450, USA}

\author[0000-0003-2244-641X]{Brian T. Welsch}
\affiliation{Natural and Applied Sciences \\
University of Wisconsin, Green Bay \\
Green Bay, WI 54311, USA}
\affiliation{Space Sciences Laboratory \\
University of California \\
7 Gauss Way \\
Berkeley, CA 94720-7450, USA}

\author[0000-0003-4043-616X]{Xudong Sun}
\affiliation{Institute for Astronomy \\
University of Hawaii at Manoa \\
Pukalani, Hawaii 96768, USA}
\affiliation{W. W. Hansen Experimental Physics Laboratory \\
Stanford University \\
Stanford, CA 94305, USA}

\author[0000-0003-2045-5320]{Erkka Lumme}
\affiliation{Department of Physics \\
University of Helsinki \\
Helsinki, Finland}

\author[0000-0001-5540-8108]{David J. Bercik}
\affiliation{Space Sciences Laboratory \\
University of California \\
7 Gauss Way \\
Berkeley, CA 94720-7450, USA}

\author[0000-0002-6338-0691]{Marc L. DeRosa}
\affiliation{Lockheed Martin Solar and Astrophysics Laboratory \\
Building 252, 3251 Hanover Street \\
Palo Alto, CA 94304, USA}

\author[0000-0003-2110-9753]{Mark C. M. Cheung}
\affiliation{Lockheed Martin Solar and Astrophysics Laboratory \\
Building 252, 3251 Hanover Street \\
Palo Alto, CA 94304, USA}



\begin{abstract}

We describe the PDFI\_SS software library, which is designed to find
the electric field at the Sun's photosphere from a sequence of vector
magnetogram and Doppler velocity measurements, and estimates of
horizontal velocities obtained from local correlation tracking using
the recently upgraded FLCT code.  The library, a collection of Fortran
subroutines, uses the ``PDFI'' technique described by
Kazachenko et al. (2014), but modified for use in spherical,
Plate-Carr\'ee geometry on a staggered grid.  The domain
over which solutions are found is a subset of the global spherical
surface, defined by user-specified limits of colatitude and
longitude.  Our staggered-grid approach, based on that of Yee (1966),
is more conservative and self-consistent compared to the centered,
Cartesian grid used by Kazachenko et al. (2014).  The library can be
used to compute an end-to-end solution for electric fields from
data taken by the HMI instrument aboard NASA's SDO Mission.
This capability has been incorporated into the HMI pipeline
processing system operating at SDO's JSOC. The library is written in a
general and modular way so that the calculations can be customized to
modify or delete electric field contributions, or used with other data
sets. Other applications include ``nudging'' numerical models of the
solar atmosphere to facilitate assimilative simulations. The library
includes an ability to compute ``global'' (whole-Sun) electric field
solutions. The library also includes an ability to compute potential magnetic
field solutions in spherical coordinates.
This distribution includes a number of test programs that
allow the user to test the software.
\end{abstract}

\keywords{Sun: magnetic fields --- 
Sun: photosphere --- Sun: corona --- Sun: activity}


\section{Introduction} \label{sec:intro}

The goal of this article is to describe the mathematical and numerical
details of our software
(\url{http://cgem.ssl.berkeley.edu/cgi-bin/cgem/PDFI\_SS}), 
which we call PDFI\_SS, to derive electric fields in 
the solar photosphere from a time sequence of vector magnetogram and Doppler
shift data (an archived version of the software, available as a gzipped tar
file, is also available from Zenodo \citet{Fisher2020pdfi}).
By reading this paper carefully, the reader should have enough information
to understand how to use the software, and also to 
understand the physical, mathematical, and numerical assumptions that 
the software employs.  For detailed usage of the software, this article is
meant to be used in combination with the source-code documentation included
within each subroutine of the library, along with additional material
distributed within the \texttt{doc} folder of the distribution.  All source
code files include a detailed description of the subroutine arguments,
along with expected dimensions and units.  For this reason, we do not
include the details of subroutine arguments within this article, but we do
discuss each important subroutine by name and describe its purpose.
It is very easy to view the source code for any subroutine in the PDFI\_SS
library in a web browser 
by first going to the above software repository URL, clicking on 
the ``Files'' link, then
clicking on the ``fortran'' folder and then clicking on the links to any of
the subroutines.

The PDFI\_SS software is based on the PDFI technique for
deriving electric fields that is described in detail in
\citet{Kazachenko2014}
(henceforth KFW14).  
The acronym ``PDFI'' stands for ``PTD plus Doppler plus
FLCT plus Ideal'' contributions to the electric field.  The physical
significance of these four electric field contributions will be elaborated
in \S \ref{sec:pdfi} of this article. 
The ``\_SS'' suffix in the name ``PDFI\_SS''
stands for ``spherical staggered'', because the fundamental difference between
the techniques described in KFW14 and those described here are that (1)
we use spherical Plate Carr\'ee coordinates instead of 
Cartesian coordinates, to allow for
realistic solar geometries for large domains of the Sun, and (2) we have
switched to a staggered-grid description of the scalar and vector
field variables in the domain.  While the basic concepts of KFW14 still apply, 
there are many differences in the details, which are described in this article.

The development of the PDFI\_SS software was motivated by
the Coronal Global Evolutionary Model (CGEM), a Strategic Capability 
Project (\url{http://cgem.ssl.berkeley.edu}) funded by NASA's 
Living With a Star (LWS) Program and by 
the National Science Foundation (NSF) \citep{Fisher2015}.  
The core activity of the CGEM project
is to drive large-scale and active-region scale magnetofrictional (MF) and 
magnetohydrodynamic (MHD)
simulations of the solar corona using time cadences of
vector magnetogram data and electric
fields inferred at the photosphere.  The PDFI\_SS software is what CGEM uses to
derive the photospheric electric fields from vector magnetogram and Doppler
data that are then used by the MF \citep{Cheung2012} and RADMHD
\citep{Abbett2007,Abbett2012,Abbett2014} models.

The PDFI\_SS software is written
as a general purpose library, which can be easily linked to other programs.
It is designed to be modular, making it easy for users
to customize the software for their own purposes, rather than being written for
a single narrow purpose.  The PDFI\_SS library is written in
Fortran, primarily because of its extensive use of the Fortran library
FISHPACK, an elliptic equation package that is well-suited to solutions 
of the two-dimensional
Poisson equations that make up the core of the PDFI technique (KFW14).  Once
compiled, it is straightforward to link the PDFI\_SS library to other 
Fortran, C/C++, and Python programs.  The SDO Joint Science Operations
Center (JSOC) magnetic field pipeline software, which is written in C, calls
one of the high-level Fortran subroutines within PDFI\_SS to compute 
electric fields within each ``CGEM patch'' 
(similar to the ``space-weather HMI active region patch'', or SHARP)
\citep{Bobra2014}).  Thus, in addition
to being a software library, many of the data products that can be computed by 
PDFI\_SS are also available to all users of the SDO JSOC.

The primary purpose of the PDFI\_SS library is to compute electric fields in
the solar photosphere from time sequences of the input magnetic and Doppler
data.  The domain of the solutions is a subset of the global solar surface,
defined by limits on colatitude and longitude, which we will refer to as
the base of a ``spherical wedge'' domain.
However, the software also includes a set of subroutines for
performing vector calculus operations on subsets of a spherical surface, it
has the ability to compute ``nudging'' electric fields in a numerical simulation
for assimilative purposes, and also includes the ability to compute 3D
potential magnetic field solutions for spherical wedge domains.  Within 
the context of the electric field inversions, the user can customize the 
electric field solutions by choosing to include or neglect the various 
contributions to
the total PDFI electric field described by KFW14.  

In \S \ref{sec:pdfi}, we discuss other recently published electric 
field inversion
methods.  We then review the PDFI equations for determining electric
fields in the solar photosphere from assumed input HMI vector magnetogram 
and Doppler
measurements, along with estimates of flows along the photospheric surface
determined from optical flow techniques.  We mention spherical geometry
corrections to expressions in KFW14 where applicable.  

In \S \ref{sec:pdfinum}, we discuss in detail the numerical implementation of
the PDFI solutions, including
the staggered  grid based on the concepts of \citet{Yee1966}, 
the finite difference
representations, the necessary coordinate transformations and interpolations,
and all the other details needed to understand and use
the PDFI\_SS software to compute electric fields in the photosphere.

In \S \ref{sec:D4stuff} we describe the data processing upstream of using
PDFI\_SS that the software expects to have done before the PDFI electric
fields are computed, including corrections for solar rotation and the temporal
evolution of the transverse magnetic field from HMI data, corrections to
the Doppler velocity from the convective blue-shift bias, and the calculation of
horizontal velocities from optical-flow methods (using FLCT).  
We include a description
of upgrades we have made to the FLCT code for computing horizontal flows.
We describe the interpolation of
the results to a Plate Carr\'ee grid, and the addition of ``padding'', which
improves the properties of the electric field solutions.  
While the discussion in \S \ref{sec:D4stuff} is specific to
HMI data, this could be used as a guideline for preparing datasets from
other instruments.

In \S \ref{sec:broad}, we describe broader applications of the 
PDFI\_SS software, beyond
the calculation of electric fields described in \S \ref{sec:pdfinum}.  These
include the use of ``nudging'' electric fields for data assimilation in
numerical models of the solar atmosphere; the use of curl-free solutions of
the electric field to match boundary conditions in other models, and the
calculation of global (whole Sun) solutions for the ``PTD'' electric field
solution.

In \S \ref{sec:potential}, we describe how to use PDFI\_SS to 
compute potential magnetic field
solutions in a spherical wedge domain with a given range in radius between
the photosphere and a ``source surface''.  This software differs from most
treatments of potential fields in spherical coordinates in that it uses a
finite difference approach rather than spherical harmonic expansion.  This
same software can be used to compute a three dimensional distribution of the
electric field due to a temporally evolving potential magnetic field
in a coronal volume lying above the photosphere, based
on the time-dependent behavior of the radial component of the magnetic field
at the photosphere.

In \S \ref{sec:cartesian}, we describe how to use the PDFI\_SS library to
compute solutions in Cartesian coordinates, by mapping a small Cartesian patch
onto the surface of a very large sphere, with the patch straddling the equator.

In \S \ref{sec:compiling}, we first lay out the development history of the
PDFI\_SS library, and then
describe in detail how to compile the PDFI\_SS
library, and then how to link the library to other Fortran, C/C++, Python, 
and legacy IDL software.

In \S \ref{sec:testing}, we describe test program calculations which
are included in the software distribution, 
including tests of the PDFI\_SS solutions 
using HMI data from NOAA AR 11158, 
an analysis of the ANMHD
test data discussed in KFW14 and \citet{Schuck2008}, 
test programs for the potential magnetic
field software, tests of the global PTD electric field solutions, and tests
of usage of the software from C and Python.  

\S \ref{sec:lists} contains an alphabetically ordered table (Table 
\ref{tab:subroutinelist}) of the most important
subroutines in the PDFI\_SS library, along with lists and descriptions of
important commonly used calling argument variables in these subroutines.  
Table \ref{tab:subroutinelist} includes a brief description of each subroutine
task, along with a link to the section of this article that describes the
subroutine in more detail.  The objective is to provide the user with an 
easy-to-use index for specific material within the article.

This article is lengthy, because it is intended to describe in detail
all the important aspects of the software.
Depending on the reader's goals, it may not be necessary to read the entire 
article.

If one is simply interested in understanding the ``big picture'' regarding
the PDFI\_SS software, one can read KFW14, \S \ref{sec:pdfi},
and the first four sub-sections of \S \ref{sec:pdfinum}.

If one is interested in using PDFI\_SS results obtained from the SDO JSOC,
namely electric field data products computed for selected active-regions, we
recommend reading KFW14, plus \S \ref{sec:pdfi}-\ref{sec:D4stuff}.

If one is interested in installing and using the PDFI\_SS software to
compute electric field solutions from magnetic field data, we recommend reading 
\S \ref{sec:pdfi}-\ref{sec:D4stuff}, and 
\S \ref{sec:cartesian}-\ref{sec:testing}.

If one is mainly
interested in using PDFI\_SS for computing ``nudging'' electric fields
for ``data driving'' applications, we recommend reading 
\S \ref{sec:pdfi}-\ref{sec:pdfinum}, \S \ref{sec:broad}, 
and \S \ref{sec:compiling}.

If one is only interested in using the potential magnetic field software,
we recommend reading \S \ref{sec:ptd}, \S \ref{sec:potential}, and
\S \ref{sec:compiling}-\ref{sec:testing}.

If one is only interested in installing and testing
the PDFI\_SS software, one can simply read 
\S \ref{sec:compiling}-\ref{sec:testing}.

\section{Review of the PDFI Electric Field Inversion Equations}
\label{sec:pdfi}

In their presentation of the PDFI method,
KFW14 reviewed the current state of electric field inversions in the literature
at the time that paper was published.  Since the publication of KFW14, 
a number of other published
efforts for electric field inversions have been done.  Here, we
first briefly summarize these efforts.

\citet{Mackay2011} and \citet{Yardley2018} solve a Poisson equation for
what is effectively
a poloidal potential using the time rate of change of the normal 
magnetic
field as a
source term, from which one can derive horizontal components of the electric
field (expressed in this case by the time derivative of a vector potential).
\citet{Weinzierl2016b, Weinzierl2016a} presented solutions for the 
horizontal components of
the electric field that combined a solution for the ``inductive'' contributions
to the horizontal electric field components,
determined from the time derivative of the radial magnetic field, with a
non-inductive contribution that was determined from surface flux
transport models.  \citet{Yeates2017} derived electric field solutions that
combine solutions for the same inductive contribution as those above, but
with the non-inductive contribution to the electric field determined from
a ``sparseness'' constraint, to minimize unphysical artifacts of the
horizontal electric field from the purely inductive solution.  
\citet{Lumme2017,Price2019} 
used solutions for all three components of the electric field
using time derivatives for all three components of $\vecB$, as described
for the ``PTD'' solutions in KFW14, using a centered grid formalism.  
For the non-inductive contribution to
$\vecE$, they used the ad-hoc treatments suggested in \citet{Cheung2012}.
The data-driven MHD simulations of \citet{Hayashi2018,Hayashi2019} used
solutions for the PTD equations derived in KFW14, evidently including
some depth-dependent information for the horizontal electric fields.
In \citet{Lumme2019}, the full PDFI solutions for all three components of
the electric field were determined using the methods
described in this article
to study the dependence of electric field solutions on time
cadence.  \citet{Lumme2019} also studied the effect of cadence on solutions
determined from the DAVE4VM method \citep{Schuck2008}.

Because of the importance of the curl operator evaluated in spherical
coordinates within PDFI\_SS, we now explicitly write out each component
of the curl before heading into the details of PDFI\_SS.  
This can be found in many standard texts in mathematics, such
as \citet{Morse1953}.  Here we use standard spherical polar coordinates, where
the unit vectors are $\unitv{\theta}$, pointing in the colatitude direction
($i.e.$ from north to south), $\unitv{\phi}$, pointing in the longitudinal or
azimuthal direction ($i.e.$ towards the right, when looking at the equator of
the Sun from outside its surface), and $\unitv{r}$, pointing
in the radial direction ($i.e.$ outward from the center of the Sun).
The quantities $\theta$ and $r$ are colatitude and radius, respectively.
The quantities $U_{\theta}$, $U_{\phi}$, and $U_r$ are the colatitudinal,
longitudinal, and radial components of $\vecU$:
\be
\unitv{\theta} \cdot \grad \times \vecU =
{1 \over r \sin \theta} {\partial U_r \over \partial \phi}\ -
{1 \over r}\ {\partial \over \partial r}\ ( r U_{\phi} )\ ,
\label{eqn:curltheta}
\ee
\be
\unitv{\phi} \cdot \grad \times \vecU =
{1 \over r}\ {\partial \over \partial r}\ ( r U_{\theta} )\ -
{1 \over r}\ {\partial U_r \over \partial \theta}\ ,
\label{eqn:curlphi}
\ee
and
\be
\unitv{r} \cdot \grad \times \vecU =
{1 \over r \sin \theta}\ {\partial \over \partial \theta}\ (\sin \theta \ 
U_{\phi})\ -
{1 \over r \sin \theta}\ {\partial U_{\theta} \over \partial \phi}\ .
\label{eqn:curlr}
\ee

Since the derivation and discussion of the equations that define the
PDFI electric field solutions have already been described in detail in KFW14,
we simply review below the equations necessary to define each 
contribution to
the PDFI electric field.  The main difference here between KFW14 and these
equations is the use of spherical coordinates.  The fact we are using
spherical coordinates makes little difference to the overall structure of
the equations, but where spherical geometry does change things from Cartesian
coordinates, we mention it.

\subsection{The PTD Contribution to the Electric Field} \label{sec:ptd}
We start with the Poloidal-Toroidal decomposition (PTD) for the magnetic field
$\vecB$ in spherical coordinates \citep{Chandrasekhar1961,Backus1986}
in terms of the Poloidal potential $P$, and
the Toroidal potential $T$:
\be
\vecB = \grad \times \grad \times P \vecrhat + \grad \times T \vecrhat ,
\label{eqn:Bptddef}
\ee
where $\vecrhat$ is the unit vector in the radial direction.
Here, in a change from the notation used in KFW14, we use $P$ for the 
poloidal potential instead of $\scrB$, and $T$ for the toroidal potential
instead of $\scrJ$.  This change was made for notational simplicity, and also 
corresponds with the notation used by \citet{Lumme2017}.
We also note another useful form for equation (\ref{eqn:Bptddef}),
namely
\be
\vecB = -\vecrhat\ \nabla_h^2 P + \grad_h \left( 
{\partial P \over \partial r} \right) + \grad \times T \vecrhat .
\label{eqn:Bptddecomp}
\ee
The operator $\nabla_h^2$ is the horizontal 
Laplacian, $i.e.$ the full Laplacian but omitting the radial derivative
contribution,
and $\grad_h (\partial P / \partial r)$ 
represents the horizontal components of the gradient of the radial
derivative of $P$.
By uncurling equation (\ref{eqn:Bptddef}), it is clear that the vector
potential $\vecA$ can be written in terms of $P$ and $T$ as
\be
\vecA = \grad \times P \vecrhat + T \vecrhat ,
\label{eqn:Aptd}
\ee
where we have omitted an explicit gauge term.

The PTD, or ``inductive'' Electric Field $\vecE^P$ is related to the 
magnetic field $\vecB$ through Faraday's Law:
\be
{\dot \vecB} = - \grad \times c \vecE^P ,
\label{eqn:Faraday}
\ee
where $c$ is the speed of light, and where we use the over-dot to denote 
a partial time derivative.  Substituting equation (\ref{eqn:Bptddef})
into Faraday's Law and uncurling, we find
\be
c \vecE^P = -\grad \times \Pdot \vecrhat -\Tdot \vecrhat,
\label{eqn:Eind}
\ee
where $\Pdot$ and $\Tdot$ are the partial time derivatives of $P$ and $T$.
The general description of the electric field will also include the gradient
of a scalar potential in addition to the inductive solution in equation
(\ref{eqn:Eind}), 
but we omit any  explicit gradient contributions to $c \vecE$ here,
and discuss the gradient contributions in subsections further below.

By evaluating the radial component of equation (\ref{eqn:Faraday}) when
substituting equation (\ref{eqn:Eind}) for $\vecE^P$, we find that 
$\Pdot$ obeys the two-dimensional Poisson equation
\be
\nabla_h^2 \Pdot = - {\dot B_r},
\label{eqn:poisson-pdot}
\ee
where $B_r$ is the radial 
magnetic field component.  
Here, the right hand side of equation
(\ref{eqn:poisson-pdot}) is viewed as a source term which can be evaluated
from the magnetogram data.
By taking the radial component of the curl of equation (\ref{eqn:Faraday}),
we find that $\Tdot$ obeys the Poisson equation
\be
\nabla_h^2 \Tdot =
-\vecrhat \cdot \left( \grad \times {\dot \vecB_h} \right) ,
\label{eqn:poisson-Tdot}
\ee
where $\vecB_h$ are the horizontal components of $\vecB$.
A third useful Poisson equation can be found by taking the divergence
of the horizontal components of equation (\ref{eqn:Faraday}):
\be
\nabla_h^2 \left( {\partial \Pdot \over \partial r} \right) = 
\grad_h \cdot {\dot \vecB_h} .
\label{eqn:poisson-dpdrdot}
\ee

The quantity $\partial \Pdot / \partial r$ is important, because it allows 
one to evaluate the radial
derivative of the horizontal electric field components.  To see this,
one can evaluate the quantity
\be
{1 \over r} {\partial \over \partial r} \left( r c \vecE^P_h \right) ,
\label{eqn:oneoverrdbydrreh}
\ee
where $c \vecE^P_h$ represents the horizontal components of $c \vecE^P$ from 
equation (\ref{eqn:Eind}):
\be
c \vecE^P_h = - \grad \times \dot P \unitv{r} .
\label{eqn:vecephdef}
\ee
The $\theta$ and $\phi$ components of the curl in equation 
(\ref{eqn:vecephdef}) both contain leading factors 
of $1/r$, as can
be seen from the first term of equation (\ref{eqn:curltheta})
and the second term of equation (\ref{eqn:curlphi}).
The radial derivative in equation (\ref{eqn:oneoverrdbydrreh}) therefore
is applied directly to $\dot P$, resulting in
\be
{1 \over r} {\partial \over \partial r} \left( r c \vecE^P_h \right)  = 
- \grad \times {\partial \dot P \over \partial r} \unitv{r} .
\label{eqn:oneoverrdbydrehptd}
\ee
Expanding the radial derivative on the left hand side (LHS) of 
equation (\ref{eqn:oneoverrdbydrehptd}), we then arrive at this
expression for the radial derivative of $\vecE^P_h$:
\be
c {\partial \vecE^P_h \over \partial r} = -\grad \times
{\partial \Pdot \over \partial r} \vecrhat - c {\vecE^P_h \over r} .
\label{eqn:dEinddr}
\ee
Here, the quantity $r$ is the radius of the surface upon which the
two-dimensional Poisson equations are solved, which for nearly all of our
purposes can be taken as the radius of the Sun $R_{\odot}$.
Equation (\ref{eqn:dEinddr}) was
not given in KFW14, but as shown here, is easy to derive.
Note that the 2nd term on the right hand side of equation
(\ref{eqn:dEinddr}) goes to zero as $r \rightarrow \infty$, 
meaning that in the Cartesian limit, this term vanishes.
The radial derivative of the horizontal inductive electric
field is useful because it allows one to compute the horizontal components of
$\grad \times \vecE$.

The availability of time cadences of vector magnetic field measurements, such
as from the HMI instrument on NASA's SDO Mission \citep{Scherrer2012}, 
enables the evaluation of
the time derivatives as the source terms in the above Poisson equations,
making such electric field solutions possible.  With the data in hand,
evaluation of $\vecE^P$ becomes a matter of solving the above Poisson equations
on a region of the Sun's surface and then evaluating equation (\ref{eqn:Eind}).

\subsection{Doppler Contributions to the Non-inductive Electric 
Field} \label{sec:doppler}

The Doppler velocity, when combined with the magnetic field measurements,
provides additional information about the electric field beyond the
inductive contribution $\vecE^P$ \citep{Ravindra2008}.  The relationship 
between the measured
line-of-sight velocity vector $\vecVlos$ and the true plasma velocity $\vecV$
is given by
\be
\vecVlos = ( \vecV \cdot \unitv{\ell} ) \unitv{\ell} ,
\label{eqn:Vlosdef}
\ee
where $\unitv{\ell}$ is the line-of-sight (LOS)
unit vector pointing toward the observer from the surface of the Sun.
Note that $\unitv{\ell}$ is a function of position
on the solar surface, since the Sun's
surface is curved.
Near a LOS polarity inversion line (PIL), the $\vecVlos$ flow carrying
transverse components of the magnetic field perpendicular to $\unitv{\ell}$ 
results
in an electric field contribution
\be
c \vecE_{PIL} = - \vecVlos \times \vecB_t ,
\label{eqn:Edopraw}
\ee
where $\vecB_t = \vecB - ( \vecB \cdot \unitv{\ell} ) \unitv{\ell}$ 
represents the components of $\vecB$ transverse to $\unitv{\ell}$.

When we are not near a LOS PIL, a non-zero Doppler velocity is less certain
to be coming from a flow that transports $\vecB_t$, and instead could be
a signature of flows parallel to $\vecB$, which have no electric field
consequences.  To account for this uncertainty away from LOS PILs, the
electric field in equation (\ref{eqn:Edopraw}) is modulated by an empirical
factor $w_{LOS}$ given by this expression:
\be
w_{LOS}= \exp \left(-\ {1 \over \sigma_{PIL}^2}
\biggl|{B_{LOS} \over \vecB_t}\biggr|^2\right) ,
\label{eqn:wlos}
\ee
where $B_{LOS}$ is the LOS component of $\vecB$, and $\sigma_{PIL}$ is 
an empirically adjustable parameter, commonly taken as unity.

To the extent that the Doppler contribution to the electric field contributes
magnetic evolution, that contribution should already have been included in
the inductive contribution described in \S \ref{sec:ptd}.  We therefore
want to include any additional curl-free contribution to the electric field
from the Doppler term.  To do this, we will represent the Doppler contribution
in equation (\ref{eqn:Edopraw}) by the gradient of a scalar potential,
which we'll call $\psi^D$:
\be
c \vecE^D = - \grad \psi^D ,
\label{eqn:edoppsi}
\ee
where we note that this form automatically results in zero curl.

The details of how the equation defining $\psi^D$ is derived are provided 
in \S 2.3.3 in KFW14.  Here, we will simply write down the result,
modified slightly to account for working in spherical, rather than
Cartesian coordinates:
\bea
\grad_h \psi^D \cdot \unitv{q}_h  +  q_r {\partial \psi^D \over \partial r}
& = w_{LOS} ( \vecVlos \times \vecB_t )_h \cdot \unitv{q}_h +\nonumber \\ 
& q_r w_{LOS} ( \vecVlos \times \vecB_t )_r ,
\label{eqn:eqnpsid}
\eea
where $\unitv{q}$ is the unit vector pointing in the same direction as
$\vecVlos \times \vecB_t$, $q_r$ is the radial component of $\unitv{q}$,
and $\unitv{q}_h$ are the horizontal components of $\unitv{q}$. 

Equation (\ref{eqn:eqnpsid}) is solved using the ``iterative'' technique
developed by co-author Brian Welsch, initially described in 
\S 3.2 of \citet{Fisher2010},
with subsequent changes discussed in \S 2.2 of KFW14.  In this article, 
the current version of the iterative technique for PDFI\_SS is described in
\S \ref{sec:relaxation}.  The iterative technique involves
repeated solutions of a two-dimensional Poisson equation that tries to
best represent the Doppler electric field from the observed data by
the gradient of $\psi^D$ in the $\unitv{q}$ direction.
\\
\\ 

\subsection{FLCT Contributions to the Non-inductive Electric 
Field} \label{sec:flct}

There are many techniques currently available to estimate velocities in the
directions parallel to the solar surface by
solving the ``optical flow'' problem on pairs of images closely adjacent in
time to estimate these flows (see $e.g.$
reviews by \citet{Welsch2007,Schuck2008,Tremblay2018}).  
Here we estimate horizontal flow
velocities $\vecV_h^F$ using the FLCT local correlation tracking 
code \citep{Fisher2008} applied to images of $B_r$ from the vector magnetogram
sequence.  The choice of FLCT is somewhat arbitrary; any other existing
technique could be used as an alternative.  We chose FLCT because we are
very familiar with the algorithm and the code, and have spent years
making the code as computationally efficient as possible.

The use of FLCT in spherical geometry introduces some complications, since
the FLCT algorithm is based strictly on assumptions of Cartesian geometry.
We adopt the solution proposed in the Appendix of \citet{Welsch2009}, in
which the $B_r$ images are mapped to a Mercator projection, FLCT is run, and
then the velocities are interpolated and re-scaled back to spherical geometry.
The FLCT code has been updated to perform this operation automatically if the
input data are specified as being equally spaced in longitude and latitude,
$i.e.$ on a Plate Carr\'ee grid.
More detail on recent versions of FLCT can be found in \S \ref{sec:horizvel}
and in the updated source code and documentation for FLCT,
available at the software repository
\url{http://cgem.ssl.berkeley.edu/cgi-bin/cgem/FLCT/home}.

Horizontal velocities $\vecV_h^F$ estimated from FLCT, and acting
on the radial component of the magnetic field, provide an additional
contribution to the electric field:
\be
c \vecE_{FLCT} = - \vecV_h^F \times B_r \vecrhat .
\label{eqn:efromvh}
\ee
We have neglected an additional term, contributing to $E_r$, coming from
$-\vecV_h^F \times \vecB_h$ in equation (\ref{eqn:efromvh}).  In KFW14, we
showed that in Cartesian coordinates, this term is already accounted for 
by the inductive contribution
to $E^P_z$.  The same argument applies here, except the contribution is to
$E^P_r$, the radial component.

In regions where the radial magnetic field component is small compared to
the horizontal magnetic field components, we trust the FLCT velocities less
because the radial magnetic field evolution is less likely to be due to
advection by horizontal flows.  Therefore, we introduce an empirical
modulation function $(1-w_r)$ that multiplies the right-hand side of
equation (\ref{eqn:efromvh})
and which reduces the amplitude of the electric field
when the magnetic field is mostly horizontal.  The quantity $w_r$ is given
by the expression
\be
w_{r}= \exp \left(-\ {1 \over \sigma_{PIL}^2}
\biggl|{B_{r} \over \vecB_h}\biggr|^2\right) ,
\label{eqn:wr}
\ee
where $B_r$ is the radial magnetic field component and $\vecB_h$ are the
horizontal magnetic field components.  The empirical factor $\sigma_{PIL}$
is the same empirical factor (typically unity) used in the above discussion
of the electric field from the Doppler velocity.

To avoid including inductive electric fields that are already
accounted for by $\vecE^P$, we remove any inductive contributions by writing
the FLCT-derived electric field in terms of the gradient of a scalar 
potential $\psi^F$:
\be
c \vecE^F = -\grad \psi^F .
\label{eqn:efrompsif}
\ee
To derive an equation for $\psi^F$, we can take the horizontal divergence of
$c \vecE^F$ and set it equal to the divergence of $(1-w_r) c \vecE_{FLCT}$ 
where $c \vecE_{FLCT}$ is taken from equation (\ref{eqn:efromvh}):
\be
\grad_h^2 \psi^F = \grad_h \cdot ( (1-w_r) \vecV_h^F \times B_r \vecrhat ) .
\label{eqn:flctpoisson}
\ee
Once this Poisson equation is solved for $\psi^F$ , 
$c \vecE^F$ can be evaluated from equation (\ref{eqn:efrompsif}).

\subsection{``Ideal'' Corrections to the Electric Field Solutions}
\label{sec:Ideal}

Most of the time, we expect electric fields in the solar photosphere
will be largely determined by the electric field in ideal MHD,
namely $c \vecE = -\vecV \times \vecB$, where $\vecV$ is the local plasma
velocity.  A consequence of this is that we
expect $\vecE \cdot \vecB = 0$.  However, we found that if the
PTD (inductive) contribution, Doppler
contribution, and FLCT contribution are added together, the resulting electric
field can have a significant component parallel to the direction of
$\vecB$.  We therefore want to find a way to add a scalar potential electric
field
\be
c \vecE^I = - \grad \psi^I,
\label{eqn:eidealdef}
\ee
such that
\be
\grad \psi^I \cdot \vecB = c \vecE^{PDF} \cdot \vecB ,
\label{eqn:eideal}
\ee
where $\vecE^{PDF}$ is the sum of the PTD (inductive), Doppler, and 
FLCT electric field contributions.  When $\vecE^I$ is added to the electric
field, the result should have $\vecE^{PDFI} \cdot \vecB \approx 0$, where
$\vecE^{PDFI}$ is now the complete PDFI electric field solution.
In \S 2.2 of KFW14, the equation that $\psi^I$ and its depth derivative obey
is given in equation (19) of that article.  The form of the equation
is changed only slightly in spherical coordinates, and is
\be
\grad_h \psi^I \cdot \unitv{b}_h + b_r {\partial \psi^I \over \partial r}
= c \vecE^{PDF}_h \cdot \unitv{b}_h + c E^{PDF}_r b_r .
\label{eqn:eidealpsi}
\ee
Here, $\unitv{b}$ is the unit vector pointing in the direction of $\vecB$,
and $b_r$ and $\unitv{b}_h$ are the radial and horizontal components of
$\unitv{b}$, respectively.
As described in \S 2.2 of KFW14, equation (\ref{eqn:eidealpsi}) is
solved using the ``iterative'' technique, also mentioned earlier in
\S \ref{sec:doppler} of this article, with further details given in
\S \ref{sec:relaxation}.
Once $\psi^I$ and $\partial \psi^I / \partial r$ have been found, the full
PDFI solution is given by
\be
c \vecE^{PDFI} = c \vecE^{PDF}_h - \grad_h \psi^I + \vecrhat \left( c E^{PDF}_r
- {\partial \psi^I \over \partial r} \right)
\label{eqn:pdfi} .
\ee
It is important to note that this same procedure to ``perpendicularize'' 
$\vecE$ with respect to $\vecB$ can be performed with any combination of
the other electric field contributions.  For example, in cases where there are
no Doppler or FLCT velocity flows available, one can substitute $\vecE^P$ for
$\vecE^{PDF}$ in equations (\ref{eqn:eidealpsi} - \ref{eqn:pdfi}) 
to generate an electric field
solution that should still minimize $\vecE \cdot \vecB$.  
This is described in detail in \S 2.3.4 of KFW14 (see also Table 1 of KFW14).

Once the PDFI electric fields have been computed, we can use them to estimate
the Poynting flux of energy in the radial direction:
\be
S_r = \unitv{r} \cdot {1 \over 4 \pi} c \vecE^{PDFI}_h \times \vecB_h .
\label{eqn:poynting}
\ee
We can also compute the Helicity Injection rate contribution function, $h_r$:
\be
h_r = \unitv{r} \cdot 2 c \vecE^{PDFI}_h \times \vecA_P ,
\label{eqn:helden}
\ee
where $\vecA_P = \grad \times P \vecrhat$, and the poloidal potential $P$
is found from equation (\ref{eqn:poisson-pdot}) but without the time 
derivative in the source term.
The relative helicity injection rate
was derived by \citet{Berger1984} in terms of a surface integral of
equation (\ref{eqn:helden}).
\citet{Schuck2019} argue that integrating over a finite area as we do here,
and neglecting the other surfaces of our spherical wedge volume
domain, may result
in a loss of gauge invariance.  For the time being, we ignore this possible
complication, and simply use equation (\ref{eqn:helden}) as an integrand to
estimate the helicity injection rate.
\citet{Berger2018} have pointed out that using the PTD formalism for describing
the magnetic field, which we employ for computing
the inductive contribution to
the electric field, has some useful properties.  The magnetic
helicity in a volume can be understood in terms of the linkage between the
contribution to the magnetic field generated by the poloidal potential with 
that generated from the toroidal potential.
That analysis also leads
to the same surface integral of the quantity in equation (\ref{eqn:helden}) for
the relative helicity injection rate, assuming that the volume integral of
$\vecE \cdot \vecB$ is zero if the coronal plasma is described by ideal MHD,
and that the contributions from the other surfaces surrounding the
volume can be neglected.

KFW14 studied the accuracy of the PDFI
electric field solutions, and the Poynting flux and
Helicity injection rates, for the case of an MHD
simulation of magnetic flux emergence in a convecting medium, originally
described by \citet{Welsch2007}.  They found that including the PTD,
Doppler, FLCT, and Ideal electric field contributions (in other words, the
full PDFI electric field) resulted in the most
accurate reconstruction of the electric field from the MHD simulation.
The comparison is described in detail in \S 4 of KFW14.  
For details of tests
of the PDFI solution from these simulation data
using PDFI\_SS, see \S \ref{sec:xanmhd}.

\section{Numerical Solution of PDFI Equations in PDFI\_SS}
\label{sec:pdfinum}

The fundamental mathematical
operations of the PDFI software are the solution of the Poisson
equation in finite domains in a two-dimensional geometry, where the source
terms depend on observed data, and then the evaluation of
electric field contributions that are either the curl of the solution 
multiplying $\vecrhat$, or the gradient of the solution in 
two dimensions.
Since solving the Poisson equation plays such a central role in PDFI\_SS, 
it is worth noting and describing the software we have chosen.

In KFW14, and in this article, we have chosen version 4.1 of the
FISHPACK Fortran library to perform the needed solutions of the Poisson 
equation.  FISHPACK is a package that was developed at NCAR many years ago
for the general solution of elliptic equations in two and three dimensions.  
We find the code is extremely efficient and accurate, and includes many 
different possible boundary conditions that are ideally suited 
for use on the PDFI problem.  In KFW14, we used subroutines from the package 
that were designed for Cartesian geometry; for PDFI\_SS, we use subroutines
from FISHPACK
designed for the solution of Helmholtz or Poisson equations in sub-domains 
placed on the surface of a sphere.  The partial
differential equations are approximated by
second-order accurate finite difference equations; the solutions FISHPACK finds
are of the corresponding finite difference equations.

Version 4.1 of FISHPACK can be downloaded from NCAR at
\url{https://www2.cisl.ucar.edu/resources/legacy/fishpack/documentation}.
A copy of the tarball for version 4.1 can also be downloaded from
\url{http://cgem.ssl.berkeley.edu/~fisher/public/software/Fishpack4.1/}.
The source code has well-documented descriptions of all the
calling arguments used by the subroutines contained in the software.  
A very useful document describing
an older version of the software is the NCAR Tech Note IA-109
\citep{Swarztrauber1975},
which contains valuable technical information about FISHPACK that is not 
described elsewhere.
The numerical technique of ``Cyclic Reduction'' for solving the Poisson 
equations in general, and on the surface of a sphere, is described by 
\citet{Sweet1974,Swarztrauber1974,Schumann1976}.

A notable feature of the FISHPACK software for Cartesian and spherical
domains is that there exist different subroutines for solving Poisson
equations that use either a centered or a staggered grid assumption, that is
whether the equations are solved at the vertices of cells, or
centers of cells, respectively.
This ability is very useful for PDFI\_SS, as will be described in further
detail below in the remainder of
\S \ref{sec:pdfinum}.

\subsection{FISHPACK Domain Assumptions and Nomenclature Used in PDFI\_SS}
\label{sec:fishconv}

The Helmholtz/Poisson equation subroutines for spherical coordinates in FISHPACK
are named \texttt{HSTSSP} (the staggered grid case), and 
\texttt{HWSSSP} (the centered grid
case).  Important input arguments to these subroutines include the source
term for the Poisson equation (a two-dimensional array), and 
boundary conditions applied to the four
edges of the problem domain (four one-dimensional arrays).  
Note that the FISHPACK software assumes that the Poisson equation
is multiplied by $r^2$ (where $r$ is the radius), meaning that the 
source terms of all Poisson equations in PDFI\_SS
must also multiplied by $r^2$ before calls to \texttt{HSTSSP} 
and \texttt{HWSSSP}.  
The boundary conditions most
useful to PDFI\_SS include the specification of derivatives of the solution
in the directions normal to the domain boundary edges 
(Neumann boundary conditions).
The problem domain is described further below.
Both of these subroutines make certain assumptions about the geometry
of the domain, the array dimensions, and how the arrays are ordered.  To avoid
confusion, we adopt exactly the same nomenclature that is used in FISHPACK
throughout our software
to describe the domain, its boundaries, and the grid spacing.

Spherical coordinates in FISHPACK assume spherical polar coordinates,
with the first independent variable $\theta$ being colatitude, and the second 
independent variable $\phi$ being azimuthal angle (longitude).
See \S \ref{sec:transpose}, and the figures in
\S \ref{sec:stagger} for a comparison between
spherical polar coordinates and the longitude-latitude coordinate system
typically used to display magnetic field data.

Both colatitude and longitude are measured in radians.  Colatitude $\theta$
ranges from $0$ (North Pole) to $\pi$ (South Pole).  Longitude $\phi$ 
ranges from
$0$ to $2 \pi$, with negative values of longitude not allowed within
these two subroutines.
The finite difference approximations to the Poisson equations
are solved in a domain where the edges of the domain are defined by lines of
constant colatitude and constant longitude.  The northern edge of the
domain is defined by $\theta = a$, and the southern edge of the domain by
$\theta = b$, where $0 < a < b < \pi$.  The left-most edge of the domain 
is defined by $\phi = c$,
and the right-most edge of the domain is defined by $\phi = d$, where
$0 < c < d < 2 \pi$.

In the colatitude direction, there are a finite number of cells, denoted
$m$.  In the longitude direction, there are a finite number of cells, denoted
$n$.  The angular extent of a cell in each direction is assumed constant,
and these angular thicknesses are given by these expressions:
\be
\Delta \theta = {(b - a) \over m} ,
\label{eqn:dtheta}
\ee
and
\be
\Delta \phi = {(d - c) \over n} .
\label{eqn:dphi}
\ee
If we are using the staggered grid case, the number of variables in each
direction is the number of cell-centers; so in this case, solution arrays
(without ghost cells)
will have dimensions of $(m,n)$.  If we are assuming the centered grid case,
the number of variables in each direction is the number of cell edges, which
is one greater than the number of cell-centers.  In that case, the solution
arrays (without ghost zones) will have dimensions of $(m+1,n+1)$.

The quantities $a$, $b$, $c$, $d$, $m$, $n$, $\Delta \theta$, and $\Delta \phi$
will retain the meanings defined here throughout the rest of this article.

\subsection{Transposing Between Solar and Spherical-Polar Array Orientation}
\label{sec:transpose}

Given the implicit assumption in FISHPACK of spherical-polar coordinates
(colatitude and longitude), and the default assumption used nearly
universally in Solar Physics of longitude-latitude array orientation, 
we are led immediately to the need for frequently transforming back and forth 
between 
longitude-latitude and colatitude-longitude array orientations.
Thus an important part of the PDFI\_SS software consists of the ability to
perform these transpose operations easily and routinely.  Detailed 
discussion of 
the subroutines that perform these operations is described in 
\S \ref{sec:interpol}
of this article; here we simply present a high-level view of where 
these transpose operations must be done.

First, if we are using vector magnetogram and Doppler data from HMI on SDO,
these data are automatically provided in longitude-latitude orientation.
Therefore, a first step is to transpose all the input data (vector magnetograms,
velocity maps, line-of-sight unit vectors) to colatitude-longitude orientation.
Then the PDFI solutions are obtained using FISHPACK software in 
colatitude-longitude order.  Essentially all mathematical operations on
the data and Electric field solutions are done in colatitude-longitude 
orientation.  Finally, because users
expect the solutions to be in the same orientation as the HMI data, we
must transpose computed results
in the other direction to provide the electric field 
solutions and other related quantities in longitude-latitude order.

For further details, see \S \ref{sec:interpol}.

\subsection{Advantages of Using a Staggered Grid Over a Centered Grid 
for PDFI}
\label{sec:advstag}

In KFW14, we used a centered grid definition for finite difference expressions
for first derivatives (equations 14-15 in KFW14)
in the horizontal directions in our definitions for the curl
and gradient.  On the other hand, we also used a standard five point expression
for the Laplacian (equation 16 in KFW14,
also used by the Cartesian FISHPACK Poisson solver), 
which uses centered grid expressions for second derivatives.
However, equation (16) from KFW14 implicitly uses first derivative finite 
difference expressions that are centered half a grid point away from the 
central point.
This means that there is an inconsistency between equation (16) and equations
(14-15) in KFW14.  This inconsistency shows up when one uses the centered
finite difference expressions to evaluate
$\veczhat \cdot ( \grad \times \grad \times \Pdot \veczhat )$ and compares
it to $-\grad_h^2 \Pdot$ using equation (16) in KFW14.  In the continuum 
limit, the two expressions should be identical, but the finite difference 
approximations are not equal; the double curl expression using centered 
finite differences for first derivatives yields an expression like 
equation (16) of KFW14, but
with the gridpoints separated by $2 \Delta x$ and $2 \Delta y$.  If the
grid resolves the solution well, the two different expressions will not differ
greatly.  This in fact is the case with the ANMHD simulation data analyzed 
in KFW14.  But if the solution has structure on the same scale as the grid
separation, the double curl expression and the horizontal Laplacian expression
can differ significantly, rendering solutions to the PDFI equations 
quite inaccurate.
This problem exists for both the Cartesian and spherical versions of the PDFI
equations.

\begin{figure}[ht!]
\includegraphics[width=3.75in]{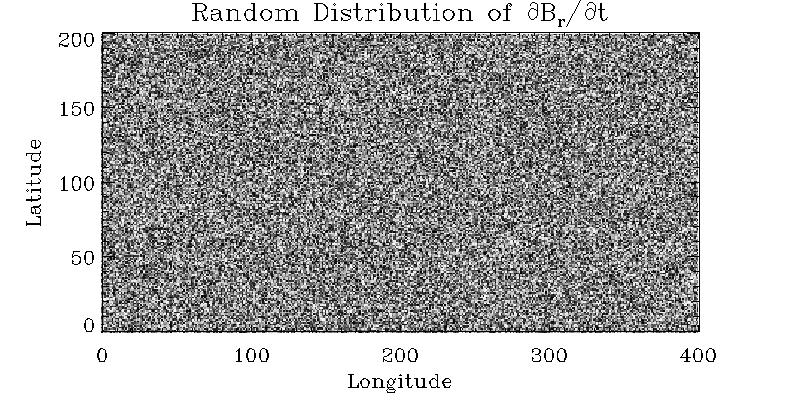}
\caption{Figure shows the input distribution of the
test $\dot B_r$ configuration, displayed in longitude-latitude order.  
This case
has $m=200$, $n=400$, $a = \pi/2 - 0.1$, $b = \pi/2 + 0.1$, $c=0$, 
and $d = 0.4$.
The input distribution is a field of random numbers distributed between $-0.5$
and $0.5$. \label{fig:brtvscurle}}
\end{figure}

\begin{figure}[ht!]
\hspace{-0.20in}
\includegraphics[width=3.7in]{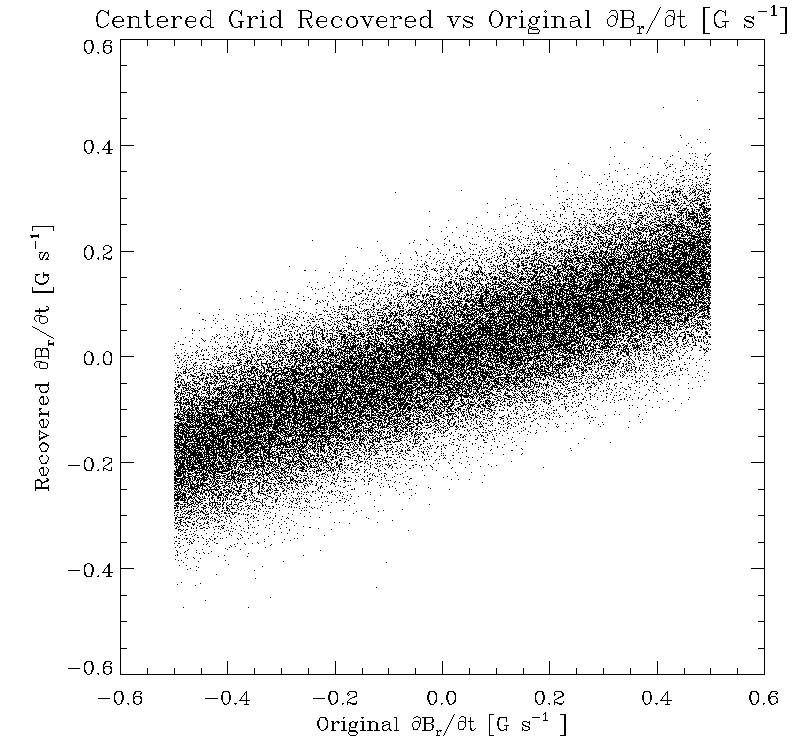}
\caption{The recovered field of $\dot B_r$ plotted versus the input field of
$\dot B_r$ for the centered grid case shown in Figure \ref{fig:brtvscurle}. \label{fig:scatterbrcenter}}
\end{figure}

\begin{figure}[ht!]
\hspace{-0.20in}
\includegraphics[width=3.7in]{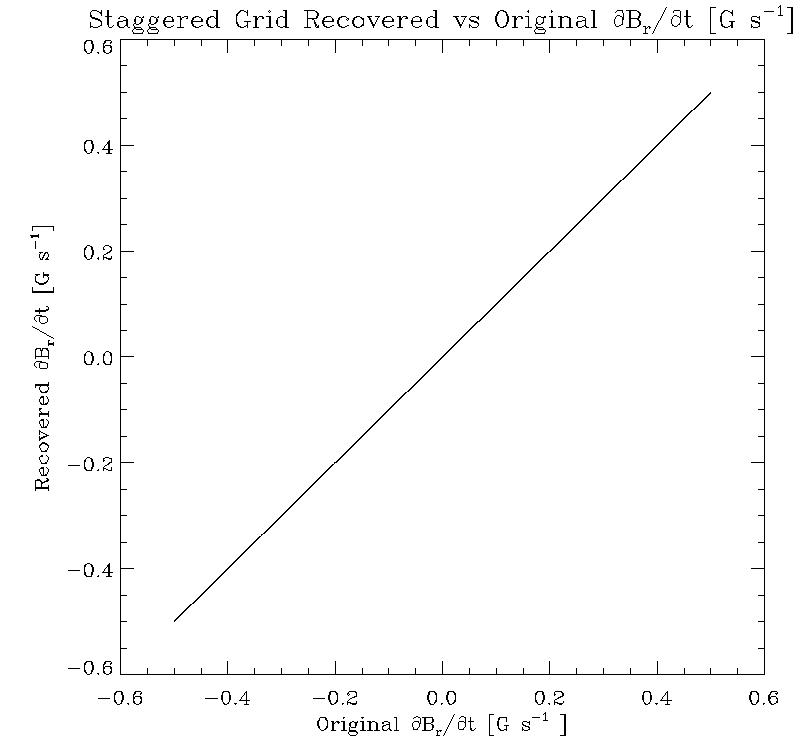}
\caption{The recovered field of $\dot B_r$ plotted versus the input field of
$\dot B_r$ for the staggered grid test case. \label{fig:brtestss}}
\end{figure}

\begin{figure}[ht!]
\hspace{-0.15in}
\includegraphics[width=3.55in]{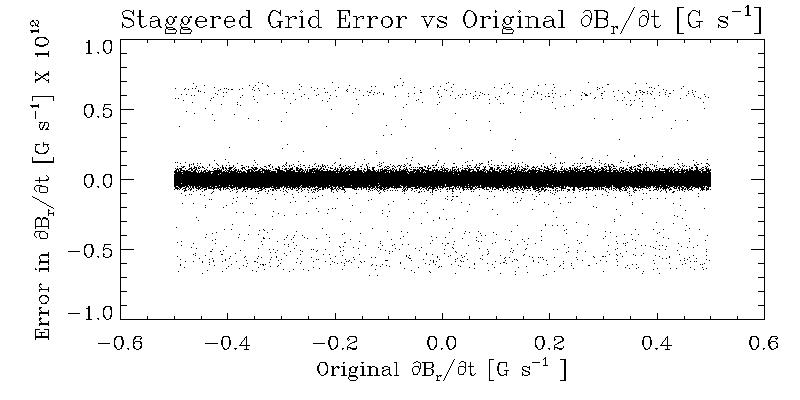}
\caption{Difference between recovered $\dot B_r$ and input $\dot B_r$ as
a function of the input $\dot B_r$ for the staggered grid case.  
\label{fig:errorbrstag}}
\end{figure}

To illustrate the problem quantitatively, we have constructed a solution 
in spherical coordinates to
the Poisson equation (\ref{eqn:poisson-pdot}) using a centered grid formulation
of the finite differences, and then evaluated $c \vecE_h$ from the horizontal
components of equation (\ref{eqn:Eind}), and then evaluated
$-\vecrhat \cdot ( \grad \times c \vecE_h )$, which should be equal to the input
source term, $\dot B_r$.  In this case, the input source term is taken to
be a field of random numbers ranging from $-0.5$ to $0.5$, which has 
significant structure on the scale of the grid.  
Figure (\ref{fig:brtvscurle}) shows an image of
the original field of the assumed
$\dot B_r$. 
Figure \ref{fig:scatterbrcenter} shows a
point-by-point scatterplot of the ``recovered'' versus original values of 
$\dot B_r$, showing about half the correct slope, and random
errors of roughly 50\%.  The recovered values of
$\dot B_r$ were computed as described above using
the centered grid finite difference expressions.  Examination of
Figure \ref{fig:scatterbrcenter}
shows that the centered grid finite difference
expressions do a poor job of describing the correct solution for this test
case.  The behavior of this test case is similar to what we might expect if
the source term has significant levels of pixel-to-pixel noise, which 
is the case with real magnetogram data in weak-field regions.

By changing the definition of how finite difference approximations to 
spatial derivatives are defined, and where different variables are located
within the grid, we can improve this behavior dramatically.  
In the centered grid case, all variables are co-located at the same grid 
points.  By defining the radial magnetic field and its time derivative to
lie at the centers of cells, with the electric field components lying on the
edges (or ``rails'') surrounding the cells, the 
finite difference approximations to the 
derivatives can be made to obey Faraday's law to floating-point roundoff
error.

Figure \ref{fig:brtestss} shows the analagous scatterplot shown in Figure
\ref{fig:scatterbrcenter}, but using the staggered grid definition described
above.  The relationship is a straight line.  Figure \ref{fig:errorbrstag}
shows the difference between the recovered and original values of $\dot B_r$.
Note that the amplitude of the error 
is multiplied by $1 \times 10^{12}$, so that the error term is visible in the
scatterplot.  
These two plots clearly show that solutions for the
electric fields in a staggered grid formulation can do a far better job of
representing the observed data than can the centered grid formulation.

These figures motivate the development of our more detailed staggered grid 
formalism, which is described in \S \ref{sec:stagger}.

\subsection{The Staggered Grid Formulation for PDFI\_SS}
\label{sec:stagger}

In three dimensions, \citet{Yee1966} worked out a second-order accurate
finite difference formulation for Maxwell's equations, pointing out that 
if one places different
variables into different locations within the grid, that the governing
continuum equations (the curls in Maxwell's equations) become conservative 
when written down in a finite 
difference form.  
In recent years, ``Mimetic Methods'' have been developed, which are higher
order analogues to the Yee grid, in
that different variables are defined at different locations within a voxel
(such as at interiors, faces or edges), and with some internal structure 
in these voxel sub-domains allowed.  The locations of the variables depend on 
which integral conservation law is being applied.
(see $e.g.$ \citet{Candelaresi2014} and references therein).
For our work, the second-order accurate Yee grid is sufficient.
The Yee grid is the basis for the numerical implementation
of the MF code described by \citet{Cheung2012}.  

In PDFI\_SS, we have a slightly
different situation, where most of the calculations are defined on a subdomain
of a spherical surface, so that the domain is two-dimensional, rather than
three-dimensional.  Nevertheless, the exercise shown in \S \ref{sec:advstag}
shows that we want to use the advantages of a staggered grid description
of the finite difference equations, which is inspired by
the Yee grid.  This is complicated by the fact that
in addition to needing curl contributions to the electric
field (see \S \ref{sec:ptd}), we also need to represent gradient 
contributions to the electric field (\S \ref{sec:doppler} - \ref{sec:Ideal}).
This must be done in such a way that both contributions are co-located along 
the rails that surround a cell-center.  
We found a way 
to satisfy these constraints with a specific staggered grid arrangement of the 
physical and mathematical variables within our two-dimensional domain,
summarized below.

\begin{figure}[ht!]
\includegraphics[width=3.5in]{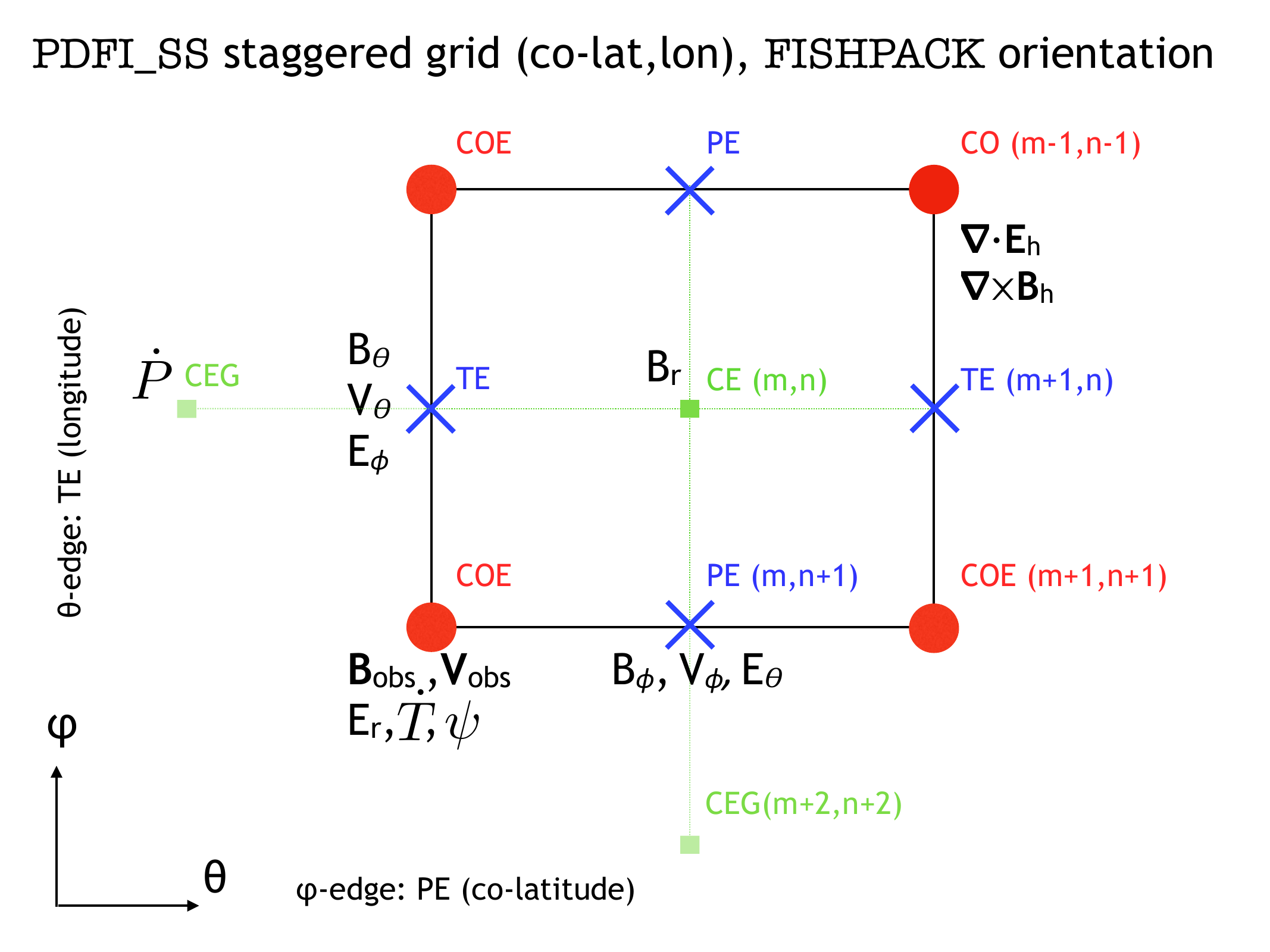}
\caption{Schematic diagram of our staggered grid, based on the Yee grid
concept, oriented in spherical polar (colatitude-longitude) orientation.  
This Figure shows the grid near the left-most, northern domain corner, oriented
in the $\theta - \phi$ (colatitude-longitude) directions.  The $x-$ axis 
increases in the colatitude direction, and the $y-$ axis increases in 
the longitude direction. The CE, CEG, CO, COE, TE,
and PE grid locations are shown, along with where some of the physical variables
are located on these grids. \label{fig:gridtp}}
\end{figure}

\begin{figure}[ht!]
\includegraphics[width=3.5in]{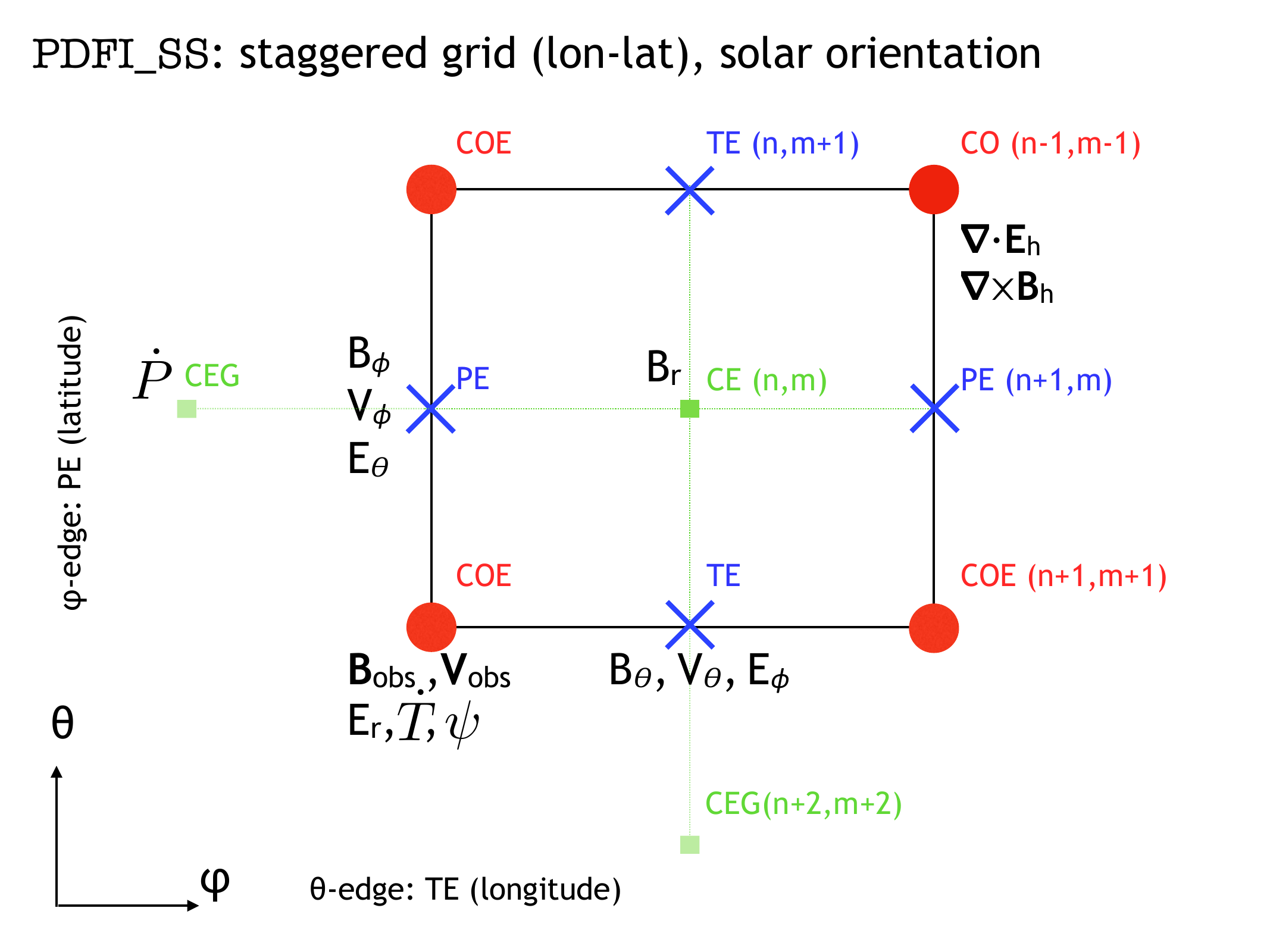}
\caption{Schematic diagram of our staggered grid, based on the Yee grid
concept, oriented in longitude-latitude ($i.e.$ ``Solar'') orientation.  
This Figure shows the grid near left-most and southern domain corner.
The $x-$ axis increases in the longitude direction, and the $y-$ axis increases
in the latitude direction.
The CE, CEG, CO, COE, TE,
and PE grid locations are shown, along with where some of the physical variables
are located on these grids. \label{fig:gridll}}
\end{figure}

%
First, we define six different grid locations for variables in PDFI\_SS,
illustrated schematically in Figures \ref{fig:gridtp} and \ref{fig:gridll}.
For the moment, we use colatitude-longitude array index order in this
discussion.  We define the CE grid locations as being the centers of
the two-dimensional cells; the CE grid variables are dimensioned $(m,n)$.
Next, we define the interior corner grid locations, the ``CO'' grid, as 
residing at all the corners, or vertices of the cells, but specifically not
including the vertices that lie along the domain edges.  Variables lying 
on the CO grid will have dimension $(m-1,n-1)$.  Next, we define the ``COE''
grid, which is also located along corners of cells, but in this case, the
corners that lie along the domain edges are included.  Variables lying on
the COE grid will have dimension $(m+1,n+1)$.  Variables that lie along
cell edges that have constant values of $\phi$ (or longitude) but are at 
midpoints in $\theta$ (or colatitude) lie on the ``PE'' (phi-edge) locations
of the domain.
Variables at PE locations have dimension $(m,n+1)$.  Variables that lie
along edges with constant $\theta$ (or colatitude) but are at midpoints in
$\phi$ lie on the ``TE'' (theta-edge) grid locations.  Variables at TE
locations have dimension $(m+1,n)$.  Finally, if we are describing these
grid locations, but using longitude-latitude index order, the dimensions
of the variables are just the reverse of the dimensions given above.

Here are some examples of where different physical and mathematical variables
are located using these grid definitions:  $B_r$ and $\dot {B_r}$ are located
on the CE grid; $B_{\theta}$ and $E_{\phi}$ are located on the TE grid;
$B_{\phi}$ and $E_{\theta}$ are located along the PE grid; $E_r$ and $\dot T$
are located along the COE grid, and $\vecrhat \cdot \grad_h \times \vecB_h$
is located along the CO grid.  The scalar potentials defined in the PDFI
equations (\S \ref{sec:doppler}-\ref{sec:Ideal}) are located on the COE grid.
The poloidal potential $\dot P$ is in principle
located along the CE grid (with dimensions $(m,n)$), but we find it convenient
to add ghost zones to $\dot P$ to implement Neumann (derivative specified)
boundary conditions.  When that
is done, we refer to this grid as a ``CEG'' grid (CE plus ghost zones).
$\dot P$ is on the CEG grid and has dimensions of $(m+2,n+2)$.

The placement of variables into the staggered grid locations described above
is very similar
to the placement used in the ``constrained transport'' MHD model of
\citet{Stone1992a,Stone1992b}, and the filament construction model of
\citet{vanBallegooijen2004}.
Table \ref{tab:gridlocations} contains a list of variables
in PDFI\_SS, and where they reside in terms of these grid locations.

\begin{table}[h!]
\renewcommand{\thetable}{\arabic{table}}
\centering
\caption{PDFI\_SS Variable Locations and Dimensions} \label{tab:gridlocations}
\begin{tabular}{c c c c}
\tablewidth{0pt}
\hline
\hline
Quantity & Grid & Dimension (tp) & Dimension (ll)\\
\hline
Input Data & COE & $(m+1,n+1)$ & $(n+1,m+1)$ \\
$B_r, \dot B_r$ & CE     & $(m,n)$  & $(n,m)$   \\
$B_{\theta}, \dot B_{\theta}$ & TE & $(m+1,n)$ & $(n,m+1)$ \\
$B_{\phi}, \dot B_{\phi}$ & PE & $(m,n+1)$ & $(n+1,m)$ \\
$V_{\theta}$ & TE & $(m+1,n)$ & $(n,m+1)$ \\
$V_{\phi}$ & PE & $(m,n+1)$ & $(n+1,m)$ \\
$V_{LOS}$ & COE & $(m+1,n+1)$ & $(n+1,m+1)$ \\
$\unitv{\ell}_{r,\theta,\phi}$ & COE & $(m+1,n+1)$ & $(n+1,m+1)$ \\
$E_r$ & COE & $(m+1,n+1)$ & $(n+1,m+1)$ \\
$E_{\theta}$ & PE & $(m,n+1)$ & $(n+1,m)$ \\
$E_{\phi}$ & TE & $(m+1,n)$ & $(n,m+1)$ \\
$\dot P$ & CEG & $(m+2,n+2)$ & $(n+2,m+2)$ \\
$\partial \dot P / \partial r$ & CEG & $(m+2,n+2)$ & $(n+2,m+2)$ \\
$\dot T$ & COE & $(m+1,n+1)$ & $(n+1,m+1)$ \\
$\psi$ & COE & $(m+1,n+1)$ & $(n+1,m+1)$ \\
$\vecrhat \cdot \grad \times \vecE$ & CE & $(m,n)$ & $(n,m)$ \\
$\vecrhat \cdot \grad \times \dot \vecB$ & CO & $(m-1,n-1)$ & $(n-1,m-1)$ \\ 
$\grad_h \cdot \dot \vecB_h$ & CE & $(m,n)$ & $(n,m)$ \\
$\grad_h \cdot \vecE_h$ & CO & $(m-1,n-1)$ & $(n-1,m-1)$ \\
$S_r$ & CE & $(m,n)$ & $(n,m)$ \\
$H_m$ & CE & $(m,n)$ & $(n,m)$ \\
$M^{COE}$ & COE & $(m+1,n+1)$ & $(n+1,m+1)$ \\
$M^{CO}$ & CO & $(m-1,n-1)$ & $(n-1,m-1)$ \\
$M^{TE}$ & TE & $(m+1,n)$ & $(n,m+1)$ \\
$M^{PE}$ & PE & $(m,n+1)$ & $(n+1,m)$ \\
$M^{CE}$ & CE & $(m,n)$ & $(n,m)$ \\
\hline
\end{tabular}
\end{table}

\subsection{Units assumed by PDFI\_SS software library}
\label{sec:units}

It is assumed by the PDFI\_SS library that all magnetic field components on
input to the library subroutines are in units of Gauss ([G]).  Units of length
are determined by the radius of the Sun, and which is assumed to be given in 
kilometers ([km]).  For solar calculations in spherical coordinates, we
expect $R_{\odot}$ to be $6.96 \times 10^5$ km, although in the software
the radius of the Sun is an input parameter that can be set by the user.
Units of time are assumed to be in seconds ([s]).  Velocities are assumed
to be expressed in units of [km s$^{-1}$].  For the ``working'' units
of the electric field, the electric field is evaluated as $c \vecE$, 
$i.e.$ the speed of light times the electric field vector, with
each component having units of [G km s$^{-1}$]. The
subroutines that compute the Poynting flux and the Helicity injection rate
contribution function are exceptions to this rule, and
assume that electric field components on input are expressed
in units of volts per cm ([V cm$^{-1}$]).  To convert from
[G km s$^{-1}$] to [V cm$^{-1}$], one can simply divide by $1000$.  To convert
units in the opposite direction, one would multiply by $1000$.

\subsection{Time derivatives, Transpose, Interpolation, and 
Masking Operations in PDFI\_SS}
\label{sec:interpol}

We mentioned in \S \ref{sec:transpose} that transpose operations from
longitude-latitude array orientation to spherical-polar coordinates (and 
the reverse) would need to be done frequently.  Now that we have 
introduced our staggered grid
definitions, we will describe in detail how these operations are done, as
well as how the interpolation from the input data grid to the 
staggered grid locations is done.  We will discuss how time derivatives 
are estimated,
and the calculation of the strong magnetic field masks,
designed to decrease the effects of noise from the magnetic
field measurements in weak-field regions on the electric field solutions.

The source terms for the PTD contribution to the electric field 
(\S \ref{sec:ptd}) consist
of time derivatives of magnetic field components.  To estimate these time
derivatives from the data, we simply difference the magnetic field values
at their staggered grid locations between two adjacent measurement times,
and divide by the cadence time period, $\Delta t$.  Thus if we have
magnetic field measurements at times $t_0$ and $t_1 = t_0+\Delta t$, 
then our electric 
field solution will be evaluated at time $t_0 + \half \Delta t$, and will
be assumed to apply over the entire time interval between $t_0$ and 
$t_1$.  Furthermore, we assume that the magnetic field values
needed to evaluate the other electric field contributions (\S \ref{sec:doppler}
- \S \ref{sec:Ideal}) will be the magnetic field values at 
$t_0 + \half \Delta t$, which will be an average of the input values at the
two times.  Similarly, the other input variables that affect the calculation
of $\vecE$ will also be an average of the variables at the two adjacent times.
If our electric field solutions are conservative, and accurately obey Faraday's
Law, then the computed electric field solutions should correctly evolve
$\vecB$ from $t_0$ to $t_0 + \Delta t = t_1$ with minimal error.

Thus for a single time step, the needed 
input data to evaluate the PDFI
solutions are arrays of $B_r$, $B_{\theta}$, $B_{\phi}$, $V_{LOS}$, 
$V_{\theta}$, $V_{\phi}$, $\ell_r$, $\ell_{\theta}$, and $\ell_{\phi}$
at two adjacent measurement times, for a total of 18 input arrays.  
Because of the FISHPACK spherical
coordinate solution constraints, the data will have to be evaluated using
equally spaced colatitude and longitude grid separations, meaning 
constant spacing
in $\Delta \theta$ and $\Delta \phi$, referred to as a ``Plate Carr\'ee'' 
grid.  In the case of HMI data from SDO,
this is one of the standard mapping outputs for the magnetic field and Doppler
measurements.  For CGEM calculations of the electric field supported by
the SDO JSOC, the values of $\Delta \theta$ and $\Delta \phi$ are set to
$0.03 ^\circ$ in heliographic coordinates (converted to radians), coinciding 
closely with an HMI
pixel size near disk center.  The PDFI\_SS software can accommodate
values of $\Delta \theta$ and $\Delta \phi$ that differ, but the FLCT software 
used upstream of PDFI\_SS needs to have these values equal to one another.  The
JSOC software produces Plate Carr\'ee data with $\Delta \theta = \Delta \phi$.

We now briefly digress to describe the relationship between the mathematical
coordinate system used by PDFI\_SS, with angular domain
limits $a$, $b$, $c$, and $d$,
and the standard WCS keywords \texttt{CRPIX1}, \texttt{CRPIX2},
\texttt{CRVAL1}, \texttt{CRVAL2}, \texttt{CDELT1}, and
\texttt{CDELT2} that describe the
position of the HMI data on the solar disk \citep{Thompson2006}.  
We want the ability to
concisely relate these two descriptions to each other.
The quantities \texttt{CRPIX1} and \texttt{CRPIX2} denote longitude and
latitude reference pixel 
locations
(the center of the Field of View measured from the lower left pixel at (1,1)),
\texttt{CRVAL1} and \texttt{CRVAL2} the
longitude and latitude (in degrees)
of the reference pixel, and \texttt{CDELT1} and
\texttt{CDELT2}, the number of degrees in longitude and latitude between 
adjacent pixels.  From the above description, we expect that \texttt{CDELT1}
and \texttt{CDELT2} will be equal to $0.03^{\circ}$ per pixel.
We have written three subroutines,\\
\texttt{abcd2wcs\_ss},\\
\texttt{wcs2mn\_ss}, and\\
\texttt{wcs2abcd\_ss},\\ 
the first of which converts $a$, $b$, $c$, $d$, $m$, and $n$ to the
WCS keywords \texttt{CRPIX1}, \texttt{CRPIX2}, \texttt{CRVAL1}, \texttt{CRVAL2},
\texttt{CDELT1}, and \texttt{CDELT2}; and in the reverse direction, 
\texttt{wcs2mn\_ss}
which finds $m$ and $n$ 
from \texttt{CRPIX1} and \texttt{CRPIX2} for the COE grid, and 
\texttt{wcs2abcd\_ss} which converts
the keywords \texttt{CRVAL1}, \texttt{CRVAL2}, \texttt{CDELT1}, and
\texttt{CDELT2} to $a$, $b$, $c$, and $d$.
The subroutine \texttt{abcd2wcs\_ss} computes the reference pixel locations
\texttt{CRPIX1} and 
\texttt{CRPIX2} for all 6 grid cases, namely the COE, CO, CE, CEG,
TE, and PE grids.  These results for the reference pixel locations
are returned as six-element arrays, in the order given above.

Returning the discussion to how the input data arrays are processed, 
the data arrays, in longitude-latitude order, are assumed to be
dimensioned $(n+1,m+1)$, with all 9 input arrays for each of the two
times being co-located in space.  The parameter $a$ is the colatitude
of the northernmost points in these arrays, and the parameter $b$ is
the colatitude of the southernmost points in the arrays.  The parameters
$c$ and $d$ are the left-most and right-most longitudes of the input arrays.

The first task is to transpose all 18 arrays from longitude-latitude to
colatitude-longitude (spherical polar coordinates, or $\theta-\phi$ order.)
Basically, the transpose operation looks like
\be
A_{tp}(i,j) = A_{\ell \ell} (j,m-i) ,
\label{eqn:transpose}
\ee
where $j \in [0,n]$, and $i \in [0,m]$, and where 
$A_{tp}$ is the array in $\theta-\phi$
index order, and $A_{\ell \ell}$ is the array in longitude-latitude order.
Here ``tp'' in the subscript is meant as a short-hand for ``theta-phi'',
and ``$\ell \ell$'' is meant as short-hand for ``longitude-latitude''.
An exception is for those arrays that represent the latitude components of
a vector (like $B_{lat}$), in which case when transforming to $B_{\theta}$
the overall sign must also 
be changed since the unit vectors in latitude and colatitude directions point 
in opposite directions.

PDFI\_SS has several
subroutines to perform these transpose
operations (and their reverse operations) on
the COE grid, namely\\
\texttt{brll2tp\_ss}\\
\texttt{bhll2tp\_ss}\\
\texttt{brtp2ll\_ss}\\
\texttt{bhtp2ll\_ss}.\\
Here the subroutines starting with ``\texttt{br}'' perform 
the transpose operation
on scalar fields, while the subroutines starting with ``\texttt{bh}'' perform
the transpose operations on pairs of arrays of the horizontal components
of vectors.  Subroutines containing
the sub-string ``\texttt{ll2tp}'' perform the transpose operation going
from longitude-latitude order to theta-phi (colatitude-longitude) order,
while those with the substring ``\texttt{tp2ll}'' go in the reverse direction.
When going from the input data to colatitude-longitude order, we use the
subroutines containing \texttt{ll2tp} within their name.  When examining
the source code, the expressions will differ slightly from that in equation 
(\ref{eqn:transpose}) to conform with the default Fortran index range (where
index numbering starts from $1$.)

Once the input data arrays on the COE grid have been transposed to
colatitude-longitude order, we then interpolate the data to their staggered
grid locations.  $B_r$ is interpolated to the CE grid, $B_{\theta}$ and
$V_{\theta}$ to the TE grid, $B_{\phi}$ and $V_{\phi}$ to the PE grid.  In
addition, to evaluate the FLCT electric field contribution, we also need to
have $B_r$ and $|\vecB_h|$ interpolated to both the TE and PE grids.  Here,
we use a simple linear interpolation, as given in these
examples for the magnetic field components:
\bea
\lefteqn{B_r (i+\half,j+\half) = {1 \over \sin \theta_i + \sin \theta_{i+1}}
\times \nonumber} \\
& & \left( \half \sin \theta_i ( B_r(i,j) + B_r(i,j+1) ) \right. \nonumber \\
& & \left. + \half \sin \theta_{i+1} ( B_r(i+1,j) + B_r(i+1,j+1) ) \right) ,
\label{eqn:brce}
\eea
\be
B_{\theta}(i,j+\half) = \half \left( B_{\theta}(i,j) + B_{\theta} (i,j+1) 
\right) ,
\label{eqn:btte}
\ee
and
\be
B_{\phi} (i+\half,j) = \half \left( B_{\phi} (i,j) + B_{\phi} (i+1,j) \right) .
\label{eqn:bppe}
\ee
The interpolations from the input data arrays on the COE grid to the
staggered grid locations can be accomplished with the subroutines\\
\texttt{interp\_data\_ss}\\
\texttt{interp\_var\_ss}.\\
The linear interpolation is a conservative choice, and results in a slight
increase in signal to noise if there is a high level of pixel-to-pixel noise
variation.  This interpolation slightly decouples the PDFI\_SS electric field
from the original input data on the COE grid:  The near perfect reproduction
of $\dot B_r$ applies for the interpolations to cell-center, but not 
necessarily for the original input $B_r$ at COE locations.


The Doppler velocity and the LOS unit vector input data arrays are kept
at the COE grid locations, so no interpolation of these data arrays
is necessary.

In addition to interpolating the input data to the staggered grid locations,
we must also construct masks, based on the input data, that reflect regions
of the domain where we expect noise in the magnetic field measurements
will make the electric field calculation unreliable.  In PDFI\_SS
the criterion for masks on the magnetic field variables is determined by
a threshold on the absolute magnetic field strength, including radial
and horizontal components.  The mask value is set to unity if the absolute
value of the magnetic field in the input data is greater than a chosen 
threshold for {\it both} of the timesteps; otherwise the mask value is set 
to zero.
This calculation is done on the COE grid, after the transpose from
longitude-latitude to theta-phi array order.  The subroutine that does this
is\\
\texttt{find\_mask\_ss}.

Subroutine \texttt{find\_mask\_ss} was originally written assuming we were using
data from three separate timesteps, rather than the two timesteps we now use.  
We now simply repeat the array inputs for one of the two timesteps which then
results in the correct behavior.
For HMI vector magnetogram data, we currently use a threshold value
\texttt{bmin} of
250G.  The threshold value is a calling argument to the subroutine, and thus can
be controlled by the user.

We need to have mask arrays for all the staggered grid locations, not just the
COE grid.  To get mask arrays for the CE, TE, and PE locations and array
sizes, we use a two-step process.  First, we use the subroutine
\texttt{interp\_var\_ss} to interpolate the COE mask array to the other
staggered grid locations.  Those interpolated points where input mask values 
transition between zero and one will have mask values that are
between zero and one.  The subroutine\\
\texttt{fix\_mask\_ss}\\
can then be used to set intermediate mask values to either zero or one,
depending on the value of a ``\texttt{flag}'' argument to the subroutine, 
which can be either zero or one.  Setting \texttt{flag} to $0$ is the more 
conservative choice; while setting \texttt{flag} to $1$ is more trusting
of the data near the mask edge values.

Once the strong magnetic field mask arrays have been computed, 
they can be used to multiply
the corresponding magnetic field or magnetic field time derivative arrays
on input to the subroutines that calculate electric field contributions.  This
can significantly reduce the impact of magnetogram noise on the electric field
solutions in weak magnetic field regions of the domain.

We denote the mask arrays coinciding with different grid locations with
the following notation:  $M^{COE}$ is the mask on the COE grid, $M^{CO}$
denotes the mask on the CO grid, and $M^{TE}$, $M^{PE}$, and $M^{CE}$ denote
the masks for the TE, PE, and CE grid locations, respectively.  The mask
arrays are also shown in Table \ref{tab:gridlocations}.

Once electric field solutions have been computed in spherical polar
coordinates, we need the ability to transpose these
arrays, as well as the staggered-grid magnetic field arrays, back to
longitude-latitude order.  Because the array sizes are all slightly different
depending on which grid is used for a given variable, we have written a
series of subroutines designed to perform the transpose operations on our
staggered grid, depending on variable type and grid location.  
There are subroutines to go from theta-phi (colatitude-longitude)  order to
longitude-latitude order, as well as those that go
in the reverse direction.  The subroutines that perform the transpose
operations on staggered grid locations all have the substring ``\texttt{yee}''
in their name.  As before, subroutine 
names that include a substring of ``\texttt{tp2ll}'' transpose the arrays
from theta-phi to longitude-latitude array order, while those with
``\texttt{ll2tp}'' go in the reverse direction.  The subroutines are:\\
\texttt{bhyeell2tp\_ss}\\
\texttt{bryeell2tp\_ss}\\
\texttt{bhyeetp2ll\_ss}\\
\texttt{bryeetp2ll\_ss}\\
\texttt{ehyeell2tp\_ss}\\
\texttt{eryeell2tp\_ss}\\
\texttt{ehyeetp2ll\_ss}\\
\texttt{eryeetp2ll\_ss}.\\

\subsection{Vector Calculus Operations Using the PDFI\_SS Staggered Grid}
\label{sec:veccalc}

Now that we have established how to generate input data on the staggered grid
locations in spherical polar coordinates, we are ready to discuss how to
perform vector calculus operations on that data.  These operations are used
inside the software that evaluates various electric field contributions,
and can also be used to perform other calculations using the electric field
solutions.

The following expressions are the continuum vector calculus operations 
in spherical polar coordinates that
are important in evaluating the PDFI equations of \S \ref{sec:pdfi}:

\be
\grad_h^2 \Psi = {1 \over r^2 \sin\theta} {\partial \over \partial \theta}
\left( {\sin\theta {\partial \Psi \over \partial \theta}} \right) 
+ {1 \over r^2 \sin^2\theta} {\partial^2 \Psi \over \partial \phi^2} ,
\label{eqn:delh2def}
\ee
where $\Psi$ is a scalar function defined in the $\theta, \phi$ domain,
\be
\grad \times {\Psi \vecrhat} =  {1 \over r \sin \theta}
{\partial \Psi \over \partial \phi} \unitv{\theta} - {1 \over r}
{\partial \Psi \over \partial \theta} \unitv{\phi},
\label{eqn:curlpsirhat}
\ee
\be
\grad_h \Psi = {1 \over r}{\partial \Psi \over \partial \theta} \unitv{\theta}
+{1 \over r \sin \theta} {\partial \Psi \over \partial \phi} \unitv{\phi} ,
\label{eqn:gradhpsi}
\ee
\be
\grad_h \cdot \vecU = {1 \over r \sin \theta} {\partial \over \partial \theta}
\left( \sin \theta \  U_{\theta} \right) + {1 \over r \sin \theta}
{\partial U_{\phi} \over \partial \phi} ,
\label{eqn:divhuh}
\ee
where $\vecU$ is an arbitrary vector field and $\grad_h \cdot$ 
represents the divergence in the horizontal directions, and finally
\be
\vecrhat \cdot \grad \times \vecU = {1 \over r \sin \theta}
{\partial \over \partial \theta} \left( \sin \theta \ U_{\phi} \right)
- {1 \over r \sin \theta} {\partial U_{\theta} \over \partial \phi} .
\label{eqn:curlhuh}
\ee
Here $U_{\theta}$ and $U_{\phi}$ are the $\theta$ and $\phi$ components of
$\vecU$.

We now convert these differential expressions to finite difference
expressions, evaluated at various different grid locations in our staggered
grid system.  Many of these expressions must be evaluated separately depending
on where the variables are located, or where we want the expression to be
centered.  For example, we will need to evaluate equation (\ref{eqn:curlhuh})
at both cell centers (the CE grid), and at interior corners (the CO grid),
and the exact expressions will differ depending on where the equations are
centered.

The subroutines that evaluate the finite difference expressions corresponding
to the above equations are:\\
\texttt{curl\_psi\_rhat\_co\_ss}\\
\texttt{curl\_psi\_rhat\_ce\_ss}\\
\texttt{gradh\_co\_ss}\\
\texttt{gradh\_ce\_ss}\\
\texttt{divh\_co\_ss}\\
\texttt{divh\_ce\_ss}\\
\texttt{curlh\_co\_ss}\\
\texttt{curlh\_ce\_ss}\\
\texttt{delh2\_ce\_ss}\\
\texttt{delh2\_co\_ss}.\\
These subroutines are discussed in more detail below.

In the following equations, we'll use this notation to distinguish
quantities lying along an edge, versus halfway between edges:  An index
denoted $i$ or $j$ denotes a location at a $\theta$ or $\phi$ edge, 
respectively, while an index denoted $i+\half$ or $j+\half$ denotes a location
half-way between edges.  For example, a quantity defined on the CO grid
will have indices $i,j$, while a quantity defined on the CE grid will have
indices $i+\half,j+\half$. Similarly, a quantity defined on the TE grid
will have the mixed index notation $i,j+\half$, and one along the PE grid
would have an index notation of $i+\half,j$.

We will start by evaluating equation (\ref{eqn:curlpsirhat}) assuming that
the scalar field $\Psi$ lies on the COE grid.  An example of this case is
evaluating the curl of the toroidal potential $T$ times $\vecrhat$
to find its contribution
to the horizontal components of the magnetic field 
(see equation (\ref{eqn:Bptddef})).  Setting 
$\vecU = \grad \times \Psi \vecrhat$, we find
\be
U_{\theta}(i,j+\half) = { \Psi^{COE}(i,j+1) - \Psi^{COE}(i,j) \over r
\ \sin \theta_i\  \Delta \phi}
\label{eqn:curlpsirhatcotheta}
\ee
for $i \in [0,m]$ and for $j+\half \in [\half,n-\half]$; and
\be
U_{\phi}(i+\half,j) = -\ { \Psi^{COE}(i+1,j) - \Psi^{COE}(i,j) 
\over r \Delta \theta} ,
\label{eqn:curlpsirhatcophi}
\ee
for $i+\half \in [\half,m-\half]$, and for $j \in [0,n]$.
Here, $U_{\theta}$ is dimensioned $(m+1,n)$, and is defined on the TE grid, 
while $U_{\phi}$ is dimensioned $(m,n+1)$ and lies on the PE grid.
To evaluate equations (\ref{eqn:curlpsirhatcotheta}) and
(\ref{eqn:curlpsirhatcophi}),
one can use subroutine \texttt{{curl\_psi\_rhat\_co\_ss}} 
from the PDFI\_SS library.

If $\Psi$ is defined on the CEG grid, this has an array size of
$(m+2,n+2)$, and we then have
\bea
\lefteqn{ U_{\theta}(i+\half,j)=  \nonumber } \\
& & \Psi^{CEG} (i+\half,j+\half) - \Psi^{CEG} 
(i+\half,j-\half)  \over r\ \sin\ \theta_{i+\half} \Delta \phi
\label{eqn:curlpsirhatcetheta}
\eea
for $i+\half \in [\half,m-\half]$, and for $j \in [0,n]$; and
\bea
\lefteqn{U_{\phi}(i,j+\half)=    \nonumber } \\
& & -\ {\Psi^{CEG} (i+\half,j+\half) - \Psi^{CEG} 
(i-\half,j+\half ) \over r \Delta \theta}
\label{eqn:curlpsirhatcephi}
\eea
for $i \in [0,m]$, and for $j+\half \in [\half, n-\half]$.

Here, $U_{\theta}$ is dimensioned $(m,n+1)$, and lies on the PE grid, while
$U_{\phi}$ is dimensioned $(m+1,n)$ lies on the TE grid.
In these expressions, array values of $\Psi^{CEG}$ at $j+\half = - \half$, and
$j+\half = n + \half$ refer to the ghost zone values (with similar
expressions for $i+\half$.)  To evaluate equations 
(\ref{eqn:curlpsirhatcetheta}) and (\ref{eqn:curlpsirhatcephi}), 
one can use subroutine \texttt{curl\_psi\_rhat\_ce\_ss}.

Note that these expressions require the evaluation of $\sin \theta$ at
both edge locations and at cell centers in $\theta$.  The need for these
geometric factors is ubiquitous in PDFI\_SS, so we have written a subroutine\\
\texttt{sinthta\_ss}\\
to pre-compute these array values before calling many of the vector calculus
subroutines.

Turning now to the discretization of
equation (\ref{eqn:gradhpsi}), namely evaluating
$\vecU = \grad_h \Psi$, the finite difference expressions for
$\Psi$ lying on the COE grid are
\be
U_{\theta}(i+\half,j) = 
{ \Psi^{COE}(i+1,j) - \Psi^{COE}(i,j) \over r \Delta \theta} ,
\label{eqn:gradhpsicoet}
\ee
where $i+\half \in [\half,m-\half]$, and $j \in [0,n]$, and
\be
U_{\phi}(i,j+\half) = { \Psi^{COE}(i,j+1) - \Psi^{COE}(i,j) \over r
\ \sin \theta_i\  \Delta \phi} ,
\label{eqn:gradhpsicoep}
\ee
where $i \in [0,m]$, and $j+\half \in [\half,n-\half]$.
In this case $U_{\theta}$ is dimensioned $(m,n+1)$ and lies on the PE grid,
while $U_{\phi}$ is dimensioned $(m+1,n)$ and lies on the TE grid.  These
two equations are relevant for computing electric field contributions from
the gradients of scalar potentials; subroutine \texttt{gradh\_co\_ss} can
be used to compute these arrays.

When $\Psi$ lies on the CEG grid, we have
\bea
\lefteqn{ U_{\theta}(i,j+\half)=  \nonumber } \\
& &  {\Psi^{CEG} (i+\half,j+\half) - \Psi^{CEG}
(i-\half,j+\half ) \over r \Delta \theta}
\label{eqn:gradhpsicegt}
\eea
for $i \in [0,m]$, and $j+\half \in [\half,n-\half]$, and
\bea
\lefteqn{U_{\phi}(i+\half,j)=    \nonumber } \\
& & \Psi^{CEG} (i+\half,j+\half) - \Psi^{CEG}
(i+\half,j-\half)  \over r\ \sin\ \theta_{i+\half} \Delta \phi
\label{eqn:gradhpsicegp}
\eea
for $i+\half \in [\half,m-\half]$, and 
$j \in [0,n]$.  Here $U_{\theta}$ is dimensioned
$m+1,n$ and lies on the TE grid, and $U_{\phi}$ is dimensioned
$m,n+1$ and lies on the PE grid.  To compute $U_{\theta}$ and $U_{\phi}$
from $\Psi$ lying on the CEG grid, one can use subroutine
\texttt{gradh\_ce\_ss}.

Moving on now to the discretization of equation (\ref{eqn:divhuh}),
the horizontal divergence of a vector field,
this can be evaluated on either the CO grid or the CE grid.
Setting $\Phi = \grad_h \cdot \vecU$, we find for the CO grid locations,
\bea
\lefteqn{\Phi(i,j) =
{1 \over r \sin \theta_i \Delta \theta} \times \nonumber} \\
& & \left(U_{\theta}(i+\half,j) \sin \theta_{i+\half} - U_{\theta}(i-\half,j) 
\sin\theta_{i-\half}\right) \nonumber \\
& & + {1 \over r \sin \theta_i \Delta \phi} \left( U_{\phi}(i,j+\half)
- U_{\phi}(i,j-\half) \right)
\label{eqn:divhcoss}
\eea
where $i \in [0,m-2]$, and $j \in [0,n-2]$.  
Here, $\Phi$ lies on the CO grid, is
dimensioned $(m-1,n-1)$;
the input array $U_{\theta}$ lies on the PE grid and is dimensioned
$(m,n+1)$, while $U_{\phi}$
lies on the TE grid, and is dimensioned $(m+1,n)$.  
Subroutine \texttt{divh\_co\_ss} can be used to
evaluate the horizontal divergence on the CO grid.

To evaluate the horizontal divergence on the CE grid, we have
\bea
\lefteqn{\Phi(i+\half,j+\half) = 
{1 \over r \sin \theta_{i+\half} \Delta \theta} \times \nonumber} \\
& & \left( U_{\theta}(i+1,j+\half) \sin \theta_{i+1} - U_{\theta}(i,j+\half)
\sin \theta_{i}\right) \nonumber \\
& & + {1 \over r \sin \theta_{i+\half} \Delta \phi} \times \nonumber \\
& & \left( U_{\phi}(i+\half,j+1) - U_{\phi}(i+\half,j) \right)
\label{eqn:divhcess}
\eea
where $i+\half \in [\half,m-\half]$, and $j+\half \in [\half,n-\half]$.  
Here, $\Phi$ lies
on the CE grid, and has dimensions of $(m,n)$.  The arrays $U_{\theta}$ and
$U_{\phi}$ lie on the TE and PE grids, respectively, with dimensions
$(m+1,n)$ and $(m,n+1)$.  Subroutine \texttt{divh\_ce\_ss} can be used
to evaluate the horizontal divergence on the CE grid.

Finally, we address the discretization of equation (\ref{eqn:curlhuh}).
Setting $\Phi = \vecrhat \cdot \grad \times \vecU$, we can evaluate $\Phi$ on
the CO or the CE grid.  For the CO grid, we have
\bea
\lefteqn{\Phi(i,j) =
{1 \over r \sin \theta_i \Delta \theta} \times \nonumber} \\
& & \left(U_{\phi}(i+\half,j) \sin \theta_{i+\half} - U_{\phi}(i-\half,j)
\sin\theta_{i-\half}\right) \nonumber \\
& & -\ {1 \over r \sin \theta_i \Delta \phi} \left( U_{\theta}(i,j+\half)
- U_{\theta}(i,j-\half) \right)
\label{eqn:curlhcoss}
\eea
for $i \in [0,m-2]$ and $j \in [0,n-2]$, with the output 
dimensioned $(m-1,n-1)$.  
The input array $U_{\theta}$ is dimensioned $(m+1,n)$ and lies on the TE
grid, and $U_{\phi}$ is dimensioned $(m,n+1)$ and lies on the PE grid.
Subroutine \texttt{curlh\_co\_ss} can be used to evaluate
equation (\ref{eqn:curlhcoss}).

Evaluating $\Phi$ on the CE grid, we have
\bea
\lefteqn{\Phi(i+\half,j+\half) =
{1 \over r \sin \theta_{i+\half} \Delta \theta} \times \nonumber} \\
& & \left( U_{\phi}(i+1,j+\half) \sin \theta_{i+1} - U_{\phi}(i,j+\half)
\sin \theta_{i}\right) \nonumber \\
& & -\ {1 \over r \sin \theta_{i+\half} \Delta \phi} \times \nonumber \\
& & \left( U_{\theta}(i+\half,j+1) - U_{\theta}(i+\half,j) \right)
\label{eqn:curlhcess}
\eea
for $i+\half \in [\half,m-\half]$ and $j+\half \in [\half,n-\half]$, 
with the output
dimensioned $(m,n)$.  The input array $U_\theta$ is dimensioned
$(m,n+1)$ and lies on the PE grid, and $U_{\phi}$ is dimensioned $(m+1,n)$
and lies on the TE grid.  Subroutine \texttt{curlh\_ce\_ss} can be used to 
evaluate equation (\ref{eqn:curlhcess}).

Note that we have not written down a finite difference expression for
the horizontal Laplacian, equation (\ref{eqn:delh2def}).  The finite
difference expression can be found in FISHPACK documentation, in
\citet{Swarztrauber1975}, and in \S \ref{sec:potential}.
The horizontal Laplacian of $\Psi$, when $\Psi$ is on the COE grid,
can be computed with subroutine\\
\texttt{delh2\_co\_ss}.\\  
This subroutine just uses
a call to \texttt{gradh\_co\_ss}, followed by a call to \texttt{divh\_co\_ss}.
The result is the Laplacian of $\Psi$ evaluated on the CO grid.
Similarly, the horizontal Laplacian of $P$, which lies on 
the CEG grid, can be computed with subroutine\\
\texttt{delh2\_ce\_ss}.\\
This just uses a call to \texttt{gradh\_ce\_ss}, followed by a call to
\texttt{divh\_ce\_ss}.  The result is the Laplacian of $P$ evaluated on the CE
grid.  Solutions of Poisson's equation found from FISHPACK
can be tested with these subroutines in PDFI\_SS
and then the the results compared with the source terms used on input
to the Poisson equation.  In all cases tested, we find agreement that is
close to floating point roundoff error.

When closely examining the source code for the above subroutines, one may
find that the index range differs from the ranges mentioned above; this
is done to adhere to default array index ranges in Fortran.  However, the
array dimensions and grid locations will be consistent with those described
above.

\subsection{Consistency With Applied Neumann Boundary Conditions Using Ghost 
Zones}
\label{sec:ghosties}

When normal derivative (Neumann) boundary conditions are input into
FISHPACK subroutines, the solution is computed only on the ``active'' part
of the grid.  To ensure that the finite difference expressions given
in \S \ref{sec:veccalc} are consistent with the boundary conditions, we
add extra ``ghost zones'' to the solutions
that ensure that the boundary conditions are
obeyed within these expressions.  The clearest example of how this is done
are the two
extra ghost zones added in $\theta$ and in $\phi$ to get values of $\dot P$ on
the CEG grid from the solution returned by FISHPACK on the CE grid.
In this case, we're using subroutine \texttt{HSTSSP} for which the
first active point in $\phi$ is at $c + \half \Delta \phi$, while the
boundary is at $\phi=c$.  Thus for $\dot P$, we add a row of $m$ cells centered
at
$\phi = c - \half \Delta \phi$, for which
\bea
\lefteqn{\dot P (\theta_{i+\half}, c - \half \Delta \phi) = \nonumber} \\
& & \dot P(\theta_{i+\half},c+\half \Delta \phi)
- \Delta \phi {\left( \partial \dot P \over \partial \phi \right)} 
\bigg|_{\phi=c}(\theta_{i+\half}) ,
\label{eqn:ghost}
\eea
where ${\left( \partial \dot P \over \partial \phi \right)} 
\big|_{\phi=c}(\theta_{i+\half})$
is the derivative of $\dot P$ specified at $\phi=c$ for the $m$ points along
the left boundary in the call to
\texttt{HSTSSP}.  There is a similar operation to determine ghost cell
values for the other three domain boundaries.  See \S \ref{sec:ptdsolve}
for a discussion of the Neumann boundary conditions for
$\dot P$ and $\partial \dot P / \partial r$.

Because the CEG grid is
dimensioned $(m+2,n+2)$, there are four unused ``corner'' values for
the $\dot P$ array at CEG locations.  These four values could be set to
any value, but for display purposes, we set the corner values to be the average
of the two closest neighbor points, so that the $\dot P$
array can be viewed as a continuous function when visualized.

\subsection{Computing the Contributions to the PDFI Electric Field}
\label{sec:econtributions}

In this section, we will describe in detail how the four
different electric field contributions to the PDFI electric field can be
computed with various subroutines within PDFI\_SS.  We will first describe 
the calculation of the PTD
(inductive) electric field.  Next, we will discuss the software for
performing the ``iterative'' method, necessary to evaluate the Doppler
and Ideal contributions to the electric field.  Following this, we
discuss the calculation of the Doppler electric field, followed by the
contribution from FLCT (or other ``optical flow'' derived horizontal velocities)
to the electric field.  Finally, we discuss the calculation of the ``ideal''
contribution to $\vecE$.

\subsubsection{Numerical Solution for $\vecE^P$ in PDFI\_SS}
\label{sec:ptdsolve}

The calculation of $\vecE^P$ (the PTD, or inductive contribution to $\vecE$)
depends exclusively on time derivatives of
$\vecB$.  In \S \ref{sec:interpol} we described the estimation of time
derivatives in terms of simple differences in the magnetic field components
that take place between two successive times in an assumed time cadence.
Once the data have been interpolated to the staggered grid locations, we
will simply define $\dot B_r$ from the data as
\be
\dot B_r (t_0 + \half \Delta t) = {B_r (t_0+ \Delta t) - B_r (t_0) 
\over \Delta t} ,
\label{eqn:Brdot}
\ee
where $\Delta t$ is the assumed time separation of the cadence.
Here, the subtraction is understood to apply
in a whole array sense, $i.e.$ to all the points in the CE grid locations
for both arrays of $B_r$, with similar expressions for
$\dot B_{\theta}$ and $\dot B_{\phi}$, defined for their own array sizes
and locations (see Table \ref{tab:gridlocations}).
If no masking for weak magnetic field regions is
desired, this definition for the time derivatives can then be used to derive the
PTD solution.  However, we have found that in weak field regions, the PTD
solution $\vecE^P$ can be strongly affected by noise.  One can suppress much
of this noise by multiplying $\dot B_r$, $\dot B_{\theta}$, and $\dot B_{\phi}$
by their respective strong magnetic field mask arrays 
$M^{CE}$, $M^{TE}$, and $M^{PE}$ as defined in \S \ref{sec:interpol}.
Whether the input is masked or not, the resulting electric field contributions
for $\vecE^P$ are computed by two subroutines, called in succession:\\
\texttt{ptdsolve\_ss}\\
\texttt{e\_ptd\_ss}.\\

Subroutine \texttt{ptdsolve\_ss} solves the 3 Poisson equations 
(\ref{eqn:poisson-pdot},\ref{eqn:poisson-Tdot},\ref{eqn:poisson-dpdrdot}) 
for $\dot P$, $\dot T$,
and $(\partial \dot P / \partial r)$.  Once these have been computed,
subroutine \texttt{e\_ptd\_ss} uses $\dot P$ and $\dot T$ to compute the
three components of $\vecE^P$.  If desired, one can then use a third
subroutine\\
\texttt{dehdr\_ss}\\
to use $(\partial \dot P / \partial r)$ to compute the radial derivatives 
of $\vecE^P_h$.

There are a lot of assumptions made about boundary conditions and
equation centering in subroutine \texttt{ptdsolve\_ss} that need to be 
mentioned.  First, the Poisson equations for $\dot P$ and
$\partial \dot P / \partial r$ are centered on the CE grid, since their
source terms $\dot B_r$ and $\grad_h \cdot \dot \vecB_h$ are both centered on
the CE grid, meaning that FISHPACK subroutine \texttt{HSTSSP} will be
used for the solution.  Second, the Poisson equation for $\dot T$ is
centered on the CO grid, since its source term
$-\vecrhat \cdot \grad_h \times \dot \vecB_h$ is located on the CO grid.  
This means
that FISHPACK subroutine \texttt{HWSSSP} will be used for its solution.

A physically meaningful boundary condition for $\vecE^P_h$ is to specify
the electric field component tangential to the domain boundary.  The
tangential electric field is directly related to the derivatives of $\dot P$
in the directions normal to the boundary.  During the development phase of
the PDFI\_SS software, we initially assumed zero tangential electric
field along the domain boundary.  However, this assumption was in conflict
with the actual HMI data:  for many regions, the net radial
magnetic flux is not balanced, and can increase or decrease in time, which
was inconsistent with our assumptions.  Therefore, we have modified
our assumed boundary conditions for $\dot P$, and now specify a boundary
condition on $E_t$ (the electric field tangential to the boundary) that is
consistent with
the observed increase or decrease in the net radial magnetic flux.  

We first evaluate the time rate of change of the net radial magnetic flux
in the domain:
\be
{\partial \Phi \over \partial t} = \Delta \phi \Delta \theta\ r^2
\sum_{i+\half,j+\half} \dot B_r (i+\half,j+\half)\ \sin\ \theta_{i+\half},
\label{eqn:dphidt}
\ee
where the sum is over the cell centers of the domain ($i.e.$ over the CE grid).
Next, we evaluate the perimeter length of the domain along the north and
south edges:
\be
L_{perim} = r \left[ (d-c) (\sin\,a + \sin\,b) \right] .
\label{eqn:lperim}
\ee
Assuming that the tangential electric field is zero along the left and right
edges of the domain,
we can then use Stokes' theorem to integrate Faraday's law over the domain
to find a constant
amplitude of the electric field on the north and south edges, $c E_{perim}$:
\be
c E_{perim} = -\ {\partial \Phi \over \partial t} / L_{perim} ,
\label{eqn:eperim}
\ee
where it is understood that $c E_{perim}$ along the domain edges points in
the counter-clockwise direction if it is positive.  The minus sign in
equation (\ref{eqn:eperim}) comes from the minus sign in Faraday's law.  
We assign the normal
derivatives of $\dot P$ to 
either zero (left and right boundaries)  or
$c E_{perim}$ (north and south boundaries) in FISHPACK subroutine
\texttt{HSTSSP} after accounting for the spherical 
geometry factors (see equation (\ref{eqn:curlpsirhat})) 
and the sign of $c E_{perim}$ relative to the 
$\phi$ unit 
vector 
along the domain boundaries:
\be
{\partial \dot P \over \partial \theta} \bigg|_{\theta=a}(j+\half) = 
-\, c E_{perim}\,r ,
\label{eqn:bca}
\ee
\be
{\partial \dot P \over \partial \theta} \bigg|_{\theta=b}(j+\half) = 
c E_{perim}\,r ,
\label{eqn:bcb}
\ee
\be
{\partial \dot P \over \partial \phi} \bigg|_{\phi=c}(i+\half) = 
0 ,
\label{eqn:bcc}
\ee
\be
{\partial \dot P \over \partial \phi} \bigg|_{\phi=d}(i+\half) = 
0 .
\label{eqn:bcd}
\ee

We still have,
in the PDFI\_SS library, the subroutine that assumes zero tangential electric
field along boundary edges if the user wants to use this assumption.  
That subroutine is named \texttt{ptdsolve\_eb0\_ss}.

Assuming that the boundary is far away from the most rapid evolution of
the data, and that there is no significant change in the net radial current
density in the domain, we set the electric field $E^P_r$ to zero at
the boundary.  Since this is proportional to $\dot T$, we use homogenous
Dirichlet boundary conditions for $\dot T$ ($\dot T$ is set to zero
at $\theta=a$, $\theta=b$, $\phi=c$, and $\phi=d$).

The physical boundary condition for $\partial \dot P / \partial r$ is to
specify the component of $\dot \vecB_h$ normal to the domain boundary, since 
the gradient
of $\partial \dot P / \partial r$ is equal to $\dot \vecB_h$ when $\dot T$
is zero on the boundary, as
can be seen from equation (\ref{eqn:Bptddecomp}).  This leads to
these Neumann boundary conditions (see equation (\ref{eqn:gradhpsi}))
for this Poisson equation:
\be
{\partial \over \partial \theta} 
\bigg|_{\theta=a} 
{\partial \dot P \over \partial r} =
r \dot B_{\theta}( \theta=a, \phi_{j+\half} ) ,
\label{eqn:bcadpdotdr}
\ee
\be
{\partial \over \partial \theta} 
\bigg|_{\theta=b} 
{\partial \dot P \over \partial r} =
r \dot B_{\theta}( \theta=b, \phi_{j+\half} ) ,
\label{eqn:bcbdpdotdr}
\ee
\be
{\partial \over \partial \phi} 
\bigg|_{\phi=c} 
{\partial \dot P \over \partial r} =
r \sin\, \theta_{i+\half}\,\dot B_{\phi}( \phi=c, \theta_{i+\half} ) ,
\label{eqn:bccdpdotdr}
\ee
\be
{\partial \over \partial \phi} 
\bigg|_{\phi=d} 
{\partial \dot P \over \partial r} =
r \sin\, \theta_{i+\half}\,\dot B_{\phi}( \phi=d, \theta_{i+\half} ) .
\label{eqn:bcddpdotdr}
\ee
Note that the use of a staggered grid decouples the boundary conditions
for $\dot T$ and $\partial \dot P / \partial r$ that existed for the centered
grid, as described in \citet{Fisher2010} and KFW14.

The solutions for $\dot P$ and $\partial \dot P / \partial r$ are returned
by \texttt{ptdsolve\_ss} on the CEG grid, which means there is an extra
row of ghost zones returned along each of the four sides of the boundary
(see discussion above in \S \ref{sec:ghosties}).

Subroutine \texttt{ptdsolve\_ss} can also be used to find the poloidal
potential $P$, the toroidal potential $T$, and the radial derivative of the
poloidal potential $\partial P / \partial r$, if instead of inputting the
time derivatives of the magnetic field components, one inputs the magnetic
field components themselves.  The vector potential $\vecA$
can be computed from
subroutine \texttt{e\_ptd\_ss}, but a minus sign must be applied to the output.


\subsubsection{Implementation of the ``Iterative'' Method in PDFI\_SS}
\label{sec:relaxation}

The ``iterative'' method for finding a scalar potential whose gradient
is designed to closely match a given vector field, was developed by
co-author Brian Welsch and initially described in \S 3.2 of \citet{Fisher2010}.
In KFW14, the method was used to derive the scalar potential representing
the Doppler electric field, as well as the ideal electric field contribution,
which has the goal of setting $\vecE \cdot \vecB = 0$.  Applying the technique
was relatively simple, because using the centered grid formalism, $\vecE$
and $\vecB$ were co-located.  In PDFI\_SS, however, the various components of
$\vecB$ and $\vecE$ all lie in different locations, making the algorithm
less straightforward to implement directly.  We now describe
our current solution to the problem.

We describe the iterative technique in terms of generating the ideal
component of the electric field (rather than the Doppler contribution),
because we think the logic is easier to follow in that case, but the
solution method applies equally well to both the Doppler and ideal
electric field contributions.

In PDFI\_SS the subroutine that computes solutions using the iterative method
is called\\
\texttt{relax\_psi\_3d\_ss}.\\
Our approach is to perform the iterative procedure on a temporary, centered
grid, coinciding with the CO grid, but computed
using centered grid finite difference
expressions, centered on cell vertices.  On input to 
\texttt{relax\_psi\_3d\_ss}, we need values
of $\vecE^{PDF}$ and $\vecB$ lying on the CO grid.  To construct
values of $\vecE^{PDF}$ on that grid from their native staggered grid locations,
we can use linear interpolation, computed from subroutine\\
\texttt{interp\_eh\_ss},\\
to interpolate the 
horizontal electric field contributions to the CO grid.  The solutions
for $E_r$ are already given on the COE grid, from which the CO grid is just
a subset.  For magnetic field values on the CO grid, we note
that the original input magnetogram data was initially provided
on the COE grid, so no interpolation to CO is necessary.
Subroutine \texttt{relax\_psi\_3d\_ss} assumes a domain size that is smaller
in extent than our overall domain boundary, with its boundaries given by
$a' = a+\Delta \theta$, $b'=b-\Delta \theta$, $c'=c+\Delta \phi$, and
$d' = d-\Delta \phi$.  Internal to \texttt{relax\_psi\_3d\_ss}, boundary
conditions assumed at $a'$, $b'$, $c'$, and $d'$ are that the normal derivatives
of the scalar potential $\psi$ are zero (homogenous Neumann boundary
conditions).  
This is implemented by assigning ghost zone 
values for $\psi$
that enforce this boundary condition under the centered grid assumption,
resulting in the array for $\psi$, including the ghost zones, lying on the 
COE grid.
Once the iterative procedure has been completed, but before exiting the
subroutine, a final set of boundary conditions are applied using ghost zones
for $\psi$ implemented on the edges of the COE grid, in which the homogenous
Neumann boundary conditions for $E_n$ (the normal components of $\grad_h \psi$)
are applied using the {\it staggered} grid formalism,
at a boundary half-way between the edges of the CO and COE grids, at
$a+\half \Delta \theta$, 
$b-\half \Delta \theta$, $c+\half \Delta \phi$, and $d-\half \Delta \phi$.
This results in slight changes to the ghost zone values of $\psi$ located at the
edges of the COE grid, compared to their values computed as ghost zones using
the centered grid formalism.
We then also set boundary conditions for 
$\partial \psi / \partial r = 0$ at the regular domain boundaries, 
$a$, $b$, $c$, and $d$.
After exiting \texttt{relax\_psi\_3d\_ss}, gradients of $\psi$
are then computed using the staggered grid formalism of \S \ref{sec:veccalc}, 
rather than using the centered grid description that is used internally
within that subroutine.

We now summarize the details of how $\psi$ and $\partial \psi / \partial r$
are computed in the subroutine, combining material
originally in \S 3.2 of \citet{Fisher2010} and \S 2.2 of KFW14, and using
spherical coordinates.  The procedures described here are slight updates
of the original iterative method, as the code has evolved from its original
formulation:

{\bf Step 1:}\\
Decompose $\grad \psi$ as
\be
\grad \psi = s_1 \unitv{b} + s_2\ \vecrhat \times \unitv{b} +
s_3\ \unitv{b} \times \left( \vecrhat \times \unitv{b} \right) ,
\label{eqn:step1}
\ee
where $s_1$, $s_2$, and $s_3$ are understood to be functions of
$\theta$ and $\phi$.  Here $\unitv{b}$ is the unit vector pointing in
the same direction as $\vecB$.  Quantities $\unitv{b}_h$ and $b_r$, used
below, denote the horizontal and radial components of $\unitv{b}$, respectively,
and $b_h^2$ is the square of the amplitude of $\unitv{b}_h$.

{\bf Step 2:}\\
Set
\be
s_1 = \vecE^{PDF} \cdot \unitv{b} ,
\label{eqn:step2a}
\ee
\be
s_2 = 0 ,
\label{eqn:step2b}
\ee
\be
s_3 = 0 ,
\label{eqn:step2c}
\ee
\be
\left( {\partial \psi \over \partial r} \right)_0 = s_1\ b_r ,
\label{eqn:step2d}
\ee
and
\be
\grad_h \psi_0 = s_1 \unitv{b}_h .
\label{eqn:step2e}
\ee
The quantities $\psi$ and $\partial \psi / \partial r$ are both regarded as
functions of $\theta$ and $\phi$, and subscript $0$ denotes the ``zeroth''
iterative approximation to $\psi$.  Equation (\ref{eqn:step2e}) should result
in cancellation of the component of $\vecE^{PDF}$ parallel to $\vecB$
if $-\grad \psi$ is added to it.
To obtain the guess for $\psi_0$, we
can take the divergence of equation (\ref{eqn:step2e}):
\be
\grad_h^2 \psi_0 = \grad_h \cdot ( s_1\ \unitv{b}_h ) .
\label{eqn:step2f}
\ee
The horizontal divergence operation on the right-hand side of equation
(\ref{eqn:step2f}) is computed using a centered grid formalism using
subroutine\\
\texttt{divh\_sc}.\\
We can solve this Poisson equation for $\psi_0$ using FISHPACK subroutine
\texttt{HWSSSP}, subject to homogenous Neumann boundary conditions
at the ``primed'' values for $a$, $b$, $c$, and $d$ noted above.
The quantity $s_1$ will be held fixed throughout the iterative sequence.
Once we have a solution for $\psi_0$, we can evaluate $\grad_h \psi_0$ using
the centered grid subroutine\\
\texttt{gradh\_sc}.\\

{\bf Step 3:} (the beginning of the iterative sequence)\\
Given the current guess for $\psi$ and $\grad \psi$, evaluate
\be
s_2 = {\vecrhat \cdot \left( \unitv{b}_h \times \grad_h \psi \right) \over
b_h^2} ,
\label{eqn:step3a}
\ee
and
\be
s_3 = {\partial \psi \over \partial r} - 
{b_r ( \grad_h \psi \cdot \unitv{b}_h) \over b_h^2} . 
\label{eqn:step3b}
\ee
Equations (\ref{eqn:step3a}) and (\ref{eqn:step3b}) can be derived by dotting
both sides of equation (\ref{eqn:step1}) with the vectors 
$\unitv{r} \times \unitv{b}$ and 
$\unitv{b} \times (\unitv{r} \times \unitv{b})$, respectively.

{\bf Step 4:}\\
Given $s_2$ and $s_3$ from Step 3,
evaluate the horizontal divergence of equation (\ref{eqn:step1}) 
\be
\grad_h^2 \psi = \grad_h \cdot \left( s_1 \unitv{b}_h + 
s_2\ \vecrhat \times \unitv{b} - s_3\ b_r \unitv{b}_h \right) 
\label{eqn:step4a}
\ee
and then solve this Poisson equation for the next iterative solution for $\psi$.
The update for $\partial \psi / \partial r$ is given by this expression:
\be
{\partial \psi \over \partial r} = s_1 b_r + s_3 b_h^2 .
\label{eqn:step4b}
\ee
 
{\bf Step 5:}\\
If the number of iterations is less than the maximum number (current 
default value
is 25), go back to Step 3.  If maximum number is exceeded, then exit
the iteration procedure.  The resulting arrays of $\psi$ and
$\partial \psi / \partial r$ on the CO grid
are the final arrays for the iterative solution.  We note again that the ghost
zone values, located on the edges of the COE grid, are adjusted from the
values computed from the centered grid finite differences to make them
consistent with the use of the staggered grid finite differences.

We now remark on several properties of the iterative technique described
above.  First, as noted in \S 3.2 of \citet{Fisher2010} and
\S 2.2 of KFW14, the mathematical problem that
the iterative method is designed to solve has no unique solution; nevertheless
this method appears to find a unique solution, meaning that
most likely the method imposes other hidden constraints which makes
it behave like a unique solution.  See \S 2.2 of KFW14 for further discussion.

Second, the iterative improvement in the solution is rapid for the
first few iterations, then improvement slows dramatically.  We have
found that implementing an error convergence criteria, as originally suggested
in \citet{Fisher2010} has proven unreliable and difficult.
We follow the suggestion of KFW14 that setting a fixed number of
iterations is a better implementation.  We adopt the suggestion from KFW14 of
25 iterations.  Experiments we have done have shown that using much larger
numbers of iterations (100 or 1000, for example) actually makes the solution
worse.

\begin{figure}[ht!]
\includegraphics[width=3.4in]{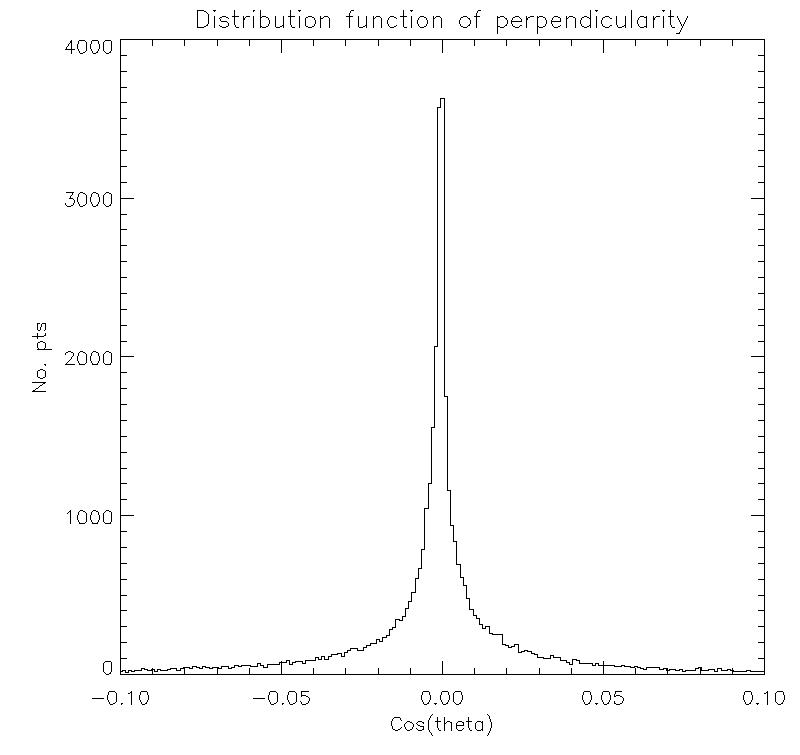}
\caption{The distribution function for the cosine of the angle between
$\vecE$ and $\vecB$ for test PDFI solution for AR11158 at 
2011.02.14\_23:35-23:47.\label{fig:cosang}}
\end{figure}

Third, in contrast to the error in $e.g.$ Faraday's law, which is near
floating point roundoff error, the ability of the iterative technique
to exactly cancel the component of $\vecE$ in the direction of $\vecB$ is much
less precise.  We find in PDFI\_SS that the typical angle between
$\vecE$ and $\vecB$ once $-\grad \psi$ has been added to $\vecE^{PDF}$
for a number of different test cases is within $2^{\circ}$ of
$90^{\circ}$.  Histograms of the cosine of the angle between the two
vectors is sharply peaked at zero (see Figure \ref{fig:cosang}), but 
with significant tails in the
distribution.  Examination of where the outlier points are located shows
a concentration near the boundaries of the strong magnetic field mask,
suggesting that the iterative procedure has its worst performance near
the mask boundaries.

Fourth, the iterative technique is sensitive to noisy input data in
$\vecE^{PDF}$.  For example, if we compute a solution to $\vecE^P$ using
unmasked magnetic field time derivatives, and then try to find an electric
field contribution that attempts to make $\vecE$ and $\vecB$ perpendicular,
the iterative technique can diverge, rather than converge.  For this reason,
we strongly recommend using masking for the magnetic field time derivative
arrays on input to \texttt{ptdsolve\_ss}, if the iterative method will then
be used to compute the ``ideal'' contribution.

Finally, we note again that once the solution for $\psi$ and 
$\partial \psi / \partial r$ are obtained after exiting
\texttt{relax\_psi\_3d\_ss}, the resulting contribution from $- \grad \psi$
to $\vecE$ is evaluated at the staggered grid locations.  This has the
advantage of producing a curl-free contribution to $\vecE$, but at the cost
of losing the direct connection to the angle between $\vecE$ and 
$\vecB$, since the vectors are no longer co-located.  To evaluate the angle
between $\vecE$ and $\vecB$, the electric fields must once again be
interpolated to the CO grid before this comparison can be done.  Subroutine\\
\texttt{angle\_be\_ss}\\
can be used to interpolate electric field components
to the CO grid, and then evaluate the cosine of the angle between the
$\vecE$ and $\vecB$ vectors on the CO grid.

\subsubsection{Implementing the Doppler Electric Field Contribution}
\label{sec:dopplerrelax}

Given the existence of subroutine \texttt{relax\_psi\_3d\_ss}, the
evaluation of equations (\ref{eqn:edoppsi}-\ref{eqn:eqnpsid}) is fairly 
straightforward, given the input line-of-sight unit vector information,
the magnetic field arrays (as input on the COE grid), and the Doppler velocity,
also on the COE grid.  The
Doppler electric field contribution is computed by subroutine\\
\texttt{e\_doppler\_ss}.\\
Once all the terms in equation (\ref{eqn:eqnpsid}) have been evaluated,
subroutine \texttt{relax\_psi\_3d\_ss} is called, and then the gradient
of the resulting scalar potential is evaluated.

A few remarks are in order on the resulting electric field contribution.
First, in weak-field regions, the behavior of the $\unitv{q}$ unit vector
can be very noisy, if the underlying magnetic field is noisy.  For this reason,
we find much better results if the Doppler velocity and the magnetic field
components are multiplied by strong field masks on input to 
subroutine \texttt{e\_doppler\_ss}.

Second, if the region on the Sun being studied is significantly away from
disk-center, we sometimes find strong contamination of the Doppler velocity
from Evershed flows, which then can result in spurious Doppler electric field
contributions near LOS polarity inversion lines at the edges of sunspots.

We have written an alternative subroutine to compute Doppler electric fields
based on a different concept, subroutine\\
\texttt{e\_doppler\_rpils\_ss},\\
which uses the locations of radial and LOS PILs to try to eliminate these
artifacts.  Early tests of the effectiveness and accuracy of this subroutine
were inconclusive, but it is available for experimentation in the software
library.

For now, we retain the original version of \texttt{e\_doppler\_ss} as the
default version, in spite of the above mentioned defects, as it seems to 
work well near disk center, and behaves correctly for the \texttt{ANMHD} 
test case described in KFW14 (see also \S \ref{sec:xanmhd}).

\subsubsection{Computing the FLCT Contribution}
\label{sec:flctsubroutine}

The subroutine that computes the FLCT contribution to the PDFI electric
field is\\
\texttt{e\_flct\_ss}.\\
The input data for computing the source terms for equation 
(\ref{eqn:flctpoisson}) are the $\vecV_h \times B_r \vecrhat$ 
electric field contributions located on the 
PE and TE grids, along with interpolated values of $B_r$ and $|\vecB_h|$.
We find that it is frequently
useful to multiply the $B_r$ and $|\vecB_h|$ input arrays
by the strong field mask to reduce the role of noise in the weak magnetic
field regions.
The divergence on the right hand side of equation (\ref{eqn:flctpoisson})
is then evaluated from the input data onto the CO grid using subroutine 
\texttt{divh\_co\_ss}.  The Poisson equation for
$\psi^F$ is then solved on the CO grid,
but with the Poisson equation domain boundaries
defined to be half-way between the
edges of the CO and COE grids: $a''=a+\half \Delta \theta$, 
$b''=b-\half \Delta \theta$, $c''=c+\half \Delta \phi$, and 
$d''=d-\half \Delta \phi$.  Applying a zero normal-gradient boundary 
condition at this boundary then allows us to compute ghost-zone values 
for $\psi^F$
along the edges of the COE grid, resulting in the output scalar potential
being defined on the COE grid.  Because the Poisson equation boundary is
staggered relative to the variables on the
corners or vertices, we use the staggered grid version for FISHPACK,
subroutine \texttt{HSTSSP} in subroutine \texttt{e\_flct\_ss}.
Once $\psi^F$ has been computed, the electric field components are computed
on the TE and PE grids by taking $-\grad_h \psi^F$ using subroutine
\texttt{gradh\_co\_ss}.
\\
\\

\subsubsection{Computing the Ideal Contribution to the PDFI Electric Field}
\label{sec:idealsubroutine}

The calculation of the ``Ideal'' contribution to the PDFI electric field
is computed by subroutine\\
\texttt{e\_ideal\_ss}.\\
On input, the values of $B_{\theta}$, $B_{\phi}$, and $B_r$ are provided
on the COE grid.  Input values of $E^{PDF}_{\theta}$, $E^{PDF}_{\phi}$, and
$E^{PDF}_r$ are also provided on their staggered grid locations.
The horizontal components of $\vecE^{PDF}$ are then interpolated to CO
grid locations using subroutine \texttt{interp\_eh\_ss}.  
To reduce the impact of noise from the weak field regions, we strongly
recommend multiplying the input magnetic field arrays by the strong field
mask array $M^{COE}$ when calling \texttt{e\_ideal\_ss}.
Next, the hard work for computing the Ideal contribution to PDFI
is handled by subroutine \texttt{relax\_psi\_3d\_ss}, described earlier
in \S \ref{sec:relaxation}, which returns the scalar potential $\psi^I$
and its radial derivative $\partial \psi^I / \partial r$, both computed on
the COE grid.  Finally, the electric field contribution $-\grad \psi^I$
is computed on the staggered grid locations by calling subroutine
\texttt{gradh\_co\_ss} for the horizontal components, and using the
array $- \partial \psi^I / \partial r$ for the radial component.

\subsection{Poynting Flux and Helicity Injection From PDFI Solutions}
\label{sec:efieldproducts}

Once the PDFI electric field has been computed, there are a number of
other useful quantities that can be computed with it, including the radial
component of the Poynting flux, as well as the contribution
function to the relative helicity injection rate.

These quantities are computed by the subroutines\\
\texttt{sr\_ss} and\\
\texttt{hm\_ss}.\\

The subroutine \texttt{sr\_ss} takes as input the horizontal components of
both the electric field and magnetic field in their staggered grid locations
on the TE and PE grids, and computes the radial component of the Poynting flux
at cell centers (the CE grid).  While most of the PDFI\_SS software assumes
that electric fields are computed as $c \vecE$, in units of 
$[{\rm G\ km\ s}^{-1}]$, subroutine \texttt{sr\_ss} assumes that the 
input electric fields don't include the factor of $c$, and are given
in units of $[{\rm V\ cm}^{-1}]$.
To convert from $c \vecE$ in units of $[{\rm G\ km\ s}^{-1}]$ to 
$\vecE$ in units
of $[{\rm V\ cm}^{-1}]$, one can simply divide by a factor of $1000$.

We find that in the weak field regions, the Poynting flux can be quite
unreliable, so we recommend that after output from subroutine \texttt{sr\_ss},
that the resulting Poynting flux array be multiplied by the strong magnetic
field mask for the CE grid, $M^{CE}$.  If the strong field masks have been
used to compute the electric field contributions, then for consistency,
the masks should
also be applied on either the input horizontal magnetic fields, or on the 
Poynting flux output (which is what we do).  The assumed units on output from
\texttt{sr\_ss} for the Poynting flux 
are $[{\rm erg\ cm}^{-2}\ {\rm s}^{-1}]$.

To compute the total magnetic energy input rate from the radial component of
the Poynting flux, one can use subroutine\\
\texttt{srtot\_ss}\\
to integrate the radial Poynting flux contribution over area.  The output is
a single value, computed in units of $[{\rm erg\ s}^{-1}]$.

The subroutine\\
\texttt{hm\_ss}\\ 
is used to compute the contribution function for the relative helicity
injection rate. 
On input, it uses the poloidal potential $P$ computed from
the radial component of the magnetic field using subroutine 
\texttt{ptdsolve\_ss}, and the horizontal components
of $\vecE$ from the PDFI solution.  We typically compute $P$ using arrays
of $B_r$ that are multiplied by the strong field mask $M^{CE}$ before 
\texttt{ptdsolve\_ss} is called, so that
the vector potential for the potential magnetic field $\vecA_P$
does not contain contributions from the weak field
regions.  The vector potential $\vecA_P$ is computed from $P$ using
subroutine \texttt{curl\_psi\_rhat\_ce\_ss} within \texttt{hm\_ss}. 
Values of $\vecA_P$ and $\vecE$ components are then interpolated to the CE grid,
and then the quantity $\unitv{r} \cdot 2 c \vecE_h \times \vecA_P$ 
is computed on the
CE grid.  The input units for the horizontal electric field are assumed to
be in units of $[{\rm V\ cm}^{-1}]$, and the units for $P$ are assumed to
be $[{\rm G\ km}^2]$.  The output array is in computed in units of
$[{\rm Mx}^2\ {\rm cm}^{-2}\ {\rm s}^{-1}]$.

One can integrate the contribution function to get a total relative helicity 
injection rate by calling subroutine\\ 
\texttt{hmtot\_ss}\\
using the output from \texttt{hm\_ss} as input.  The output is a single value,
given in units of $[{\rm Mx}^2\ {\rm s}^{-1}]$.

\subsection{Putting It All Together:  
subroutine {\rm \texttt{pdfi\_wrapper4jsoc\_ss}}}
\label{sec:pdfiwrapper4jsoc}

The preceding parts of this section have described in detail how the
HMI input data is transposed, interpolated, and then used to compute the 
various contributions to the PDFI electric
field, and how that can then be used to create maps of the Poynting flux and
the contribution function for the relative helicity injection rate.
We have written a subroutine in the PDFI\_SS library,\\ 
\texttt{pdfi\_wrapper4jsoc\_ss},\\
that combines all of
these pieces together.  
The SDO JSOC calls this subroutine to compute the electric field and 
related variables to create the CGEM data series which is distributed by
the JSOC.  The subroutine is also useful, in that it can serve as a template
for a customized calculation of the electric field, allowing a user to
eliminate unwanted terms, experiment with various masking strategies for
input data, or experiment with new electric field contributions.

The list of major tasks performed by\\
\texttt{pdfi\_wrapper4jsoc\_ss}, along with
the subroutines used for these tasks, is given in order below:
\begin{itemize}
\item Transpose the 18 input arrays from longitude-latitude order to
colatitude-longitude (theta-phi) order (\texttt{brll2tp\_ss},
\texttt{bhll2tp\_ss})
\item Convert Doppler velocities from (m/sec) to km/sec and change sign 
convention to positive for upflows
\item Compute strong-field mask arrays for staggered grid locations 
from arrays of input magnetic field
arrays on the COE grid (\texttt{find\_mask\_ss}, \texttt{fix\_mask\_ss})
\item Interpolate input data to staggered grid locations
(\texttt{interp\_data\_ss}, \texttt{interp\_var\_ss})
\item Compute $\sin \theta$ arrays at colatitude edges and cell centers
(\texttt{sinthta\_ss})
\item Compute $\dot P$, $\dot T$, $\partial \dot P / \partial r$ and
$P$, $T$, and $\partial P / \partial r$ (\texttt{ptdsolve\_ss})
\item Compute PTD electric field contribution $\vecE^P$ (\texttt{e\_ptd\_ss})
\item Compute Doppler electric field contribution $\vecE^D$
(\texttt{e\_doppler\_ss}, \texttt{relax\_psi\_3d\_ss})
\item Compute FLCT electric field contribution $\vecE^F$
(\texttt{e\_flct\_ss})
\item Compute Ideal electric field contribution $\vecE^I$
(\texttt{e\_ideal\_ss}, \texttt{relax\_psi\_3d\_ss})
\item Add all four contributions for $\vecE^{PDFI}$, convert units to
$[{\rm V\ cm}^{-1}]$
\item Compute radial derivatives of horizontal components of electric field
(\texttt{dehdr\_ss})
\item Compute Poynting flux, and its area integral (\texttt{sr\_ss},
\texttt{srtot\_ss})
\item Compute contribution function for Helicity Injection and its area integral
(\texttt{hm\_ss}, \texttt{hmtot\_ss})
\item Transpose all output arrays to longitude-latitude array order
(\texttt{bhyeetp2ll\_ss},\texttt{bryeetp2ll\_ss},\\
\texttt{ehyeetp2ll\_ss}, \texttt{eryeetp2ll\_ss})
\item Return as calling arguments the staggered grid arrays of all three
magnetic field components, all three electric field components, the
radial derivative of the horizontal electric field components, the radial
component of the Poynting flux, the Relative Helicity injection contribution
function, the energy input rate into the upper atmosphere, and the relative
helicity injection rate.  Note that for the radial electric field component,
we output both the total radial electric field, and also the purely inductive
component.  The inductive component is used when computing the horizontal
components of the curl of $\vecE$, whereas the total radial electric field
would be used for the evaluation of $e.g.$ $\vecE \times \vecB$, or 
for evaluating the angle between $\vecE$ and $\vecB$ (subroutine 
\texttt{angle\_be\_ss}).
The strong field mask arrays for the COE, CO, CE,
TE, and PE grids are also returned.  All returned arrays are oriented in
longitude-latitude index order.
\end{itemize}
The input datasets to, and the output datasets
from \texttt{pdfi\_wrapper4jsoc\_ss}, are archived and publicly available
through the SDO data center with the series name \texttt{cgem.pdfi\_input} and
\texttt{cgem.pdfi\_output}, respectively.  They can be directly accessed
through the SDO JSOC website \url{http://jsoc.stanford.edu} as are
all SDO/HMI and
AIA data, or through a variety of other means including the {\it Solarsoft}
IDL packages or the {\it SunPy} Python package.  Users are referred to the SDO
data analysis guides for data query and retrieval methods, such as
\url{http://jsoc.stanford.edu/How\_toget\_data.html} and
\url{https://www.lmsal.com/sdouserguide.html}.
Each record in these two data series can be uniquely identifield via two
keywords, \texttt{CGEMNUM} and \texttt{T\_REC}, which indicates the CGEM
identification number and the nominal observation time, respectively.
The \texttt{CGEMNUM} is currently defined to be identical to the NOAA active
region (AR) number when the CGEM region coincides with a single named 
active region, and 100,000 plus the SHARP number \citep{Bobra2014} when
the CGEM region coincides with a SHARP region consisting of multiple
active regions or no named active region.  
For \texttt{cgem.pdfi\_output}, the nominal \texttt{T\_REC}
is designated at 06, 18, 30, 42, and 54 minutes after the hour.  For example,
users can find a pair of input records for AR 11158 at the beginning of
2011 February 15 with 
\texttt{cgem.pdfi\_input[11158][2011.02.15\_00:00-}\\
\texttt{2011.02.15\_00:12]}, which
includes vector magnetic field, the FLCT velocity field, the Doppler velocity,
and the local unit normal vectors.  The corresponding PDFI output can be found
with \texttt{cgem.pdfi\_output[11158]}\\
\texttt{[2011.02.15\_00:06]}.  The processing necessary to define the input
data (\texttt{cgem.pdfi\_input}) is described in \S \ref{sec:D4stuff}.

\subsection{Errors in Electric Field Inversions}
\label{sec:error}

There is currently
no formal way for deriving errors in the electric fields within the PDFI\_SS
software.  The fact that the electric field solutions are derived from
solutions of elliptic equations means that any magnetic field or
Doppler velocity errors result in
non-localized errors in the resulting electric fields, making analytic error
propagation studies difficult.
The effects of random errors in the magnetic field
measurements and how these propagate into the PDFI electric field inversions
in HMI data has been studied by \citet{Kazachenko2015} and \citet{Lumme2019}.
In \citet{Kazachenko2015}, given estimated errors in the radial (30G) 
and the two horizontal
components (100G) of the magnetic feld determined from the width of 
distribution functions in the
weak field regions of NOAA AR 11158, this resulted 
in estimated relative errors of 15-20\% 
in the three electric
field components at a given pixel location for a given
pair of active region magnetograms.
These results were
derived by applying Monte Carlo techniques.
\citet{Lumme2019} performed a more detailed error analysis on the PDFI electric
fields that was focused primarily on global quantities such as the spatially
and/or temporally integrated Poynting flux and Helicity injection rate
contribution functions.  They showed that spatial averaging and 
temporal integration resulted in significantly lower relative
errors than one obtained for individual pixel values for a pair of magnetograms.

Neither of these studies addresses another source of error, the systematic 
effects to the velocity and magnetic field signals that are due to 
incomplete corrections for the daily orbital motion of the SDO spacecraft 
around the Earth. These effects appear to generate a false temporal signal 
at the first few harmonics of the orbital frequency in the magnetic 
and velocity signals. A false temporal signal in the magnetic field will 
generate a false electric field through Faraday's law. These systematic 
errors in the observed quantities from orbital artifacts are characterized 
by \cite{Hoeksema2014} and \cite{Schuck2016}.  \cite{Schuck2016}
provide a suggested correction for the Doppler velocity that appears to 
remove much of the artificial temporal signal, but as of yet, no similar 
correction for the magnetic field components is available. 
While these systematic errors can affect the electric field solutions over 
several-hour time scales, short-term variations are small, and the work 
of \cite{Lumme2019} indicates that they do not greatly  affect the 
time evolution on longer time scales. Nevertheless, the results of the 
PDFI\_SS electric field solutions would be improved if these artifacts 
could be removed.


\subsection{Interpolation of Input Data to Other Resolutions}
\label{sec:changeresolution}

It is possible that the user may wish to obtain electric field solutions
at a different resolution than the $0.03^\circ$ resolution provided by the JSOC
upstream processing (described in \S \ref{sec:D4stuff}).  One might be
tempted to simply interpolate the output electric field results to a different
resolution, but doing so will generally destroy the adherence of the solutions 
to Faraday's law  (an exception to this rule is the flux-preserving
``downsampling'' subroutines, described in \S \ref{sec:laplace}, but these
only work for certain specified cases to decrease the resolution).
We have found that if one wants electric field inversions with an arbitrary
change of the resolution, the best solution is to interpolate the input data
to the desired resolution, and then compute the solutions from scratch from
$e.g.$ subroutine \texttt{pdfi\_wrapper4jsoc\_ss}.

The interpolation technique we have used for this process is the 9th order
B-spline, a subset of interpolation solutions described by 
\citet{Thevenaz2000}.  The low-level source
code for this interpolation procedure was written by co-author
Dave Bercik, inspired by
\citet{Thevenaz2000} and the accompanying C source-code at 
\url{http://bigwww.epfl.ch/thevenaz/interpolation/}.
It is implemented in subroutine\\
\texttt{bspline\_ss}.\\  
To interpolate a single one of the 18 input data arrays 
to a different resolution (either coarser or finer), one can use
subroutine\\
\texttt{interp\_hmidata\_ll},\\
where the original and desired array dimensions can be specified.
In this subroutine, the degree of the B-spline can be specified, but we
recommend setting \texttt{degree} to 9.
To interpolate the entire 18-level stack of arrays, input as a 3D array,
interpolated to a new 18-level 3D array, one can use subroutine\\
\texttt{interp\_hmidata\_3d\_ll}.\\  
This subroutine assumes
\texttt{degree=9}.  The latter two subroutines assume that
the domain boundaries $a$, $b$, $c$, and $d$ remain the same in the output
interpolated data arrays as those values for the input arrays.

\section{Upstream Data Processing Necessary for PDFI\_SS}
\label{sec:D4stuff}

Before the PDFI\_SS software can be run, the full-disk HMI data must be 
processed into a form where PDFI\_SS can use the data.  Basically, 
five procedures are necessary to get the data into a suitable form: 
(1) Estimate the full-disk
Doppler velocity data ``convective blue-shift'' bias, 
arising because hot upwelling plasma contributes more to the observed intensity
than cooler downflowing plasma;
(2) The data
surrounding an active-region of interest must be isolated from the full disk
data, and tracked with a rotation rate defined by the center of the active 
region, and mapped into a co-rotating reference frame; 
(3) The azimuth angles of pixels' transverse magnetic fields are
smoothed in time by flipping any ambiguity choices that produce large,
short-lived azimuth changes (``top hats'' in the time series of
changes in azimuth) --  then the resulting magnetic field, Doppler, and
line-of-sight unit vector data are mapped onto a Plate Carr\'ee grid;
(4) Successive radial-field magnetograms are then used to estimate apparent
horizontal motions using the Fourier Local Correlation Tracking (FLCT)
algorithm;  
(5) We add a ribbon of data surrounding each of the input data arrays that
is set to zero.  We find that this ``zero-padding'' improves the quality of
the electric field inversions.
The source code that performs these tasks can be viewed
at \url{http://jsoc2.stanford.edu/cvs/JSOC/proj/cgem/prep/apps/}.
We now describe these five procedures in more detail.

\subsection{Doppler Velocity Correction for Convective Blueshift}
\label{sec:blueshift}
The Doppler velocity calibration software that computes the convective blueshift
\citep{Welsch2013}
was initially written in Fortran by co-author Brian Welsch, and then modified
by co-author Xudong Sun to be called from an HMI module written in C.  The
module uses full-disk vector magnetograms 
and Doppler data as input, and
estimates a ``bias'' that we later subtract from the Doppler shift measurement.
Additional output includes both LOS and radial PIL masks for the LOS magnetic
field $B_\ell$ and the radial field $B_r$.  The source code for this module
can be seen by clicking on the ``view'' link at
\url{http://jsoc.stanford.edu/cvs/JSOC/proj/cgem/prep/apps/doppcal\_estimate.f90}.
We have chosen to work with the Doppler data derived from the full spectral
inversion rather than the traditional Doppler data derived from the LOS field
pipeline, following the recommendation of the HMI Team.  We have performed
a comparison between the ``vector Doppler'' and ``LOS Doppler'' data.  The
comparison was done in a cutout that tracked NOAA AR 11158 in full disk Doppler
velocity maps, and it revealed that the two types of raw uncalibrated Doppler
maps have systematic differences, with median difference oscillating in phase
with the radial velocity of the SDO spacecraft.  The removal of the convective
blueshift using the method of \citet{Welsch2013}
reduces the median difference between the two velocities significantly, 
particularly in strong-field
pixels ($|\vecB| > 300$G).  Subsequent tests of the impact of the differences
between Doppler velocities from the two different datasets on the
calculation of the integrated Poynting flux and Helicity injection rate showed
only a modest difference.  We conclude that while there are differences in
the results using the two different datasets, our processing reduces these
differences, and there is not a substantial difference in the final results.

Once the convective blueshift has been computed, it is used to correct the
Doppler velocity measurements during the step 
described in \S \ref{sec:azimuth} below.

\subsection{Active Region Extraction}
\label{sec:extraction}

This
module extracts a series of AR vector field patches in native coordinates from 
full-disk data, with constant center latitude (rounded to the nearest pixel), 
and tracks them at a constant 
rotation rate.  These patches are given 
a unique ``CGEM number'' as an identifier, and are used as input for the 
subsequent modules.  

\subsection{Azimuth Correction and Remapping}
\label{sec:azimuth}

This module takes a series AR patches from \S \ref{sec:extraction}, 
flips ambiguity choices that create large, transient changes in azimuth
(see \citet{Welsch2013} for a detailed description),
corrects the Doppler velocity with the bias computed
in \S \ref{sec:blueshift}, computes 
the LOS unit vector, and maps these quantities onto a Plate-Carr\'ee 
(uniformly spaced in longitude and latitude) coordinate system with a pixel 
spacing of 0.03 heliographic degrees (coinciding closely with the HMI pixel size
near disk center).  
We remove differential rotation based on the fit of \citet{Snodgrass1984}, 
remove the spacecraft velocity, and then correct for the Doppler bias 
computed from \S \ref{sec:blueshift}.

\subsection{FLCT Horizontal Velocity Estimate}
\label{sec:horizvel}

We currently use the local correlation tracking code FLCT (``Fourier Local
Correlation Tracking'') \citep{Fisher2008} to estimate
horizontal flow velocities, which are then used to compute the non-inductive
contribution to the horizontal electric field described in \S \ref{sec:flct}
and \S \ref{sec:flctsubroutine}.
The original idea for local correlation tracking was first described by
\citet{November1988}.

The basic idea of the FLCT code is
to link small changes in two images taken at two closely spaced times, to
a two-dimensional flow velocity that moves features in the first image 
toward the corresponding
features in the second image.  To compute the ``optical flow'' velocity at
a given pixel location, both images are multiplied by a windowing function,
assumed to be a gaussian of width $\sigma_{FLCT}$, centered at that given pixel
location, to de-emphasize parts of
the two images that are far away from the given location.  The
cross-correlation function of the resulting sub-images is computed using
Fourier Transform techniques, and the location of the peak of the
cross-correlation function is found to sub-pixel accuracy.  
The difference between the location of
the peak and the original pixel location is assigned to be the distance of the
pixel shift (in both horizontal directions), and this shift, divided by the 
time difference
between images, is identified with the horizontal flow velocity at that pixel.  
This procedure
is then repeated for all pixel locations in the two images.  To compensate
for noisy data in the images, the algorithm allows one to select a threshold
parameter \texttt{thr}.  If the average image value has an absolute value less
than \texttt{thr}, no velocity is computed, and a mask value for that pixel
is set to zero, to indicate that no value was computed.  The velocity itself
is then set to zero as well at that pixel.  The code also allows the user to
filter the images with a low-pass filter before computing the cross-correlation
function, if there is a large degree small-scale noise.

The FLCT algorithm as originally conceived was
described in \citet{Welsch2004}, with major improvements to the algorithm 
described
in \citet{Fisher2008}.  Since the publication of that article,
co-authors Fisher and Welsch have made a number of
improvements to the algorithm and the code
to increase the accuracy and speed of FLCT.  
The developer site for the FLCT code is a fossil repository, located at:
\url{http://cgem.ssl.berkeley.edu/cgi-bin/cgem/FLCT/index}.  The latest
version can always be downloaded there.  We have also published 
a recent snapshot of the FLCT software from the above repository 
as an archive on Zenodo \citep{Fisher2020flct}.

First, the original C code as described in \citet{Fisher2008} was
written as a stand-alone executable, intended to be used while running in an IDL
session.  To read in the image data, and to write out the resulting velocity
fields, the information was communicated with IDL using disk I/O.  
While this works fine for an IDL session, it is inefficient, and doesn't 
allow the FLCT method to be easily incorporated into other software.  
Therefore, the current
version of FLCT has been re-written as a library of functions, easily callable
from C, Fortran, or Python programs.  There is still also a stand-alone
FLCT executable that has the same user interface as the original version, but 
this stand-alone code now consists mainly of I/O tasks, and calls functions 
from the FLCT library to perform the main computation.  The construction of the
library was done in consultation with co-authors Erkka Lumme and Xudong Sun
to make sure it could be used from the ELECTRICIT \citep{Lumme2017,Lumme2019} 
Python software, the JSOC's
HMI software, and from other Fortran test programs.

Second, the FLCT algorithm was rewritten so that the means of the 
sub-images described above are subtracted from the sub-images before the
cross-correlation function was computed.  We found this resulted in more 
accurate results.

Third, while the FLCT algorithm as written strictly only applies in
Cartesian coordinates, \citet{Welsch2009} described in an Appendix of that
article how data on a spherical
surface can be mapped into a conformal Mercator projection.  FLCT can then
be run in this projection, and once the velocities are derived, they can be
scaled and mapped back onto the spherical surface.  We have now modified the
FLCT code so that if the input images are given on a Plate Carr\'ee grid, the
code itself handles the mapping to the Mercator projection, runs the FLCT
algorithm to find the velocities on the Mercator map, and then re-scales and
remaps the data back to the Plate Carr\'ee grid.

Fourth, we have we performed a study of biases in the calculation of velocities
using the FLCT code.  A number of published studies have shown that FLCT tends
to underestimate flow velocities in cases where the flow velocities are known.
Two especially insightful articles on this topic are \citet{Freed2016} and
\citet{Loptien2016}.  The Appendix of \cite{Freed2016} quantifies this behavior
as a function of FLCT input parameters.
Our own study identifies a likely reason for the systematic velocity 
underestimates, in that the gaussian windowing function at the heart of
the algorithm is centered at the same pixel location in both images, even
though the second image has been slightly shifted.  
We have developed an experimental
technique to correct for this bias, which is an input option to the FLCT
library functions.

Further details regarding these changes can be viewed in the README file
in the latest FLCT distribution, along with documentation files in the
\texttt{doc} folder within the distribution.  A more complete discussion of the
updated FLCT code will be described in a future article.

To compute the FLCT flow velocities in the Plate Carr\'ee data for input to
PDFI\_SS, for each time, we use pairs of images of $B_r$ that are one timestep
behind of and one timestep ahead of the current time.  So, for
the nominal HMI cadence of 12 minutes, the images are 24 minutes apart.  The
parameter $\sigma_{FLCT}$ is chosen to be 5 pixels.  The value of the
threshold \texttt{thr} is set to 200G.  We also have chosen not to apply
any low-pass filtering of the images in the FLCT code, as we find we get
better results overall.  For now, we have not implemented the
experimental bias correction, but may apply it in the future.

\subsection{Zero Padding the Input Data}
\label{sec:padding}

We have found that the properties of the electric field solutions are improved
by adding a region of ``padding'' around the input data, in which a ribbon
of data with a width approximately 50-60 pixels is added to each of the four
boundaries, with the values of the padded
data for all 18 input arrays set to zero.  The exact width for each padded
region varies slightly, such that the resulting values of $m$ and $n$ are 
each divisible by 12.  This property of the resulting data arrays facilitates 
the use of the electric field data by the CGEM magnetofrictional model
\citet{Cheung2012}, because this property of $m$ and $n$ makes it easier to
set up computational runs that use many processors.

Adding the padding is done as the last step before defining the input data
for the electric field inversions, and is performed as part of the HMI
magnetic pipeline.  To mimic the padding operation within the
PDFI\_SS library, we have written several Fortran subroutines which do the
same thing as the JSOC padding process.  The subroutine\\
\texttt{pad\_int\_gen\_ss}\\
takes as input the unpadded values of $m$ and $n$, and ``first guess'' values
of the amounts of latitude and longitude padding, \texttt{mpad0}
and \texttt{npad0}, and computes output values of $m$ and $n$, and also
outputs the exact amounts
of padding that will be applied along each of the four boundaries, such that
the output values of $m$ and $n$ are divisible by 12.

The adjusted values of $a$, $b$, $c$, and $d$ are computed from the original
values of $a$, $b$, $c$, and $d$, plus the four padding amounts returned
by \texttt{pad\_int\_gen\_ss}, by subroutine\\
\texttt{pad\_abcd\_as\_ss}.  The padded arrays themselves can then have
their interiors filled with the original, unpadded input data, by
calling subroutine\\
\texttt{add\_padding\_as\_ss}.

In the \texttt{test\_wrapper.f} test program (see \S \ref{sec:testing}), 
which mimics the call
of \texttt{pdfi\_wrapper4jsoc\_ss} from the JSOC software, these three padding
subroutines are called to mimic the same padding procedure performed by the JSOC
software.  The trial padding values, \texttt{mpad0} and \texttt{npad0} are
set to 50 pixels.

\section{Other Applications of the PDFI\_SS Electric Field Software} 
\label{sec:broad}

Beyond the calculation of the PDFI electric field solutions in active regions,
described in \S \ref{sec:pdfinum} and \S \ref{sec:D4stuff},
there are a number of other uses for electric field solutions that use the
PDFI\_SS library.  These can be summarized as (1) curl-free electric field
solutions, useful for boundary condition matching, (2) ``Nudging'' electric
field solutions for both one and three component data-driving in numerical
simulations, (3)
global ($4 \pi$ steradian) PTD electric field solutions, and (4) Evaluation
of the curl of $\vecE$, useful for checking electric field distributions
for their fidelity in the solution of Faraday's law.  These topics will be
addressed in this section of the article.

\subsection{Curl-free Electric Field Solutions For Boundary Condition Matching}
\label{sec:laplace}

One of the important components of the CGEM project is an electric-field based
Surface Flux-Transport Model (SFTM), developed by co-authors DeRosa and Cheung,
the details of which will be described in a future publication.
A summary can be found in the CGEM Final report at
\url{http://cgem.ssl.berkeley.edu}.
The SFTM computes the global horizontal
electric field in spherical coordinates based on differential
rotation and meridional flows acting on the radial component of the magnetic
field, along with a term that describes the dispersal of magnetic flux by
supergranular motions.  The electric field in the two horizontal directions
is then used to evolve the radial magnetic field at the photosphere.
The SFTM is used in regions of the Sun for which no PDFI electric fields have
been computed, mainly outside of active regions.  Where PDFI solutions are
computed with PDFI\_SS, the model inserts the PDFI solutions into the
global domain, and evolves $B_r$ by using the PDFI solutions, rather than the
SFTM solutions.  There are two complications to doing this:  First, the
SFTM generally uses a coarser grid than is used by the PDFI solutions, and 
second, there will generally be a solution mis-match
at the boundary between the PDFI\_SS domain and the global SFTM model.  
Such a mis-match, if not corrected, results in a large, spurious curl of
$\vecE$ at the boundaries, which will then result in a spurious evolution of
$B_r$ at the boundaries or ``seams'' where the PDFI electric fields are 
inserted.  We now describe how we cope with these two complications.

In general, the SFTM is run with considerably coarser
resolution than the $0.03^{\circ}$ resolution computed by default with
$e.g.$ \texttt{pdfi\_wrapper4jsoc\_ss} in PDFI\_SS.  
Before the PDFI\_SS solutions can be inserted into the SFTM model,
both solutions must have the same grid resolution.
Our approach is to perform a flux-preserving ``downsampling'' of the
higher resolution electric field results to the same grid resolution that
is used by the SFTM.  This must be done in such a way that the magnetic flux
evolution in the coarser grid is physically consistent with that in the finer 
grid.

\begin{figure}[ht!]
\hspace{-0.25in}
\includegraphics[width=3.7in]{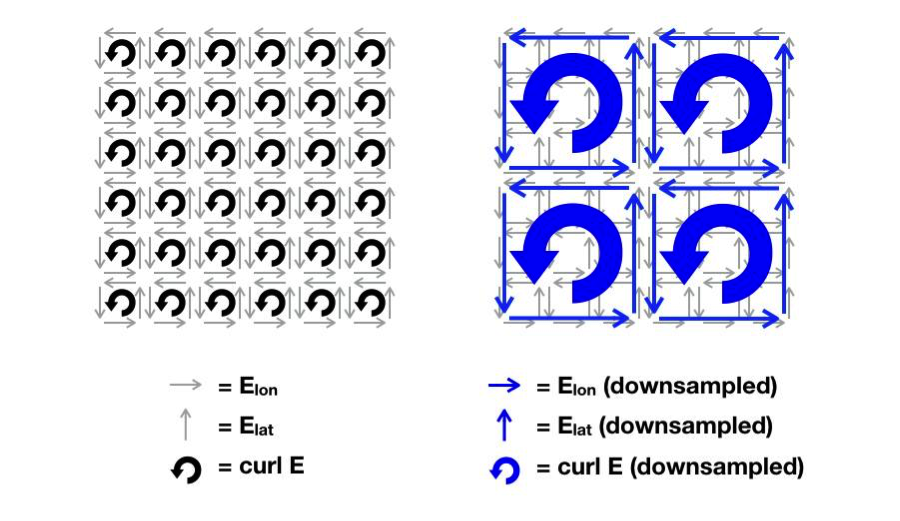}
\caption{Illustration of downsampling from the high resolution grid
to a coarser resolution grid that is used by the Surface Flux Transport Model
(SFTM), where the ``circulation''
symbol $\circlearrowleft$ represents the curl of $\vecE$ as calculated by
taking the line integral of $E_{lon}$ and $E_{lat}$ around the corresponding
cell boundary.
The electric field on the boundaries of the macro-pixels is defined
such that the line integral of $\vecE$ is the same as that from the high
resolution grid, and the evolution of $\vecB$ is consistent between the
coarse grid and the high resolution grid.
\label{fig:downsampling}}
\end{figure}

Our solution is to define ``macro pixels'' for the coarser grid in terms
of the fine grid, such that there is a whole integer number of fine grid edges
fitting within the macro pixel edges, in both horizontal directions, and that
the line integral of the electric field around the edges of a macro pixel is
equal to the line integral of the electric field along those fine grid pixels
that touch the macro pixel boundary.  This condition is illustrated 
schematically in Figure \ref{fig:downsampling}.

The downsampling, considering only horizontal components of the electric field,
can be accomplished with subroutine\\
\texttt{downsample\_ss}\\ 
when using colatitude-longitude array orientation, or subroutine\\
\texttt{downsample\_ll}\\ 
when using longitude-latitude array orientation, which is the relevant case for
the SFTM model.
Once \texttt{downsample\_ll} has been called, then the coarse-resolution
PDFI\_SS horizontal electric fields can be inserted into the SFTM model results.

For completeness, we have also written two additional subroutines,\\
\texttt{downsample3d\_ss} and\\
\texttt{downsample3d\_ll},
which downsample not only the horizontal components of the electric field
but also the radial component $E_r$ and the radial derivatives of the 
horizontal components of the electric field.  This additional information is
needed to create a downsampled three-component electric field that can be used
to compute all 3 components of Faraday's law in the coarser grid in a way that
is consistent with the solutions on the original finer grid.

Now we discuss the problem of the mismatch between electric fields in the SFTM
and the PDFI solutions, once the latter have been downsampled to the same
grid resolution in SFTM.  The
idea is to add a solution to the PDFI results which has zero curl, but which
then matches the SFTM results at the PDFI domain boundaries.

For a curl-free electric field with specified values of the tangential
electric field on its boundaries, $\dot P$ obeys the Laplace equation
\be
\grad_h^2 \dot P = 0, 
\label{eqn:laplace}
\ee
where the tangential component of the horizontal electric field on the 
boundaries is related to $\dot P$ by equation (\ref{eqn:vecephdef}).
It thus follows that the Neumann boundary conditions needed by
the FISHPACK subroutine \texttt{HSTSSP} are given by
\be
{\partial \dot P \over \partial \theta} \bigg|_{\theta=a,b} = 
r c E_{\phi}\bigg|_{\theta=a,b}
\label{eqn:elapa}
\ee
for the $n$ points along the north and south boundaries at 
$\theta=a$ and $\theta=b$, respectively,
and
\be
{\partial \dot P \over \partial \phi} \bigg|_{\phi=c,d} = 
-r \sin \theta_{i+\half}\, c E_{\theta}\bigg|_{\phi=c,d}
\label{eqn:elapc}
\ee
for the $m$ points along the
left and right boundaries at $\phi=c$ and $\phi=d$, respectively.
Once the Laplace equation for $\dot P$ is solved with these boundary 
conditions, the horizontal components of the electric field within the
domain are evaluated by taking minus the curl of $\dot P \vecrhat$.
These operations are performed by subroutine\\
\texttt{e\_laplace\_ss}\\
for arrays in colatitude-longitude orientation, and by\\
\texttt{e\_laplace\_ll},\\
where the input electric field components
at the boundaries and the output electric fields within the domain are
computed in longitude-latitude orientation.  The latter case is the one
relevant to SFTM, which uses longitude-latitude orientation exclusively.

In the SFTM model, the electric field components at the boundaries on input
to these subroutines in equations (\ref{eqn:elapa}-\ref{eqn:elapc})
are defined by the difference between the initial
SFTM electric field values and the downsampled PDFI electric field values 
at the boundary locations.

Another useful application of our curl-free electric field solutions is
to match boundary conditions assumed in computational models for the solar
atmosphere.  The PDFI electric fields solutions computed by subroutine
\texttt{pdfi\_wrapper4jsoc\_ss} can have non-zero electric field components
parallel to the domain boundaries, originating
from the contributions from gradients
in the scalar potentials.  If a computational model requires that the
electric field parallel to the boundary is zero, and if the
radial magnetic field data is flux balanced, then subroutines
\texttt{e\_laplace\_ss} or \texttt{e\_laplace\_ll} can be used to compute
a curl-free electric field solution which exactly matches the PDFI solution
for the tangential component of $\vecE$ on the boundary.  That solution can then
be subtracted from the PDFI solution, yielding solutions for $\vecE$ which
have zero tangential electric field on the boundaries, but still obey all
three components of Faraday's law.

\subsection{Nudging Electric Field Solutions}
\label{sec:nudging}

Imagine that we have a computational model for the temporal evolution of
$\vecB$ in a volume, with the lower
boundary surface of the volume coinciding with the photosphere, 
where we have evaluated $B_r$ at the centers of cells in a Plate Carr\'ee grid, 
and for which we've computed electric field solutions on the edges or rails 
that surround the cells, using PDFI\_SS solutions.  The HMI data and the 
electric field
solutions together define a time sequence of magnetic field and electric field
solutions that are consistent with one another, at least in terms of Faraday's
law.  However, the computational model will in general be based on an additional
set of physical or mathematical assumptions that can contain far more 
constraints
on how $\vecB$ behaves in the model.  Given some initial condition for
$\vecB$ at $t=0$ that matches $B_r$ at the photosphere, is there any 
guarantee that the model's evolution for $\vecB$
will be consistent with how the HMI magnetic field behaves at the photosphere?  
In general, the answer to this is no.  Given that sooner or later,
the computational model
will ``go off the rails'' as compared to how the observed
magnetic field 
changes over time, what can we do to ``nudge'' the model to get back on track?

In PDFI\_SS, we have developed a series of subroutines that are designed to
compute a nudging electric field, in effect giving the computational model a
``kick'' to make its evolution behave more consistently with the observed
magnetic field data.  The idea is to use the mis-match between the
computational model and the data to compute an electric field that is designed
to return the model's magnetic field evolution to match the photospheric
magnetic field evolution.

To illustrate this in the simplest way, we consider the computational model
to be the spherical version of the magnetofrictional coronal model developed
by \citet{Cheung2012}.  In this model, the observed values of the horizontal
components of the magnetic field are not used, and the model is constrained to
match the observed evolution of $B_r$ at the centers of the photospheric cells,
$i.e.$ on the CE grid at the photosphere:  
\be
{\delta B_r(t+\Delta t) - \delta B_r(t) \over \Delta t}
= - \grad \times \delta c \vecE_h,
\label{eqn:brtnudge}
\ee
where $\delta B_r(t) = B_r^{\,target}(t) - B_r^{\,model}(t)$.
We can use the PTD approximation to compute $\delta c\vecE_h$, where
\be
\delta c \vecE_h = -\grad \times \dot P \vecrhat , 
\label{eqn:Eerror}
\ee
and $\dot P$ obeys the
2D Poisson equation (\ref{eqn:poisson-pdot}), with the source term
equal to the LHS of equation (\ref{eqn:brtnudge}).  The boundary conditions
assumed are the same as those employed in determining $\dot P$ in
subroutine \texttt{ptdsolve\_ss} (\S \ref{sec:ptdsolve}).

A useful way to think about this is to imagine what happens
over a single timestep $\Delta t$ taken by the model, assuming that both the
target and model magnetic field values are equal at time $t$:
\be
{ B_r^{\,target}(t+\Delta t) - B_r^{\,model}(t+\Delta t) \over \Delta t}
= - \grad \times \delta c \vecE_h
\label{eqn:onetimestepnudge}
\ee
The vector quantity $\delta c\vecE_h$ is the electric field that must be 
added to
the model's electric field to return $B_r$ to its observed value at
time $t+\Delta t$.

Depending on the details of the computational model, such
a nudging step could be taken within the model's own time-advancing 
algorithm, or
alternatively, if the error is small, it can just be added to the model's
electric field on output, and applied to the calculation of $B_r$ for
the next time step
evolution.  The latter case is how the nudging electric field is used within
CGEM's spherical magnetofrictional model.

We must add an additional comment on the use of nudging when
using electric fields determined from PDFI solutions, as described in
\S \ref{sec:pdfinum}, to drive numerical models.  The procedure described in
that section recommends using strong field masks for computing solutions for
the electric field.  From the perspective of deriving electric fields from
the data that are physically meaningful in the presence of magnetic field
noise, this is the correct thing to do.
However, using these electric fields to drive a numerical model without also
using nudging can result in
inconsistencies when particular regions of the domain move from being within
the strong-field mask region to being in the weak-field region, as time evolves:
A given pixel initially within the strong-field region which has a non-zero
curl of $\vecE$ will
suddenly have zero curl, meaning that the magnetic field at that point will
no longer evolve forward in time if only the PDFI solutions of \S 
\ref{sec:pdfinum}
are used to drive the model.  The use of an additional 
nudging electric field step,
with no strong-field masks applied, will then
allow regions of the domain which move between strong-field and
weak-field regions
to evolve in a way that is consistent with the magnetic
field data in both regions.  This is how the CGEM magnetofrictional
model uses nudging electric fields to address this particular problem.

The nudging electric field in equation (\ref{eqn:brtnudge}) is formally
just the horizontal components of the PTD (inductive) electric field
from $\dot P$.  It can be computed
with subroutine\\
\texttt{enudge\_ss}\\
for input $B_r$ error terms in equation (\ref{eqn:onetimestepnudge}), and with
outputs $E_{\theta}$ and $E_{\phi}$.  If the $B_r$ error term is oriented
in longitude-latitude array order, one can use subroutine\\
\texttt{enudge\_ll}\\
to compute the corresponding components $E_{lon}$ and $E_{lat}$ in 
longitude-latitude array order.

Another useful application of \texttt{enudge\_ll} is within the CGEM SFTM
for handling the magnetic field evolution of active regions rotating onto
the disk from the east limb of the Sun.  The magnetic field observations,
if incorporated directly into the SFTM, have unwanted impacts on global
magnetic flux balance and other distortions due to the extreme viewing angle.  
Once the magnetic structure
of the rotating active region becomes clearer a couple of days after
the active region has rotated onto the disk, the nudging software can be used
to reconstruct an artificial, but physically reasonable evolution, by using
as input to \texttt{enudge\_ll} the term on the LHS of
equation (\ref{eqn:brtnudge}) equal to the difference of the
magnetic field two days after rotating onto the disk and an initial 
$\delta B_r$ of $0$.  The value of $\Delta t$ in equation 
(\ref{eqn:brtnudge}) is then set to two days.  
This allows the active region to grow in a natural way within the SFTM
without having the unwanted global impacts on the SFTM solution.  At that
point, the PDFI\_SS solutions can begin to be inserted directly into the SFTM
as described in \S \ref{sec:laplace}.

The concept of nudging can be generalized to include the calculation of
all three components of $\vecE$, in response to evolution in a computational
model which computes
all three components of $\vecB(t)$, instead of just the evolution of
$B_r (t)$.  The same general concept is used:  Differences between a model's
temporal evolution of $\vecB$ versus a ``target'' observed evolution can be
used to derive corrective values for all three components of $\vecE$ using
the PTD solutions.  The primary difference is that in the latter case, the
electric field components are computed on all the edges (rails) in a 3D layer
of voxels bisected by the photosphere, in contrast with the 2D case, in which
the horizontal electric fields are computed along the edges surrounding the
photospheric face with $B_r$ computed on the CE grid.
The subroutine\\
\texttt{enudge3d\_ss}\\
can be used to compute the electric fields on all the edges of the voxels,
given source term time derivatives for $B_r$, $B_{\theta}$, and $B_{\phi}$
and the radial thickness of the voxels.
This subroutine assumes all arrays are in colatitude-longitude order.

\subsection{Global PTD (Nudging) Solutions for $\vecE$}
\label{sec:global}

The emphasis of most of the software in PDFI\_SS is for spherical wedge
domains that subtend only a subset of the full spherical domain.  However, for
completeness, we have written some electric field software for the global
domain ($4 \pi$ steradians), including PTD solutions which could also be used
for computing nudging solutions in a global domain.  This software takes
advantage of the special case of ``global'' boundary conditions available in
some of the FISHPACK Helmholtz/Poisson equation subroutines, plus some
additional constraints to be applied at the north and south poles in
PDFI\_SS subroutines.  A good discussion of the ``global'' boundary conditions
at the poles can be found in the description of the
FISHPACK subroutine \texttt{PWSSSP} in \citet{Swarztrauber1975}.

The global versions of \texttt{enudge\_ss} and \texttt{enudge\_ll}, which
compute horizontal electric field components from a global distribution of
$\dot B_r$, are computed by subroutines\\
\texttt{enudge\_gl\_ss}, and\\
\texttt{enudge\_gl\_ll},\\
for arrays oriented in colatitude-longitude, and longitude-latitude order,
respectively.  Note that with $\dot B_r$ defined on the CE grid, there are no
values of $B_r$ defined at the north or south poles.  On the other hand, the 
output
arrays of the azimuthal or longitudinal component of the electric field, are
defined at the poles.  Note, however that
physical considerations demand that this
component of $\vecE$ must be zero at the poles, or else the behavior of
$B_r$ would become singular.  The co-latitudinal (or latitudinal) component
of $\vecE$ is not defined at the poles.  The global solutions 
for the poloidal potential $\dot P$ within these two subroutines
do not include ghost zones, in contrast to the spherical wedge solutions.

While localized spherical wedge solutions can have a flux imbalance,
the global solutions {\it must} be flux balanced, to avoid a monopole term
in $B_r$ or $\dot B_r$.  Any existing monopole term in the input data
is removed before the electric fields are computed.  The subroutines\\
\texttt{fluxbal\_ss} and\\ 
\texttt{fluxbal\_ll}\\
are used to compute a corrected input field that is flux balanced.  The flux
balance is corrected in such a way that the locations of
pre-existing polarity inversion lines are not moved.  The algorithm can be
summarized as follows:  The positive and negative magnetic fluxes within
the domain are summed separately.  Whichever polarity is the minority polarity
then has the flux in each of its constituent pixels enhanced by a constant
relative factor such that the total net radial flux is zero.  This
is similar to a technique proposed by \citet{Yeates2017}, except that in the
latter case, both polarity regions are adjusted.

In an analogy with the subroutine \texttt{enudge3d\_ss}, we can also compute
global PTD solutions for all three components of $\vecE$, given the time
derivatives $\dot B_r$, $\dot B_{\theta}$, and $\dot B_{\phi}$ given on a
global, staggered grid.  The electric field components are computed on all
the edges (rails) of a global set of spherical voxels, similar to the
geometry assumed in \texttt{enudge3d\_ss}.  The solutions are computed by
subroutines\\
\texttt{enudge3d\_gl\_ss}, and\\
\texttt{enudge3d\_gl\_ll},\\
for arrays oriented in colatitude-longitude, and longitude-latitude order,
respectively.  The poloidal and toroidal potentials are solved using the
``global'' FISHPACK boundary conditions mentioned earlier.  

There are a number
of specific considerations for the north and south poles and the left and right
boundaries that must be mentioned.
First, the toroidal potential $\dot T$ is defined at the north and south poles.
Physically, $\dot T$ is related to $\dot J_r$ at the poles through the Poisson
equation (\ref{eqn:poisson-Tdot}), so we need to evaluate the radial current 
density
at the poles.  We estimate this quantity by using Ampere's law for 
$\dot B_{\phi}$ along the highest latitudes and then dividing by the area
subtended by this small disk to estimate $\dot J_r$ at the poles.  Similarly, we
can use periodic boundary conditions for $\dot B_{\phi}$ to evaluate $\dot J_r$ 
at the left and right boundaries in $\phi$.
Second, the quantity $\dot B_{\theta}$ can be defined at the north and south
poles, but its value has no effect on the calculation of electric fields
from Faraday's law, because the amount of magnetic flux across the $\theta$
faces at the poles is zero.  Therefore, we assume $\dot B_{\theta}$ is zero
at the north and south poles, for simplicity.  Internal to these
subroutines, there are no ghost zones used in the solutions for the poloidal
or toroidal potentials.

\subsection{Evaluating the Curl of Electric Field Solutions}
\label{sec:curlofe}

In order to test the accuracy with which electric field solutions obey
Faraday's Law, we need to be able to calculate the curl of $\vecE$.  Here
we describe a number of subroutines we have written to do this.

One simple and common example is taking the radial component of the
curl of the horizontal components of $\vecE$.  Given arrays of $E_{\theta}$
and $E_{\phi}$ on the PE and TE grids, respectively, subroutine\\
\texttt{curlehr\_ss}\\
will compute $\vecrhat \cdot \grad \times c \vecE_h$ evaluated on the CE grid.
This can be compared directly to $\dot B_r$, the radial time derivative
of $B_r$ (they should be equal and opposite.)  This subroutine assumes the
arrays are all in colatitude-longitude order.

There are several approaches to computing the curl of $\vecE$ for all three
components.  If the components of $\vecE$ are computed on all the rails of
a layer of voxels, subroutines\\
\texttt{curle3d\_ss}, and\\
\texttt{curle3d\_ll}\\
can compute all three components of the curl of $\vecE$.  The quantities
returned by these subroutines are actually minus the curl of $\vecE$, so
they can be compared directly with time derivative of the three components
of the magnetic field, and should be equal.  It is important that the radial
component of $\vecE$ contain {\it only} the inductive contribution to $E_r$.
The subroutine \texttt{curle3d\_ss} assumes arrays are in colatitude-longitude
order, while \texttt{curle3d\_ll} assumes arrays are in longitude-latitude
order.  Both of these subroutines can handle either spherical wedge electric
field solutions, or global electric field solutions.

If one is dealing strictly with electric fields defined at the photosphere,
and not in a layer of spherical voxels bisected by the photosphere, there is
a different approach which can be used.  First, if radial derivatives of the
horizontal electric field components have been computed at the photosphere,
as is the case when using $e.g.$ subroutine
\texttt{pdfi\_wrapper4jsoc\_ss}, or when downloading the electric field
solutions from the JSOC, one can use subroutine\\
\texttt{curle3dphot\_ss} to compute all three components of the curl of $\vecE$,
evaluated at the photosphere.  If using quantities downloaded from the JSOC,
you will need to (1) transpose the arrays from longitude-latitude to colatitude-
longitude array order and (2) convert the units of the radial and
horizontal E-fields
from [V cm$^{-1}$] to [G km sec$^{-1}$] by multiplying by 1000, and (3) convert the units
of $\partial E_{\theta} / \partial r$ and $\partial E_{\phi} / \partial r$ 
from [V cm$^{-2}$] to 
[G sec$^{-1}$] by multiplying
by $10^8$.  It is essential that $E_r$ include {\it only} the inductive
contribution to $E_r$ when evaluating the curl.  The three components returned
by the subroutine are actually minus the curl of $\vecE$, so they can be 
compared directly with the time derivatives of the three magnetic field 
components.

If the radial derivatives of $E_{\theta}$ and $E_{\phi}$ are not available, they
can be computed with subroutine\\
\texttt{dehdr\_ss},\\
which uses as input the solutions for the poloidal potential and its
radial derivative, as returned from $e.g.$ \texttt{ptdsolve\_ss}.

\section{A Potential Magnetic Field Model for Spherical Subdomains with
PDFI\_SS}
\label{sec:potential}

For many reasons, it is useful to compute solutions for Potential
(current-free) Magnetic Field Models in a 3D domain that is consistent with
the domain we use for our electric field solutions at the photosphere.  
Because of our need for
these solutions in spherical coordinates for the CGEM project, we include
the ability to compute them within the PDFI\_SS library.  We now discuss
the equations for a Potential Magnetic Field Model using the same PTD formalism
we use to compute the inductive electric field solution at the Photosphere.

The electric current density,
$\vecJ$, can be derived from the magnetic field $\vecB$, by taking its curl:
\be
{4 \pi \over c} \vecJ = \grad \times \vecB .
\label{eqn:jdef}
\ee
We can then substitute the decomposition for $\vecB$ in terms of the poloidal
and toroidal potentials $P$ and $T$, using equation (\ref{eqn:Bptddecomp}), 
yielding
\be
{4 \pi \over c} \vecJ = \grad \times (-\grad_h^2 P \vecrhat + \grad_h
( {\partial P / \partial r} ) + \grad \times \vecrhat T ) .
\label{eqn:jptd}
\ee
Focusing for the moment on the middle term on the right hand side of equation
(\ref{eqn:jptd}), we note that the net contributions to the curl from horizontal
derivatives of $\grad_h ( \partial P / \partial r )$ are zero, but there are
contributions to the curl of $\grad_h ( \partial P / \partial r )$ from
radial derivatives (see equations (\ref{eqn:curltheta}) 
and (\ref{eqn:curlphi})).  Evaluating these contributions explicitly yields
\be
\grad \times \grad_h \bigl( {\partial P \over \partial r} \bigr) 
= - \grad \times
\vecrhat \bigl( {\partial^2 P \over \partial r^2} \bigr) .
\label{eqn:curlgradhP}
\ee
Using this result, the expression for $\vecJ$ becomes
\bea
\lefteqn{{4 \pi \over c} \vecJ = }\nonumber\\
& & -\grad \times \vecrhat 
\bigl( \grad_h^2 P + {\partial ^2 P
\over \partial r ^2} \bigr) + \grad_h \bigl( {\partial T \over \partial r} 
\bigr) -\vecrhat \grad_h^2 T .
\label{eqn:jdecomp}
\eea
The first two terms on the RHS of equation (\ref{eqn:jdecomp}) represent the 
horizontal components of the current density $\vecJ$, and the last term
represents the radial component of the current density.

For a current-free magnetic field distribution, both the horizontal and radial
contributions to $\vecJ$ must be zero.  Although a number of solutions
to this condition involving both $P$ and $T$
are possible, we choose a particularly simple one, namely:
\be
\grad_h^2 P + {\partial ^2 P \over \partial r ^2} = 0 ,
\label{eqn:bercik}
\ee
and
\be
T = 0 .
\label{eqn:laplaceT}
\ee
Thus the magnetic field solution is determined entirely by the
poloidal potential $P$, which obeys equation (\ref{eqn:bercik}).  Note that this
equation is not Laplace's Equation, in contrast to the case in
Cartesian coordinates, where the poloidal potential for a current-free field
does obey Laplace's equation (Appendix A of \citet{Fisher2010}).  We call
equation (\ref{eqn:bercik}) ``Bercik's Equation'', since to our knowledge
it was first derived by co-author Dave Bercik.  This solution will also be
a potential magnetic field solution, since a magnetic field distribution with
no currents can also be expressed as the gradient of a scalar potential.

It is useful to compare our formulation for the potential magnetic
field in terms of $P$ with the similar PTD formulation of \citet{Backus1986}
in spherical coordinates.
He defines a poloidal potential which we'll call $\scrP$ here, but which differs
from our $P$ by a factor of r:  $\scrP = P / r$.  \citet{Backus1986} shows 
in \S 4.4 of his article that
for a potential magnetic field, $\nabla^2 \scrP = 0$, ie $\scrP$ obeys the
Laplace equation.  Substituting $P / r$ for $\scrP$, one finds
that $P$ obeys equation (\ref{eqn:bercik}), showing that the 
two PTD formulations for a potential magnetic field are consistent.

Deriving the potential magnetic field distribution from the poloidal potential
$P$ has this useful property: If one needs to know either the scalar
potential or the vector potential, both are easy to derive from $P$, whereas
converting directly from the scalar potential to the vector potential, or
visa-versa, can be cumbersome.

Getting the vector potential from $P$ is particularly straightforward:  When
$T=0$, equation (\ref{eqn:Aptd}) results in
\be
\vecA^P = \grad \times P \vecrhat ,
\label{eqn:apot}
\ee
where $\vecA^P$ denotes the vector potential for the potential magnetic field.

To derive the scalar potential, we first note that the first term in Bercik's
equation, $\grad_h^2 P$, is equal to $-B_r$.  Since the left hand side of
Bercik's equation must be zero, it follows that
\be
{\partial \over \partial r} \bigl({\partial P \over \partial r}\bigr) = B_r .
\label{eqn:pdoubleprime}
\ee
Note also from equation (\ref{eqn:Bptddecomp}) that with $T=0$, the horizontal
components of $\vecB$ are given by the horizontal gradient of
$\partial P / \partial r$.  Therefore, all 3 components of $\vecB$ can be
expressed as the gradient of $\partial P / \partial r$, meaning that
the scalar potential $\Psi$ is given by
\be
\Psi = - {\partial P \over \partial r} ,
\label{eqn:psipot}
\ee
where we use the conventional definition for the scalar potential $\Psi$,
$\vecB^P = - \grad \Psi$.  In contrast
to the poloidal potential $P$, $\Psi$ {\it does} obey the Laplace equation, 
as can be seen by setting $\grad \cdot \vecB^P = 0$ when using
$\vecB^P = - \grad \Psi$.

The volume domain over which we will find a solution for $P$ will be defined at
the bottom by the photospheric boundary at $r=R_{\odot}$, and will
extend up to a radial height of an assumed ``Source Surface'' ($r=R_{SS}$), 
where the
horizontal components of the magnetic field will go to zero.  
The side walls of the
volume will coincide with the same colatitude and longitude boundaries we
use for the electric field solutions, $\theta=a$, $\theta=b$,
$\phi=c$, and $\phi=d$.  At the photospheric surface, the radial component
of the magnetic field will be defined by the observed photospheric radial
field.  For boundary conditions on the north and south side-walls
($\theta=a$, and $\theta=b$, respectively) we assume homogenous Neumann
(zero-gradient) boundary conditions for $P$.  For the left and right side-walls 
($\phi=c$, and $\phi=d$, respectively), we provide two possible boundary 
conditions: (1) periodic
boundary conditions in $P$, or (2) homogenous Neumann
boundary conditions in $P$.  The side
wall boundary conditions are tantamount to defining the behavior of the
vector potential at the boundaries.

In contrast to most spherical Potential-Field models, our solutions use
a finite difference methodology, rather than the more commonly used
spherical harmonic
decomposition.  The spherical harmonic decomposition method is not 
well-suited to high-resolution data such as that from HMI, and would 
require the spherical harmonic number $\ell$ to be several thousand to resolve 
the 360 km pixels that HMI provides.

Other existing techniques for potential field models in spherical coordinates
that do not use spherical harmonic decomposition include the
potential field model of Appendix B in \citet{vanBallegooijen2000}, the
FDIPS finite
difference code \citep{Toth2011}, the method described by 
\citet{Jiang2012a}, the Green's function approach of 
\citet{Sadykov2014}, and the finite difference model of \citet{Yeates2019}.  
The FDIPS code uses an iterative approach that applies a Krylov
technique, the method of \citet{Jiang2012a} uses a combination of spectral
derivatives in the azimuthal direction along with the \texttt{BLKTRI}
subroutine from FISHPACK for handling the other two dimensions, and the
method of \citet{Sadykov2014} derives a Green's function for the Laplace
equation in a portion of the sphere, and then integrates this with the
observed radial field on the photosphere.  Our own technique, described below
in detail,
resembles that of \citet{Jiang2012a}, except that instead of
using spectral derivatives in the azimuthal direction, we use second order
accurate finite differences in azimuth.  The finite difference code of
\citet{Yeates2019} also appears to be very similar to our approach, 
employing the poloidal potential $P$.  \citet{Yeates2019} 
assumes that gridpoints in $r$ are distributed logarithmically, rather 
than linearly.

The FISHPACK library has a capability, through the subroutine \texttt{BLKTRI},
for solving general second order elliptic
finite difference equations in two dimensions, when they can be expressed in a
block-triadiagonal form.
This turns out to be the key for deriving a 3D potential field solution using
the Poloidal Potential $P$ in a computationally efficient manner.  By
Fourier transforming the finite difference contribution to the horizontal
Laplacian from
the azimuthal term, the 3D potential field problem can be converted to a 
series of $n$ 2D finite difference equations, each of which can then
be solved with
\texttt{BLKTRI}.  Here $n$ is the number of cells in the azimuthal (longitude)
direction.

\subsection{The Solution for the Poloidal Potential $P$}
\label{sec:scrbpotss}

The broad outline of
the procedure for finding the poloidal potential $P$ is: 
(1) Convert the continuum version of Bercik's equation
(\ref{eqn:bercik}) to a second-order accurate finite difference equation for
$P$; (2) Convert the finite difference version of the azimuthal second
derivative term in the horizontal Laplacian to an eigenvalue problem, by
Fourier transforming the 2nd order finite difference contribution in the azimuth
(longitude) direction; (3) Derive a 2D finite difference expression as a
function of colatitude and radius for the amplitude of each Fourier mode,
with each mode obeying specified boundary conditions in $r$ and $\theta$;
(4) Solve each one
of these resulting 2D elliptic problems using the \texttt{BLKTRI} subroutine in
FISHPACK, and (5) inverse transform the resulting solution back to a
function in 3D space.  

These tasks are all performed within subroutine\\
\texttt{scrbpot\_ss},\\ 
which returns the solution $P$ as a three-dimensional
array.  We now describe these steps in detail.

\subsubsection{The Continuum Equation for P and Defining the Finite 
Difference Grid}
\label{sec:Lcont}

The model is based on a solution for $P$ in a 3D
domain, where $P$ obeys Bercik's Equation (\ref{eqn:bercik}).  We first
multiply equation (\ref{eqn:bercik}) by $r^2$, and can then write the resulting
equation as
\be
L_{\phi}(P) + L_{\theta}(P) + L_r (P) = 0 ,
\label{eqn:bercikr2op}
\ee
where we've decomposed the left hand side of the equation into three operators
acting on the poloidal potential $P$.
Using equation (\ref{eqn:delh2def}) for $\grad_h^2 P$, we can write these
operators as
\be
L_{\phi}(P) = {1 \over \sin^2\theta} {\partial^2 P \over \partial \phi^2},
\label{eqn:lphicont}
\ee
\be
L_{\theta}(P) = {1 \over \sin\theta} {\partial \over \partial \theta}
\left( {\sin\theta {\partial P \over \partial \theta}} \right) ,
\label{eqn:lthetacont}
\ee
and
\be
L_r (P) = r^2 \left( {\partial ^2 P \over \partial r^2} \right) .
\label{eqn:lrcont}
\ee

\begin{figure}[ht!]
\includegraphics[width=3.5in]{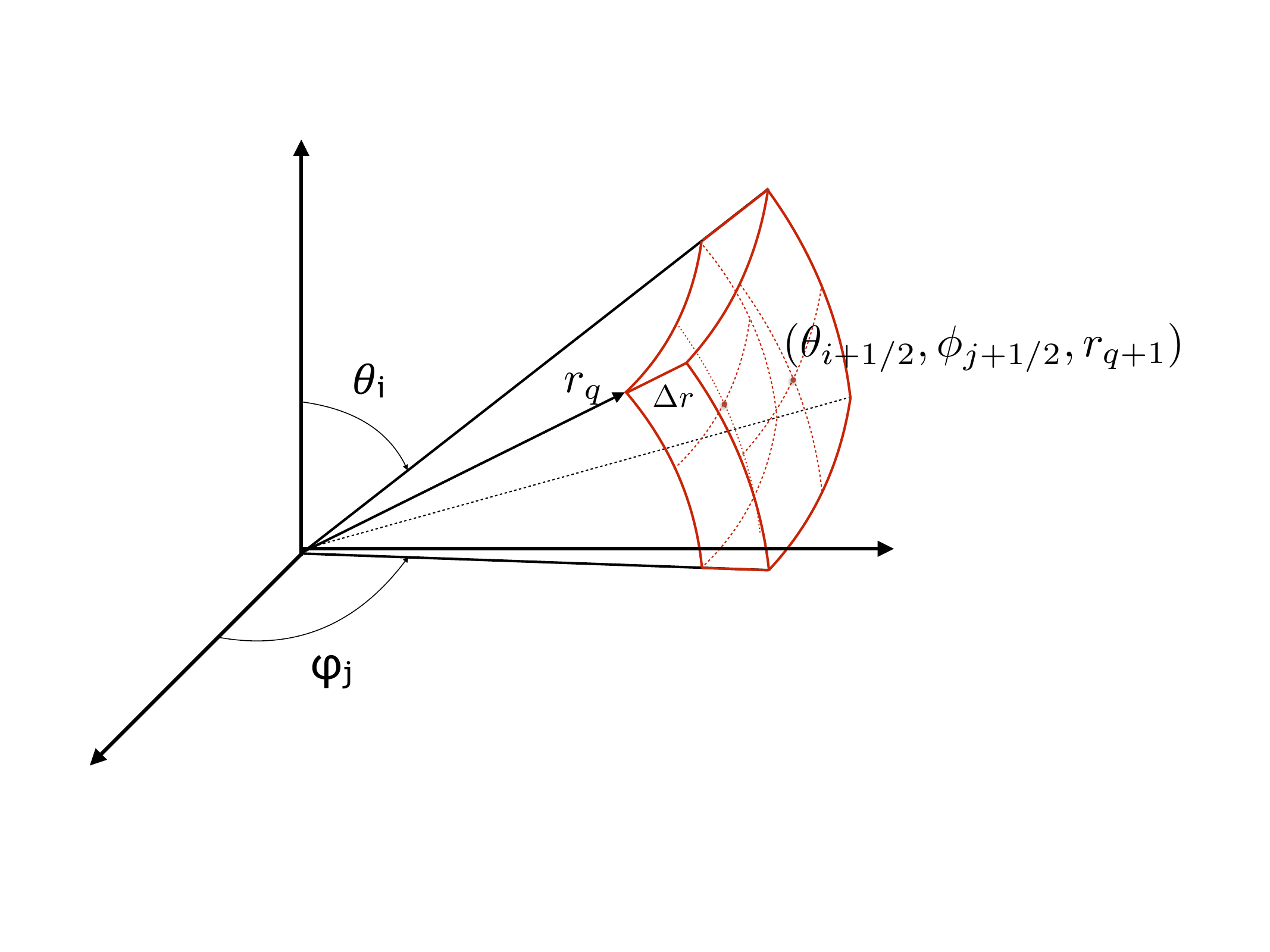}
\caption{Schematic diagram showing one voxel of our staggered 3D spherical grid
for the potential field solutions,
based on the Yee grid
concept. The Poloidal potential $P$ lies at radial face centers of each voxel.
$B_r$ is located at radial face centers, $B_{\theta}$ at $\theta$ face centers,
and $B_{\phi}$ at $\phi$ face centers.  \label{fig:wedgevoxel}}
\end{figure}

The locations where $P$ is defined must be consistent with the 2D staggered
grid locations defined in \S \ref{sec:stagger}.  In three dimensions, we
have 3D voxels instead of 2D cells.  $P$ must be defined on the radial faces of
the voxels, and in the center of these faces, to be consistent with the
location of $P$ on the CE grid at the surface of the photosphere.  We will
therefore denote the indices for $P$ with $i+\half$ as the $\theta$ index,
$j+\half$ as the $\phi$ index, and $q$ for the index of radial faces, in
keeping with the index notation described in \S \ref{sec:veccalc}.  We
will place the source-surface as the last radial face of the active part of
the domain.  If there are $p+1$ radial faces in the $r$ direction, it means
that the radial spacing $\Delta r$ is given by
\be
\Delta r = (R_{SS}-R_{\odot})/p ,
\label{eqn:deltardef}
\ee
where $p$ is the number of voxels between $R_{\odot}$ and $R_{SS}$.
The dimension of $P$ is therefore $(m,n,p+1)$ in the $\theta$, $\phi$, and
$r$ directions, respectively.  Figure \ref{fig:wedgevoxel} shows a diagram
of a voxel in this three-dimensional grid.

In this section of the article, where describing finite difference 
expressions, we assume 
index ranges that start at $0$ and go to $p$ in the
radial direction for $P$, from $\half$ to $m-\half$ in the $\theta$ direction,
and from $\half$ to $n-\half$ in the $\phi$ direction.  But keep in mind that
when examining these
expressions in the source code, we have used default index ranges in Fortran, 
where the first index starts from $1$.
\newline
\newline

\subsubsection{Fourier Transform $P$ in Azimuth and Derive Finite Difference
Equations for each Fourier Mode}
\label{sec:FTP}
We now make the assumption that the solution for $P$ can
be separated into a product of eigenfunctions in the azimuthal direction 
multiplied by coefficients which are a function of colatitude $\theta$ and 
radius $r$, where
each eigenfunction can be enumerated by a wavenumber index, $j'$.

Let 
\be
P_{i+\half,j+\half,q}=\sum^{n-1}_{j'=0}Q_{j'}^{i+\half,q} 
\Phi_{j'} (\phi' _{j+\half}) ,
\label{eqn:phisep}
\ee
where $\Phi_{j'}(\phi')$ are a series of orthogonal basis functions in $\phi'$, 
and
$Q_{j'}^{i+\half,q}$ are the amplitudes for each one of these $n$ basis 
functions.  Here, the azimuthal variable $\phi'$ has its range normalized to
be from $0$ to $2 \pi$, instead of from $c$ to $d$:
\be
\phi'= {(\phi - c) \over (d - c)} \times 2 \pi .
\label{eqn:phiprimedef}
\ee
The expression for $L_\phi$ operating on $P$ when using second-order accurate
finite differences in $\phi$ becomes, for each Fourier mode $j'$,
\bea
\lefteqn{L_{\phi} (Q_{j'}^{i+\half,q} \Phi_{j'} (\phi'_{j+\half})) =} 
\nonumber\\ 
& &  \left( {2 \pi \over d-c}\right)^2 {Q_{j'}^{i+\half,q} 
\over \sin^2\, \theta_{i+\half} \Delta \phi'^2} \times \nonumber \\
& & \left(\Phi_{j'}(\phi'_{j+\half}+\Delta \phi')+
\Phi_{j'}(\phi'_{j+\half}-\Delta \phi')\right. \nonumber \\
& & \left.  -2 \Phi_{j'}(\phi'_{j+\half})\right) .
\label{eqn:Ldeffindiff}
\eea
Here, the factor of $(2 \pi / (d-c))^2$ accounts for the scaling of the 2nd
derivative between $\phi$ and $\phi'$, and the quantity $\Delta \phi'^2$
is the square of the corresponding spacing between gridpoints in $\phi'$
($\Delta \phi' = 2 \pi / n$).

If we let the basis functions $\Phi_{j'}(\phi')$ be complex
exponentials
\be
\Phi_{j'}(\phi') = \exp(i k(j') \phi') ,
\label{eqn:Phidef}
\ee
(or sines and cosines over the same range of $\phi'$), then 
it is straightforward to show that the above finite difference expression
becomes
\bea
\lefteqn{
L_{\phi} \left( Q_{j'}^{i+\half,q} \Phi_{j'} (\phi'_{j+\half}) \right) =} 
\nonumber\\
& & 
{-2 \left( 1 - \cos ( k(j') \Delta \phi' ) \right) \over \sin^2
\theta_{i+\half}\,\Delta \phi^2 } 
\times Q_{j'}^{i+\half,q} \Phi_{j'}(\phi'_{j+\half}) .
\label{eqn:LphiFT}
\eea
Note that the factors of $(2 \pi / (d-c))^2$ 
and the expression for 
$\Delta \phi'^2 = (2 \pi / n)^2$ 
occuring in equation (\ref{eqn:Ldeffindiff}) 
result simply in division by $\Delta \phi^2$
in equation (\ref{eqn:LphiFT}).
Equation (\ref{eqn:LphiFT}) depends explicitly on wavenumbers $k(j')$, whose
values depend on the details of the Fourier transform implementation.  
In the limit of low wavenumber, the cosine expression in equation
(\ref{eqn:LphiFT}) results in the second derivative term being proportional
to $-k(j')^2$, as one would expect, but as the wavenumber increases, the 
behavior deviates from this, also as one might expect since the finite
difference expression begins to deviate from the spectral derivative result.

The most important point is that the result of applying the $L_{\phi}$ operator
to $Q_{j'}^{i+\half,q} \Phi_{j'}(\phi')$ is simply multiplication by a factor
(the eigenvalue of the operator) times that same function.  Since the other
two operators $L_{\theta}$ and $L_r$ that define Bercik's equation do not
depend on $\phi$ at all, the result will be a common
factor of $\Phi_{j'}$ for all three operators, which can then be factored out.
Furthermore, since the $\Phi_{j'}$ are all orthogonal to each other, the sum of
the three operators for the Bercik equation acting on the solution must 
be zero not only for the entire solution, but also for each individual term
in the expansion (\ref{eqn:phisep}).  Therefore,
for each value of the Fourier mode $j'$, we
need to determine only the coefficients $Q_{j'}^{i+\half,q}$.  When
evaluating the finite difference versions of $L_{\theta}$ and $L_r$, we will
therefore consider their action only on $Q_{j'}^{i+\half,q}$.

Evaluating equation (\ref{eqn:lthetacont}) using second-order accurate finite
differences, applied to $Q_{j'}^{i+\half,q}$, we find
\bea
\lefteqn{L_{\theta} (Q_{j'}^{i+\half,q}) = \nonumber} \\
& & {\sin \theta_{i} \over \sin \theta_{i+\half} \Delta \theta^2} 
Q_{j'}^{i-\half,q} 
-\, {\sin \theta_{i+1} + \sin \theta_{i} \over \sin \theta_{i+\half}
\Delta \theta^2} Q_{j'}^{i+\half,q}\nonumber\\
& & + {\sin \theta_{i+1} \over \sin \theta_{i+\half} 
\Delta \theta^2} Q_{j'}^{i+\oneptfive,q}
\label{eqn:lthetaFD}
\eea
for $i+\half$ that is not adjacent to the $\theta=a$ or $\theta=b$ boundaries.
For $i+\half = \half$, adjacent to the $\theta=a$ boundary, the homogenous
Neumann boundary condition on $P$ means that the ghost zone value
$Q_{j'}^{-\half,q}$ must be equal to $Q_{j'}^{\half,q}$.  Since the expression
for the operator that will be input into \texttt{BLKTRI}
can't involve ghost-zones, we can make that substitution into
the operator equation to eliminate $Q_{j'}^{-\half,q}$ and then find
\bea
\lefteqn{L_{\theta} (Q_{j'}^{\half,q}) = }\nonumber\\
& & -\, {\sin \theta_{1} \over \sin \theta_{\half} 
\Delta \theta^2} Q_{j'}^{\half,q} +
{\sin \theta_{1} \over \sin \theta_{\half} \Delta \theta^2} 
Q_{j'}^{\oneptfive,q}
\label{eqn:lthetaFDa} .
\eea
Doing a similar exercise for $i+\half=m-\half$, adjacent to the $\theta=b$
boundary, after applying the homogenous Neumann boundary condition we have
\bea
\lefteqn{L_{\theta} (Q_{j'}^{m-\half,q}) = \nonumber}\\
& & {\sin \theta_{m-2} \over \sin \theta_{m-\half} \Delta \theta^2}
Q_{j'}^{m-\oneptfive,q}
-\,{\sin \theta_{m-2} \over \sin \theta_{m-\half} \Delta \theta^2} 
Q_{j'}^{m-\half,q} .
\label{eqn:lthetaFDb}
\eea

For the finite difference version of the $L_r$ operator 
equation (\ref{eqn:lrcont}) acting on
$Q_{j'}^{i+\half,q}$ we have
\bea
\lefteqn{L_r(Q_{j'}^{i+\half,q}) = }\nonumber\\
& & { r_q^2 \over \Delta r^2} 
\left( Q_{j'}^{i+\half,q-1} -2 Q_{j'}^{i+\half,q} + Q_{j'}^{i+\half,q+1} 
\right) .
\label{eqn:lrFD}
\eea
The boundary condition at the last radial point, $r_{p} = R_{SS}$ is
determined by the outer boundary condition that $\Psi$ is a constant we can
set to 0, meaning
that the radial derivative of $Q_{j'}^{i+\half,p}$ is zero.
The ghost zone value, $Q_{j'}^{i+\half,p+1}$ must therefore be equal to
$Q_{j'}^{i+\half,p}$, which then results in the equation for the operator
acting on the last radial point
\be
L_r(Q_{j'}^{i+\half,p}) = {r_{p}^2 \over \Delta r^2} 
Q_{j'}^{i+\half,p-1} - {r_{p}^2 \over \Delta r^2}
Q_{j'}^{i+\half,p} .
\label{eqn:lrFDRSS}
\ee

\subsubsection{Getting the Finite Difference Equations into Block Tri-diagonal 
Form}
\label{sec:blktri}
The finite difference equations (\ref{eqn:LphiFT}-\ref{eqn:lrFDRSS}) for
each Fourier mode $j'$ can be written in a block tri-diagonal form, which
can then be used as input for the FISHPACK subroutine \texttt{BLKTRI}.
The block tri-diagonal form means that for each of the 
given values of $i+\half$
and $q$, the finite difference expressions for $Q_{j'}^{i+\half,q}$ involve
only points at $i-\half$, $i+\half$, and $i+\oneptfive$ in the $\theta$
direction, and only points at $q-1$, $q$, and $q+1$ in the $r$ direction.
Subroutine \texttt{BLKTRI} expects the coefficients for the finite difference
equations to be input through six one-dimensional arrays, 
\texttt{am, bm, cm} (each dimensioned $m$), and \texttt{an, bn, cn} 
(each dimensioned
$p+1$).  The array \texttt{am} specifies the coefficients multiplying 
$Q_{j'}^{i-\half,q}$ for each of the $m$ values of $i+\half$, \texttt{bm}
specifies the diagonal coefficient (the one multiplying $Q_{j'}^{i+\half,q}$),
and \texttt{cm} specifies the coefficient multiplying $Q_{j'}^{i+\oneptfive,q}$.
These can be found by inspection of equations
(\ref{eqn:lthetaFD}-\ref{eqn:lthetaFDb}).  The array \texttt{bm}, the diagonal
coefficients, must also include the term from $L_{\phi}$ multiplying
$Q_{j'}^{i+\half,q} \Phi_{j'}(\phi'_{j+\half})$ 
in equation (\ref{eqn:LphiFT}).
Note that for $i+\half=\half$, \texttt{am }$=0$, and for $i+\half=m-\half$, 
\texttt{cm }$=0$.  The arrays \texttt{an, bn, cn} are the coefficients
multiplying $Q_{j'}^{i+\half,q-1}$, $Q_{j'}^{i+\half,q}$, and
$Q_{j'}^{i+\half,q+1}$ in equations (\ref{eqn:lrFD}-\ref{eqn:lrFDRSS}). 
Note that at $q=p$, \texttt{cn}$=0$.  As described in more detail in
\S \ref{sec:BrphotBC}, at the photospheric level $q=0$, the values of
\texttt{an, bn, cn} will all be zero.
\newline
\newline

\subsubsection{Fourier Transform Details:  Applying Azimuthal
Boundary Conditions and Determining Wavenumbers}
\label{sec:azimuthboundary}

So far, we have said little about the eigenfunctions $\Phi_{j'}(\phi')$.
If we use the standard Fourier Transform expansion of complex exponentials,
or equivalently pairs of sines and cosines, then the eigenfunctions obey
periodic boundary conditions, and the wavenumbers in the expansion assume
their conventional values.  This is one of the options available in our
software.  The other expansion we have assumed is the half-wave cosine 
transform, in which all of the eigenfunctions are cosines, and have zero
derivative at either end of the $\phi$ domain.  In this case, the homogenous
Neumann boundary condition is achieved, and the range of $\phi'$ goes from
$0$ to $\pi$ instead of from $0$ to $2 \pi$ (in equation 
(\ref{eqn:phiprimedef}) $2 \pi \rightarrow \pi$.)

The choice of boundary conditions in $\phi$
is made through an input argument \texttt{bcn}, to
subroutine \texttt{scrbpot\_ss}.  Periodic boundary conditions are chosen by
setting \texttt{bcn = 0}, while homogenous Neumann boundary conditions are
chosen by setting \texttt{bcn = 3}.  These values correspond with the same
boundary condition values used in other FISHPACK subroutines.

In both cases, we have adopted the Fast Fourier Transform (FFT) 
software that is already included
in FISHPACK, called FFTPACK.  We make this choice primarily for convenience.
The wavenumbers $k(j')$ needed in equation (\ref{eqn:LphiFT}) are computed 
with subroutine\\
\texttt{kfft\_ss}\\
for the periodic boundary condition case, and with subroutine\\
\texttt{kcost\_ss}\\ 
for the homogenous Neumann boundary condition case.  We find the overall
speed of \texttt{scrbpot\_ss} does depend on the choice of boundary condition:
The compute time for homogenous Neumann boundary conditions is
roughly twice that for periodic boundary conditions.

\subsubsection{Matching the Solution to Observed $B_r$ at Photosphere, 
and Photospheric Boundary Conditions for Fourier Coefficients}
\label{sec:BrphotBC}

At the photospheric layer, ($q=0$, or $r=R_{\odot}$) we specify the arrays
\texttt{an, bn, cn }so that at the first array elements, all three array
values are set to $0$.  This means that at this layer, we ignore the radial
variation of $Q_{j'}^{i+\half,q}$, and instead will set the
horizontal Laplacian (determined by the
\texttt{am, bm, cm} arrays) to match the observed values of $B_r$.  To do this,
we must determine from the observed data what the value of each Fourier
coefficient $Q_{j'}^{i+\half,q=0}$ is at the photosphere.

The procedure is straightforward.  Given the observed photospheric
array of $B_r$ on the
CE grid (at $i+\half,j+\half$), we first solve the horizontal Poisson equation
\be
R_{\odot}^2 \grad_h^2 P(q=0) = -R_{\odot}^2 B_r
\label{eqn:scrbpotphot}
\ee
using FISHPACK subroutine
\texttt{HSTSSP}.  Once we have the solution, we then Fourier transform the
solution in the $\phi$ direction.  The Fourier transform then results in the
values of $Q_{j'}^{i+\half,q=0}$.  Applying the horizontal laplacian operator
(from arrays \texttt{am, bm, cm}) will result in the
Fourier transform of $-r^2 B_r$ at the photosphere, 
$(-r^2 B_r)_{j'}^{i+\half,q=0}$.
This can be used to specify a two-dimensional source term array expected by
\texttt{BLKTRI}, \texttt{y}, which is dimensioned $(m,p+1)$.
For each Fourier mode
$j'$, we can set 
\texttt{y(0:m-1,0)}=$(-r^2 B_r)_{j'}^{i+\half,q=0}$.  All the other values of
\texttt{y} for $q>0$ are set to 0, consistent with the right hand side of
Bercik's equation being set to $0$ for all radial layers above the photosphere.
In principle, one should be able to Fourier transform the observed array
$B_r$ directly and do the same thing, but in practice this produces significant
artifacts mainly due to the effects of flux imbalance in the input data.  
The procedure as described above, on the other hand, appears to be 
robust and accurate.

\subsubsection{Assembling the 3D Solution from \texttt{BLKTRI}}
\label{sec:deconstruction}

Once the photospheric values $(-r^2 B_r)_{j'}^{i+\half,q=0}$ 
are known for each value
of $j'$, we can then perform a loop over $j'$ and call \texttt{BLKTRI}
to get the solutions for $Q_{j'}^{i+\half,q}$ for all the radii from
$R_{\odot}$ to $R_{SS}$, and all the colatitudes between $a$ and $b$.  
After each solution is obtained, we store the
results in a 3D array dimensioned $(m,p+1,n)$, which is basically
the Fourier transform of $P$, but stored with the Fourier transform index
$j'$ as the last index for the array.  We then perform a final loop
and both inverse Fourier Transform the results back to real space, and
transpose
the index order such that the result is the 3D array $P_{i+\half,j+\half,q}$.
The output array $P$ is dimensioned $(m,n,p+1)$.

\subsubsection{Testing the accuracy of \texttt{scrbpot\_ss}}
\label{sec:berciktest}

We have written a subroutine to test the accuracy with which
the finite difference version of Bercik's equation is satisfied: subroutine\\
\texttt{berciktest\_ss},\\
which computes minus the horizontal laplacian of $P$, and the radial second
derivative of $P$, and provides these two quantities as output variables.
We find that these two computed quantities agree closely with one another,
with an accuracy that approaches roundoff error.  Having this subroutine
available was extremely useful in developing and debugging the code in
subroutine \texttt{scrbpot\_ss}.

\subsection{Computing the Vector Potential and Magnetic Field Components 
from $P$}
\label{sec:magneticroutines}

We first make some general comments about the properties of the solution
for $P$ determined from subroutine \texttt{scrbpot\_ss}: 
(1) For well-resolved solutions, the resulting 3D array $P$ can be
huge; (2) If the input
radial magnetic field at the photosphere has a net flux imbalance, the
$P$ solution will not reflect the flux imbalance ($i.e.$ it is consistent
with a net radial magnetic flux of zero), and (3) the solution for $P$ includes
no ghost zones.  Part of the reason for not constructing ghost-zones is
because the array is already so large.  In addition, we find that where
ghost zones are needed to evaluate curls or gradients, we can add them
on an as-needed, layer-by-layer basis.  All of the subroutines we discuss
here perform this operation internally when necessary.

While the solution for $P$ computed by \texttt{scrbpot\_ss} has no net radial
flux, we feel it is important for the potential field solutions to include
a net flux when the user desires it.  We therefore include a
subroutine,\\
\texttt{mflux\_ss},
which can be used to compute the net radial flux from the input radial
magnetic field data.  The resulting net radial flux is then used to augment
the solution for $P$ to result in a potential magnetic field solution that
is consistent with the data.  The output from \texttt{mflux\_ss} is a single
value of the net radial flux $\Phi_M$ over the photospheric domain 
defined by the values of $R_{\odot}$, $a$, $b$, $c$, and $d$.

The vector potential $\vecA^P$ within the 3D volume can be computed in a 
straightforward way from $P$ and from $\Phi_M$.  
The radial component of $\vecA^P$ is zero, so only the horizontal components
of $\vecA^P$ are computed.

The vector potential $\vecA^P$ is computed by subroutine\\
\texttt{ahpot\_ss},\\
using $P$ and $\Phi_M$ on input, and on output computing
the two components of $\vecA^P$, $A_{\theta}$ and $A_{\phi}$, each of which
are 3D arrays of dimension $(m,n+1,p+1)$ and $(m+1,n,p+1)$, respectively.
The quantity $A_{\theta}$ is computed along radial faces of the voxels, on the
PE (phi-edge) grid locations in the horizontal directions, and $A_{\phi}$ is
also on radial faces but on the TE (theta-edge) grid locations in the horizontal
directions.  To compute $\vecA^P$, there is an outer loop over the radial
index $q$.  Then for each radial layer, ghost zones are added 
to $P$ that correspond
to the boundary conditions assumed at the $\theta$ and $\phi$ edges of the 
domain.  Then $\grad \times \vecrhat P$ is computed for that given radius
using subroutine \texttt{curl\_psi\_rhat\_ce\_ss}, populating the $A_{\theta}$
and $A_{\phi}$ arrays at that radius.  After that, an additional term is added
to $A_{\phi}$:
\be
A_{\phi}^{\Phi_M} = -B_0 {R_{\odot}^2 \over r_q} \cot(\theta_i) ,
\label{eqn:aphifluxim}
\ee
where $B_0 = \Phi_m / A_{phot}$, and $\theta_i$ are the colatitude values
of the cell edges in the $\theta$ direction.  Here, the photospheric area
of the domain is given by
\be
A_{phot} = R_{\odot}^2 (\cos(a)-\cos(b)) \times (d-c) .
\label{eqn:Aphot}
\ee
After adding $A_{\phi}^{\Phi_M}$ to $A_{\phi}$,
the vector potential preserves any radial net flux
that is included with the observed radial magnetic field data.  If zero net
radial flux is desired, one can simply set $\Phi_M=0$ on input to
\texttt{ahpot\_ss}.

Once the vector potential has been computed with \texttt{ahpot\_ss},
it can be used to compute all three components of the potential magnetic field
by using subroutine\\
\texttt{curlahpot\_ss}.\\
The output from this subroutine are $B_{\theta}$, $B_{\phi}$, and $B_r$,
with each component of $\vecB$ computed at the corresponding face centers
of each voxel:  $B_{\theta}$ is computed at the $\theta$ face centers, 
$B_{\phi}$
is computed at the $\phi$ face centers, and $B_r$ is computed at radial 
face centers. 
$B_{\theta}$ is computed from radial derivatives of $A_{\phi}$, $B_{\phi}$
from radial derivatives of $A_{\theta}$, and $B_r$ computed using
subroutine \texttt{curlh\_ce\_ss} acting on $A_{\theta}$ and $A_{\phi}$.  
The dimensions of these arrays are
$(m+1,n,p)$ for $B_{\theta}$, $(m,n+1,p)$ for $B_{\phi}$, and $(m,n,p+1)$
for $B_r$.

We also have the ability to compute the magnetic field components directly
from $P$ and $\Phi_M$, if desired, with subroutines\\
\texttt{brpot\_ss}, and\\
\texttt{bhpot\_ss}.\\
Subroutine \texttt{brpot\_ss} does an outer loop over radial index $q$.
For each radial layer, ghost-zones are added to $P$ to make the solution
consistent with the applied boundary conditions on the $\theta$ and $\phi$
edges of the domain.  Then the pair of subroutines
\texttt{curl\_psi\_rhat\_ce} and \texttt{curlh\_ce\_ss} are called in
succession to compute the horizontal Laplacian of $P$, which then results
in the values of $B_r$ within that radial layer.  An additional term is then
added to the solution, $B_0 \times R_{\odot}^2 / r_q^2$, where as before,
$B_0 = \Phi_M / A_{phot}$, to account for any net radial magnetic flux.
The radial magnetic field component lies in the center of radial voxel faces.

Subroutine \texttt{bhpot\_ss} does an outer loop over the radial index $q$,
and first differences $P$ between two adjacent levels in $r$ to evaluate
$\partial P / \partial r$, after ghost zones have been added to each of the
two layers.  This derivative is evaluated at voxel centers in radius and
also in $\theta$ and $\phi$.  Then both $B_{\theta}$ and $B_{\phi}$ are
evaluated by using subroutine \texttt{gradh\_ce\_ss} to take the 
horizontal gradient of
$\partial P / \partial r$.  Here, a net radial magnetic flux plays no role in
the result, so is ignored.  The output arrays $B_{\theta}$ and $B_{\phi}$
are computed at $\theta$ and $\phi$ face centers, respectively.

The array dimensions of $B_{\theta}$, $B_{\phi}$, and $B_r$ when computed by
\texttt{brpot\_ss} and \texttt{bhpot\_ss} are identical to those computed by
\texttt{curlahpot\_ss}.  The values of all three magnetic field components
computed using the two different methods agree with each other to a high
degree of accuracy, with an error level just slightly worse than roundoff 
error.

One defect of the staggered grid formalism used here is that the horizontal
magnetic field components computed with the model at the lowest radial layer
do not lie on the photosphere,
where we have the magnetic field measurements, but instead lie half a voxel
above the photosphere.  We would like to compare and contrast horizontal
magnetic field components from the data with their potential field counterparts
lying on the photosphere.  Fortunately, we can use the finite difference
form of Bercik's equation to infer the photospheric values of the horizontal
magnetic field components.  Equation (\ref{eqn:pdoubleprime}) shows that
\be
B_r^{phot} = {\left(\partial P / \partial r\right)_{\half \Delta r} -
\left(\partial P / \partial r \right)_{-\half \Delta r} \over \Delta r} ,
\label{eqn:psibelowphot}
\ee
where $(\partial P / \partial r)_{-\half \Delta r}$ would be the ghost zone
value for $\partial P / \partial r$ just below the photosphere.
We know $B_r$ at the photosphere from the data, and from the solution for
$P$, we can difference $P$ to find
$(\partial P / \partial r)_{\half \Delta r}$, so we can
solve for the ghost zone value below the photosphere:
\be
\left(\partial P / \partial r\right)_{-\half \Delta r} = 
\left(\partial P / \partial r \right)_{\half \Delta r} - \Delta r B_r^{phot} .
\label{eqn:delpsibelowphot}
\ee
Once this has been done, we can then interpolate (average)
$\partial P / \partial r$ between values at $r=-\half \Delta r$ and
$r=+\half \Delta r$ to get the photospheric value of $\partial P / \partial r$,
\be
\left( {\partial P \over \partial r}\right)_{phot} = 
\left({\partial P \over \partial r} \right)_{\half \Delta r}-
\left({ \Delta r \over 2} \right) B_r^{phot} .
\label{eqn:dpdrphot}
\ee
Then the horizontal gradient of this quantity yields the potential field
values of $B_{\theta}$ and $B_{\phi}$ at the photosphere.  
These operations are carried out
by subroutine\\
\texttt{bhpot\_phot\_ss}.\\
The subroutine uses the solution $P$ computed by
\texttt{scrbpot\_ss} and observed photospheric values of $B_r$ on input,
and computes $B_{\theta}$ and $B_{\phi}$ at the photosphere on output.
$B_{\theta}$ lies along the TE grid, and $B_{\phi}$ lies along the PE grid.
These arrays can be compared directly to the staggered-grid values of the
observed data, for comparisons of the differences and similarities
between the observations and what the potential field model predicts.

\begin{figure}[ht!]
\hspace{-0.1in}
\includegraphics[width=3.5in]{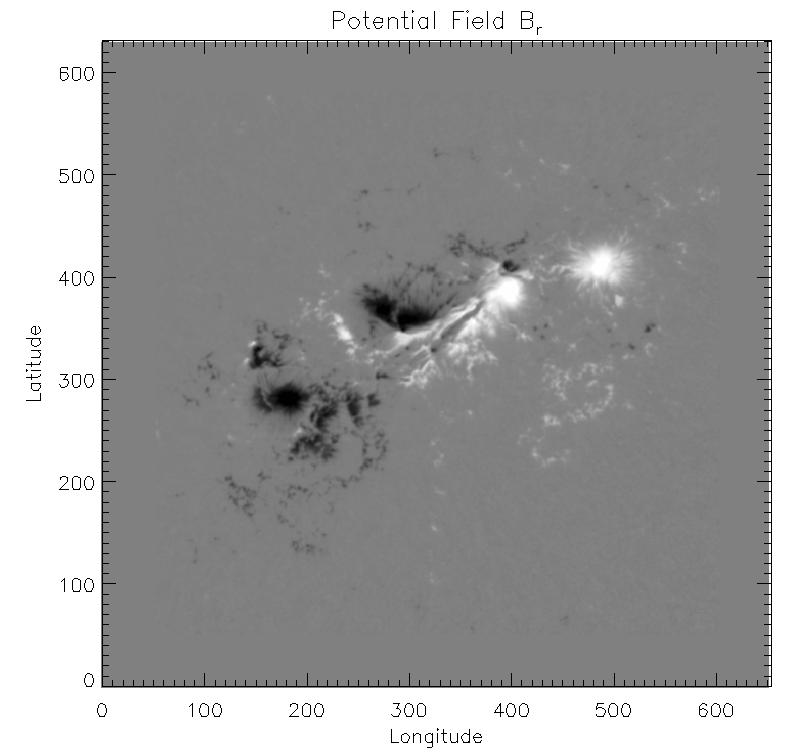}
\caption{Here we display, in longitude-latitude order, 
the image of $B_r$ from the potential field solution
at the photosphere computed from subroutines \texttt{scrbpot\_ss, ahpot\_ss}
and \texttt{curlahpot\_ss}.  The vector magnetogram data was taken from
HMI data of AR11158 on February 15, 2011, 22:47UT.
The maximum error between the given observed values of
$B_r$ and the model values are less than $10^{-6}\,$G.  The linear 
grey-scale range in the Figure is from -2000G to 2000G.}
\label{fig:brpotscrbfig}
\end{figure}

\begin{figure}[ht!]
\hspace{-0.1in}
\includegraphics[width=3.5in]{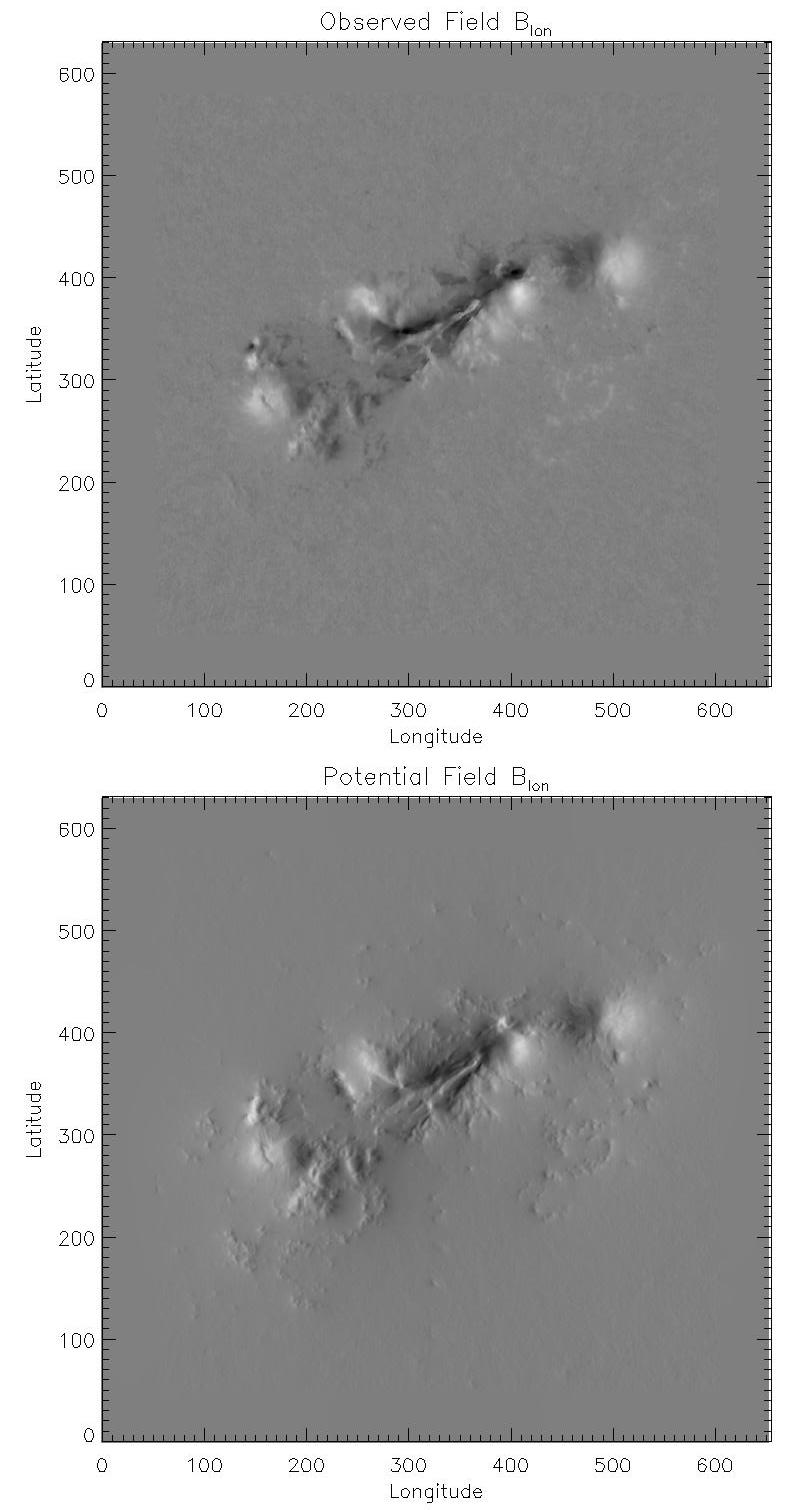}
\caption{Images of $B_{\phi}$, both from the observed
vector magnetogram data (top), and from the potential field solution at the 
photosphere (bottom), plotted in longitude-latitude
orientation.  The potential field solution
at the photosphere is computed from subroutines \texttt{scrbpot\_ss}
and \texttt{bhpot\_phot\_ss}.  The vector magnetogram data was taken from
HMI data of AR11158 on February 15, 2011, 22:47UT.
Note the significant differences between the $B_{\phi}$ values at the sheared
neutral line, and the similar behaviors at the sunspots.
The linear grey-scale range in the Figure is from -2000G to 2000G.}
\label{fig:blonplusblonpot}
\end{figure}

\begin{figure}[ht!]
\hspace{-0.1in}
\includegraphics[width=3.5in]{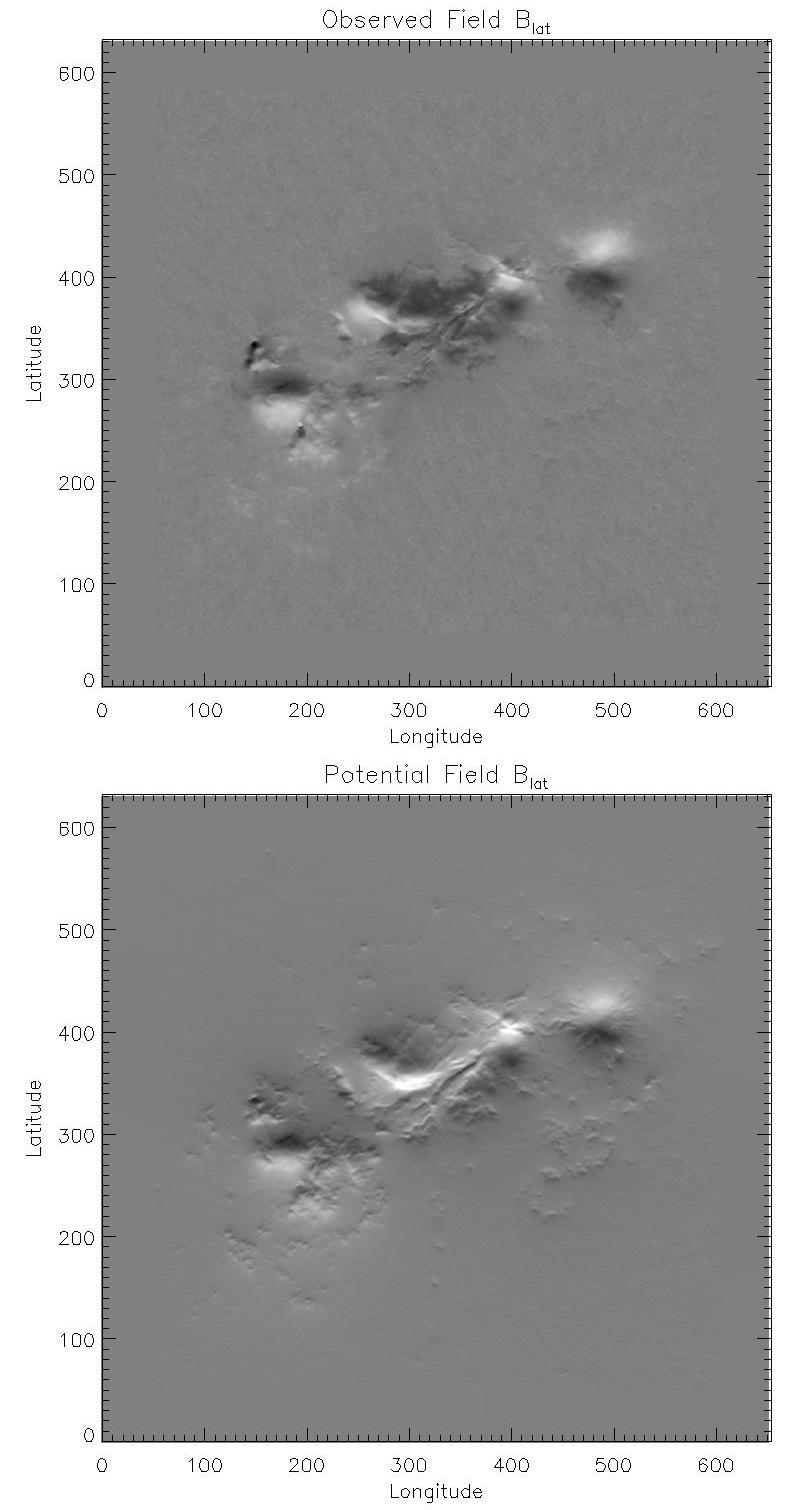}
\caption{Images of $B_{\theta}$, converted to $B_{lat}$, both from the observed
vector magnetogram data (top), and from the potential field solution at the 
photosphere (bottom), plotted in longitude-latitude
orientation.  The potential field solution
at the photosphere is computed from subroutines \texttt{scrbpot\_ss}
and \texttt{bhpot\_phot\_ss}.  The vector magnetogram data was taken from
HMI data of AR11158 on February 15, 2011, 22:47 UT.
Note the significant differences between the $B_{lat}$ values at the sheared
neutral line, and the similar behaviors at the sunspots.
The linear grey-scale range in the Figure is from -2000G to 2000G.}
\label{fig:blatplusblatpot}
\end{figure}

\begin{figure}[ht!]
\hspace{-0.1in}
\includegraphics[width=3.5in]{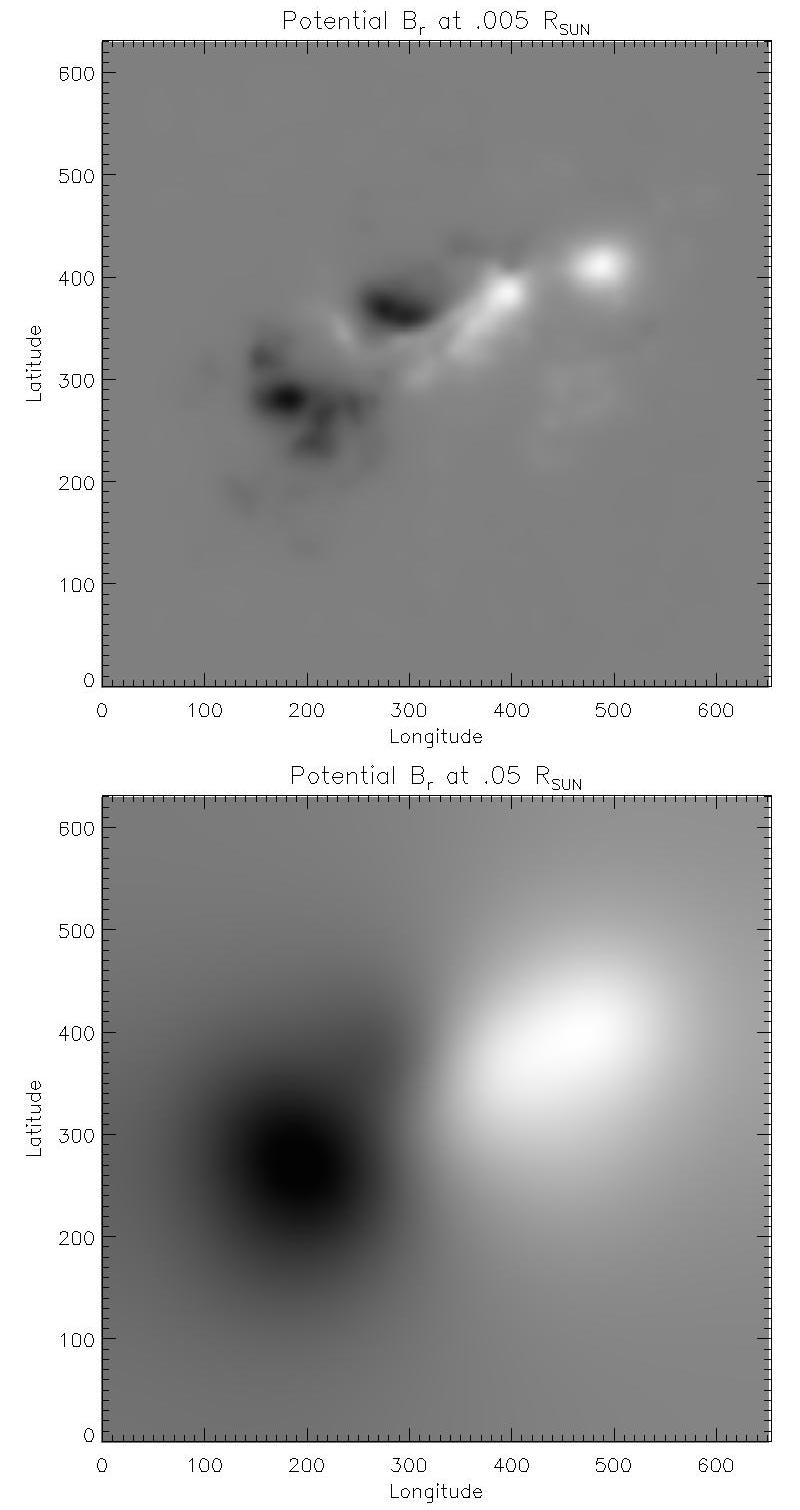}
\caption{Images of the potential field solution for $B_{r}$,
plotted in longitude-latitude
orientation, at distances above the photosphere of 0.5\% (top) 
and 5\% of $R_{\odot}$ (bottom).
The potential field solution
above the photosphere is computed from subroutines 
\texttt{scrbpot\_ss, ahpot\_ss},
and \texttt{curlahpot\_ss}.  The vector magnetogram data was taken from
HMI data of AR11158 on February 15, 2011, 22:47UT.
The linear grey-scale range in the Figure
is from
-1200G to 1200G for the top image, and from -80G to
80G for the bottom image.}
\label{fig:brpotabovephot}
\end{figure}

Figures \ref{fig:brpotscrbfig}, \ref{fig:blonplusblonpot},
\ref{fig:blatplusblatpot}, and \ref{fig:brpotabovephot} show test results of
the potential field software, using data from a vector magnetogram of
AR11158 taken on February 15, 2011, 22:47UT.  The values of $m$, $n$, and $p$
for this calculation are 632, 654, and 1000, respectively.  The value of
the assumed source-surface was $R_{SS} = 2 R_{\odot}$.  Here, the homogenous
Neumann boundary condition in $\phi$ was used to compute the solution.
On an Apple MacBook Pro (early 2015), with 16GB of RAM and an SSD disk drive,
these solutions can be derived and written to disk on timescales of roughly
ten minutes, using a single processor, with the compute time noticeably
shorter for periodic boundary conditions in $\phi$ as compared to
homogenous Neumann boundary conditions.  
The compute time is dominated by subroutine
\texttt{scrbpot\_ss}.  Once the solution for $P$ has been obtained,
evaluating the magnetic field components takes a small fraction
of the total compute time.  In the horizontal directions, the
angular resolution is close to that of HMI; in the radial direction, $\Delta r$
is roughly 700 km, about twice the horizontal spacing as that 
at the photosphere.


\subsection{Nearly Global Potential-Field Source-Surface (PFSS) Models}
\label{sec:PFSS}

While our potential field software was designed for deriving solutions on
active-region sized portions of the Sun, it can also be used for deriving
solutions that lie above a very large fraction of the solar disk.  First,
the range of $\phi$ can be extended to the entire circumference of the Sun by
simply choosing $c=0$ and $d= 2\pi$, and then choosing the periodic boundary
condition option in $\phi$ (\texttt{bcn = 0}) when 
calling \texttt{scrbpot\_ss}.  Second, we
have tested the software by choosing very small values of $a$ and values of
$b$ that approach $\pi$, with no major ill effects or artifacts
near the poles.  In particular, we've chosen $a$ and $b$ such that their
values differ from $0$ and $\pi$ by only $0.01^\circ$ without difficulty.
We then compared the morphology of the solutions at various radii computed
with the spherical harmonic based PFSS model of \cite{Bercik2019}, using 
moderate numbers for maximum spherical harmonic degree, with solutions from
the software in PDFI\_SS described here, using compatible resolution for the
finite difference equations presented here.  The solutions seemed compatible
overall at several different radii between the photosphere and the source
surface.  

\subsection{Using the Potential-Field Software to compute Electric Field
Solutions in the Coronal Volume}
\label{sec:Eptdpfss}

If instead of specifying the radial magnetic field at the photosphere, one
specifies the partial time derivative of the radial magnetic field at 
the photosphere,
subroutine \texttt{scrbpot\_ss} will find $\dot P$ instead of $P$.  In 
that case, if one then calls subroutine \texttt{ahpot\_ss} with $\dot P$ as
input, the output will be the electric field
components $c E_{\theta}$ and $c E_{\phi}$ (both with a minus sign) throughout
the coronal volume.  These are the
electric fields that correspond to the time derivative of the corresponding
potential magnetic fields in the volume.  Calling 
subroutine \texttt{curlahpot\_ss} will
then compute the time derivative of all the magnetic field components in
the volume defined by the potential field software.  If homogenous Neumann
boundary conditions in $\phi$ are chosen (\texttt{bcn = 3}), these solutions
will be compatible with the PTD solution for $c \vecE_h$ at the photosphere
computed from \texttt{ptdsolve\_ss} and \texttt{e\_ptd\_ss}, apart from the
minus sign.  It then becomes possible to perform detailed investigations of
how the electric field corresponding to the changing potential magnetic
field distributions behaves in the coronal volume.  

Electric fields computed
in this way appear to be generally consistent
with one of the contributions to the
electric field $\vecE_P$ identified with the changing potential magnetic field
in the analysis of \citet{Schuck2019}.  They identify that electric field as
coming from a solenoidal contribution, denoted $\vecSigma_P$, plus that from the
gradient of a scalar potential, denoted $\grad \Lambda_P$.  The calculation
described above will compute a solenoidal (inductive) version of
$\vecSigma_P$.  However, looking at trial test
cases indicates that along the $\unitv{\phi}$ and $\unitv{\theta}$ normal
faces of
the spherical wedge volume,
the component of the electric 
field normal to these surfaces is not zero, in contrast
to the boundary condition (32c) assumed for
$\vecSigma_P$  in \citet{Schuck2019}.
This is true for either
\texttt{bcn=3} or \texttt{bcn=0}, neither of which constrains the
behavior of the component of $\vecE$ normal to the faces.
Thus the PDFI\_SS solutions do not appear to
be consistent with the assumed boundary conditions for $\vecSigma_P$
in \citet{Schuck2019}.

Apart from this, we have not yet pursued any further detailed
studies using this possible application of 
the PDFI\_SS potential field software at this time, but simply point out 
this possibility for future work.

\subsection{Computing Energies for the Potential Magnetic Field}
\label{sec:magenergies}

Once the 3D distribution of the magnetic fields have been computed, it is
straightforward to estimate the energy in the potential magnetic field, by
either performing an integral of $B^2 / (8 \pi)$ over the computational
volume, or by estimating this quantity through a photospheric surface
integral, using Gauss' Theorem, and ignoring side and top boundaries.  
We have written three subroutines to provide such estimates.  These
subroutines are:\\
\texttt{emagpot\_ss},\\
\texttt{emagpot\_srf\_ss}, and\\
\texttt{emagpot\_psi\_ss}.\\
Subroutine \texttt{emagpot\_ss} takes as input the three 3D arrays
$B_{\theta}$, $B_{\phi}$, and $B_r$, interpolates the magnetic field components
from voxel faces to the voxel centers, and then evaluates $B^2 / (8 \pi)$
at the center of each voxel, and then sums up the magnetic energy density from
each individual voxel.  The advantage of this subroutine is that no assumptions
about side-wall boundary conditions are made; a possible disadvantage is that
magnetic field energy outside the volume is not computed and therefore
underestimated.  Energies are computed in units of [ergs].

Subroutine \texttt{emagpot\_srf\_ss} uses the fact that the volume integral
of $B^2 / (8 \pi)$ can also be written, using Gauss' Theorem, as an area
integral of $(1 / (8 \pi)) \vecA \times \vecB \cdot \vecnhat$ 
over all the surfaces
surrounding the volume, where $\vecnhat$ is the outward surface normal vector
for each surface.  However, the subroutine ignores all the surfaces
except the photosphere, and simply integrates 
$(-1 / (8 \pi)) \vecrhat \cdot (\vecA_h \times \vecB_h)$ 
over the photospheric domain.  On
input, the subroutine takes the photospheric values of $A_{\theta}$,
$A_{\phi}$, and the {\it potential field} values 
(not the observed values!) of $B_{\theta}$ and $B_{\phi}$
as $e.g.$ computed from \texttt{bhpot\_phot\_ss}.  We find that
\texttt{emagpot\_srf\_ss} tends to overestimate magnetic energies to a modest
degree when compared to the results from the volumetric integral computed by
\texttt{emagpot\_ss}.  Most likely this is due to the effects of ignoring
all the non-photospheric surfaces in the area integral.

A third subroutine for computing the magnetic energy is 
\texttt{emagpot\_psi\_ss}.  Here, the photospheric values of the potential
magnetic field $B_r$, and the scalar potential $\Psi$ are used on input
to compute the magnetic energy by integrating $\Psi B_r / (8 \pi)$ over
the photospheric surface.  This equation also results from a use of Gauss'
Theorem, when $\vecB$ is expressed as minus the gradient of the scalar potential
$\Psi$.  With our solutions derived in terms of $P$, this is less 
convenient to use
than \texttt{emagpot\_srf\_ss}, since $\Psi = - \partial P / \partial r$ is
normally not evaluated at the photosphere, but half a voxel above it.
However, equation (\ref{eqn:dpdrphot}) shows a strategy to evaluate
$\Psi = - \partial P / \partial r$ at the photosphere, which is implemented
in PDFI\_SS by calling subroutine\\
\texttt{psipot\_phot\_ss}.\\
We also find that \texttt{emag\_psi\_ss}
tends to overestimate magnetic energies compared to the volumetric
integral computed by \texttt{emagpot\_ss}, probably for similar reasons
as \texttt{emagpot\_srf\_ss}.

\subsection{Transposing Potential Field Solutions from Colatitude-Longitude to
Longitude-Latitude Order}
\label{sec:transpose3d}

The Potential Magnetic Field solutions are computed in colatitude-longitude
order for computational purposes within the PDFI\_SS software, but for
display purposes, and other applications that use longitude-latitude array
orientation, we want an efficient capability to transform the 3D solutions
from colatitude-longitude orientation to longitude-latitude orientation.

We have written several subroutines to perform these transpose operations,\\
\texttt{ahpottp2ll\_ss},\\
\texttt{bhpottp2ll\_ss}, and\\
\texttt{brpottp2ll\_ss}.\\

The subroutine \texttt{ahpottp2ll\_ss} converts the two components of the
vector potential from colatitude-longitude order to longitude-latitude order,
and also changes the sign from $A_{\theta}$ to $A_{lat}$.  The subroutine
\texttt{bhpottp2ll\_ss} does the same operation on the horizonal components
of the potential magnetic field, also changing the sign of $B_{\theta}$ when
converting to $B_{lat}$.  The subroutine \texttt{brpottp2ll\_ss} transposes
the radial magnetic field array, $B_r$.  It can also be used to transpose
the poloidal potential itself, $P$.

Since this is a very memory-intensive operation, in the potential-field
documentation file\\ 
\texttt{Potential-Fields-Spherical.txt}
located in the fossil repository 
\url{http://cgem.ssl.berkeley.edu/cgi-bin/cgem/PDFI\_SS} in the \texttt{doc} 
folder, 
there is a discussion of how one can use
a combination of C and Fortran pointers to perform the transpose operations
``in place'' if desired, which uses significantly less memory.  No change
to the existing source code for the transpose subroutines in the PDFI\_SS
library is necessary to do this; this memory-sharing operation is done entirely
in the calling program.  This same principle is also outlined in \S
\ref{sec:xbptrans}, which describes a test program for doing these transpose
operations.

\subsection{Computing Potential Magnetic Fields using $\vecB_h$}
\label{sec:psipotbh}

\citet{Welsch2016} showed an alternative method for deriving potential magnetic
fields from vector magnetogram data, where instead of matching $B_r$ at
the photosphere, one could instead match $\grad_h \cdot \vecB_h$ as measured
from the data.  They found that these solutions could result in quite
different values of $B_r$ at the photosphere as compared to the observations,
just as potential field models based on $B_r$ can have horizontal magnetic
fields that differ considerably from the observed values (see $e.g.$ Figures 
\ref{fig:blonplusblonpot} and \ref{fig:blatplusblatpot}.)  \cite{Welsch2016}
found that potential field solutions matching $\grad_h \cdot \vecB_h$ can
have substantially smaller magnetic energies than those matching $B_r$.  We
have implemented a technique for finding potential field
solutions that match $\grad_h \cdot \vecB_h$ using much of the same potential
field framework described above.  We now outline this technique, which is
included in the PDFI\_SS software.

We first note that relating the poloidal potential $P$ to 
$\grad_h \cdot \vecB_h$ in a way that can use \texttt{BLKTRI} is not as
straightforward as it was for relating $P$ to $B_r$.  However, if we solve
for the scalar potential $\Psi$ instead of $P$, then much of the mathematical
and numerical framework we use in \texttt{scrbpot\_ss} can be adapted to solve
for $\Psi$.

Writing $\vecB = -\grad \Psi$, and then taking the horizontal divergence of
$\vecB$ at the photosphere, we have
\be
R_{\odot}^2\nabla_h^2 \Psi = - R_{\odot}^2 \grad_h \cdot \vecB_h ,
\label{eqn:psiatphot}
\ee
where in the volume above the photosphere, $\Psi$ obeys the Laplace equation
\be
r^2 \nabla_h^2 \Psi + {\partial \over \partial r} \left(r^2 {\partial \Psi \over
\partial r} \right) = 0.
\label{eqn:psilaplace}
\ee
Equation (\ref{eqn:psiatphot}) is of exactly the same form as
equation (\ref{eqn:scrbpotphot}), but involving $\Psi$ and 
$\grad_h \cdot \vecB_h$ instead of $P$ and $B_r$.
Equation (\ref{eqn:psilaplace}) has a somewhat different radial term than
does Bercik's equation (\ref{eqn:bercik}), but it
is compatible with the use of \texttt{BLKTRI}.
The source-surface boundary condition at $r=R_{SS}$
will differ
between $P$ and $\Psi$.  We now describe the details of how the equation
for $\Psi$ is solved, particularly where the details differ from those in
\S \ref{sec:scrbpotss}.

\subsubsection{Finite Difference Expressions for the Laplace Equation for $\Psi$
and the Solution Procedure}
\label{sec:psipot}

Our strategy for solving the Laplace equation (\ref{eqn:psilaplace})
is identical with
that for solving Bercik's equation.  We will convert the azimuthal, colatitude,
and radial derivatives to finite differences, and then Fourier transform the
equations in the azimuthal direction, and derive finite difference equations
for the amplitude of each Fourier mode as a function of $\theta$ and $r$.
Using the same notation of \S \ref{sec:FTP}, the operators $L_{\phi}$
and $L_{\theta}$, when converted to finite difference form, will be identical
to the operators in \S \ref{sec:FTP}, where we also assume homogenous
Neumann boundary conditions at $\theta=a$ and $\theta=b$.  

As in \S \ref{sec:FTP}, we write the solution $\Psi$ as
\be
\Psi_{i+\half,j+\half,q} = \sum_{j'=0}^{n-1} Q_{j'}^{i+\half,q}
\Phi_{j'}(\phi'_{j+\half}) ,
\label{eqn:FTPsi}
\ee
where $Q_{j'}^{i+\half,q}$ is the amplitude of the coefficient of $\Phi_{j'}$
as a function of colatitude and radius indices.  
The actions of the $L_{\phi}$ and
$L_{\theta}$ operators on $Q_{j'}^{i+\half,q}$ are identical to those in
\S \ref{sec:FTP}.

Since the $L_r$ operator differs from that in \S \ref{sec:FTP}, we write
down the result:
\bea
\lefteqn{L_r (Q_{j'}^{i+\half,q}) = }\nonumber\\
& & {(r_q-\half \Delta r)^2 \over \Delta r^2} Q_{j'}^{i+\half,q-1}\nonumber\\
& & -\, {(r_q-\half \Delta r)^2 + (r_q+\half \Delta r)^2 \over \Delta r ^2}
Q_{j'}^{i+\half,q} \nonumber\\
& & + {(r_q+\half \Delta r)^2 \over \Delta r^2} Q_{j'}^{i+\half,q+1} ,
\label{eqn:lrlaplace}
\eea
for values of $q$ that are in the interior of the problem.  For the outermost
radial position $q=p$, we want to impose the boundary condition that
$\Psi \rightarrow 0$ as $r \rightarrow R_{SS}$, so that
$\vecB_h \rightarrow 0$.  This means we have a
ghost-zone value of $Q_{j'}^{i+\half,p+1} = -Q_{j'}^{i+\half,p}$.  Since
\texttt{BLKTRI} doesn't use ghost zones, we make this substitution into
equation (\ref{eqn:lrlaplace}), resulting in
\bea
\lefteqn{L_r (Q_{j'}^{i+\half,p}) = }\nonumber\\
& & {(r_p-\half \Delta r)^2 \over \Delta r^2} Q_{j'}^{i+\half,p-1}\nonumber\\
& & -\, {(r_p-\half \Delta r)^2 + 2 (r_p+\half \Delta r)^2 \over \Delta r ^2}
Q_{j'}^{i+\half,p} .
\label{eqn:lrlaplaceRSS}
\eea

From equations (\ref{eqn:lrlaplace}) and (\ref{eqn:lrlaplaceRSS}) we can
easily determine the values of the arrays \texttt{an, bn} and \texttt{cn} 
for radial
positions above the photosphere in
subroutine \texttt{BLKTRI}.

As in \S \ref{sec:BrphotBC}, the first (photospheric) values of the
\texttt{an, bn} and \texttt{cn}  arrays are set to zero.

The solution procedure for the finite difference equations for $\Psi$
is otherwise 
identical to that described in \S \ref{sec:scrbpotss} for $P$.
Getting the solution for $\Psi$ is carried out by subroutine\\
\texttt{psipot\_ss}.\\
To test the accuracy with which Laplace's equation is obeyed by $\Psi$, we have
written the subroutine\\ 
\texttt{laplacetest\_ss},\\
which outputs separately the horizontal and radial contributions to the
Laplacian.  We found this was useful in debugging \texttt{psipot\_ss}.

Finally, there is an issue that the solution to the Laplace
Equation can contain a spurious artifact in $\Psi$ that is proportional to
$r^{-1}$.  The origin of this artifact appears to be an interaction
between the source-surface boundary condition for $\Psi$, which essentially
sets $\Psi=0$ at $r=R_{SS}$, plus the homogenous
Neumann boundary conditions in $\theta$ for the $k=0$ mode at the photosphere, 
when \texttt{BLKTRI} is called for this particular mode.  This in effect allows
$\Psi$ at the photosphere to ``float'', $i.e.$ to have an arbitrary 
constant added to it.  
At radii in-between the photosphere and source-surface,
the solution is connected by an $r^{-1}$ dependence which satisfies the
Laplace equation.  This solution, while mathematically legitimate, has no
physical basis, and results in a sometimes large, horizontally uniform 
radial component $B_r$ which must be removed from the solution.  We have
written a subroutine\\
\texttt{psi\_fix\_ss},\\
which evaluates and removes this artifact, by evaluating the $B_r$ term at
the photosphere, and then removing it from the entire solution volume.  This
subroutine is called just before exiting subroutine \texttt{psipot\_ss}.
Subroutine \texttt{psi\_fix\_ss} can also be used to {\it impose} an observed
nonzero net radial flux to the solution for $\Psi$, if desired.

\subsubsection{Getting The Potential Magnetic Field Components from $\Psi$:
Subroutine \texttt{bpot\_psi\_ss}}
\label{sec:bpotpsiss}

In principle, once $\Psi$ is known, the magnetic field components can be found
by simply taking minus the gradient of $\Psi$.  In practice,
there is a major challenge to deriving the magnetic field components from
$\Psi$: gradients of $\Psi$ put the magnetic field values in a different
location in the grid from where we need it.  We now describe how we evaluate
the magnetic field components at the grid locations where we need them.

The scalar potential $\Psi$ lies at the centers of the radial faces of our
voxels. In the horizontal directions, $\Psi$ lies on the CE grid (see Figure
\ref{fig:wedgevoxel}).  The grid location of $\Psi$ in the horizontal 
directions is different from the grid locations of the scalar
potentials $\psi$ for computing the electric field contributions
at the photosphere, where
these various scalar potentials all lie on the COE grid.  This is why we have
used the notation of the upper-case $\Psi$ to distinguish the scalar potential
for the potential magnetic field derived from $\vecB_h$ from the notation of the
lower-case $\psi$ contributions to the photospheric electric field.  

When we
take horizontal gradients of $\Psi$, the results lie on the
$\theta$ and $\phi$ edges of the radial faces; but we
need them at mid-points in radius (mid-way in $r$ between radial faces)
at the centers of
the horizontal faces of the voxels in $\theta$ and $\phi$).  Similarly, when
we take the radial component of the gradient of $\Psi$ to get $B_r$, it is
evaluated at mid-points in $r$ within the voxels, but we need $B_r$ on radial
face centers.

We now describe how we interpolate the horizontal magnetic field components
from the $\theta$ and $\phi$ edges of radial faces to the $\theta$ and $\phi$
face centers of our voxels.  

If we imagine that we have another set of voxels
(``offset voxels'') that are offset by $\half \Delta r$ from our grid voxels, 
then we also imagine that each of our voxels contains the upper 
half of an offset voxel in the bottom half of our
given voxel, and the bottom half of the next highest offset voxel 
in the top half of our voxel.  
We want
the $\theta$ and $\phi$ magnetic fluxes from our voxel to match the flux
from the top half of the lower offset voxel, plus the flux from the bottom half
our the upper offset voxel.  These considerations result in the following
expression for the interpolated horizontal magnetic field components:
\bea
\lefteqn{B_{\theta}^{i,j+\half,q+\half} = {1 \over r_{q+1}^2 - r_{q}^2} 
\times \nonumber}\\
& & \left( (r_{q+\half}^2-r_{q}^2) 
B_{\theta}^{i,j+\half,q} \right. \nonumber\\ 
& &  + \left. (r_{q+1}^2-r_{q+\half}^2) 
B_{\theta}^{i,j+\half,q+1} \right) ,
\label{eqn:btinterp}
\eea
and
\bea
\lefteqn{B_{\phi}^{i+\half,j,q+\half} = {1 \over r_{q+1}^2 - r_{q}^2}
\times \nonumber}\\
& & \left( (r_{q+\half}^2-r_{q}^2)
B_{\phi}^{i+\half,j,q} \right. \nonumber\\
& &  + \left. (r_{q+1}^2-r_{q+\half}^2)
B_{\phi}^{i+\half,j,q+1} \right) ,
\label{eqn:btinterp}
\eea
where we use the fact that the area of the side faces of a voxel are
proportional to $r_{q+1}^2-r_{q}^2$ if the bottom and top
radial faces of the voxel are
located at $r_q$ and $r_{q+1}$.

Now that we have values of $B_{\theta}$ and $B_{\phi}$ interpolated to
horizontal face centers, we can evaluate $\grad_h \cdot \vecB_h$ at voxel
centers, using subroutine \texttt{divh\_ce\_ss}, where the result projected
onto the horizontal directions lies on the CE grid.  We can then use the
constraint $\grad \cdot \vecB = 0$ to derive the radial derivative of
$r^2 B_r$:
\be
{\partial \over \partial r} ( r^2 B_r ) = - r^2 \grad_h \cdot \vecB_h .
\label{eqn:dbrdr}
\ee
Since $B_r$ evaluated from $-\grad \Psi$ is also co-located with
$\grad_h \cdot \vecB_h$, we can use equation (\ref{eqn:dbrdr}) to extrapolate
$B_r$ to the upper and lower radial faces, where we want the values.  
The evaluation of $\grad_h \cdot \vecB_h$ and the
extrapolation to radial faces is done within subroutine\\
\texttt{br\_voxels3d\_ss}.\\

All of these tasks are accomplished within subroutine\\
\texttt{bpot\_psi\_ss}.\\
If a non-zero net radial flux is desired for the potential field, one
can specify its value in the input variable \texttt{mflux} in the call
to \texttt{bpot\_psi\_ss}.

If one wants to compute a solution for $\Psi$ itself which is consistent with 
an imposed
net radial flux, one can call subroutine \texttt{psi\_fix\_ss} with a non-zero
value of the net radial flux \texttt{mflux}.
Doing this is adviseable if using
$\Psi$ to compute magnetic energies with subroutine \texttt{emagpot\_psi\_ss}.

\subsubsection{Testing Potential Field Models that use $\vecB_h$ at Photosphere}
\label{sec:bhtest}

How can we characterize the accuracy of solutions to the potential field models
that use photospheric values of $\vecB_h$?  The interpolation
and extrapolation steps described in \S \ref{sec:bpotpsiss} will introduce some
amount of error into the magnetic field solutions, as compared to the
solutions based on $B_r$, where these steps are not needed. Our objective here
is to provide a method for estimating errors in the solution obtained 
from $\vecB_h$.

The test described here is based on the following procedure: (1) First, obtain
the potential-field solution that matches observed photospheric values of
$B_r$ using subroutines \texttt{scrbpot\_ss}, \texttt{ahpot\_ss}, and
\texttt{curlahpot\_ss}, along with subroutine \texttt{bhpot\_phot\_ss}
to compute the horizontal potential field components at the photosphere.
(2) Using the photospheric potential field components $B_{\theta}$ and
$B_{\phi}$ computed from the above, compute the scalar potential
$\Psi$ to match $\grad_h \cdot \vecB_h$ at the photosphere by calling
subroutine \texttt{psipot\_ss}.  (3) Compute the magnetic field components
by calling subroutine \texttt{bpot\_psi\_ss}.  (4) Compare the original
solutions for $B_{\theta}$, $B_{\phi}$, and $B_r$ with those computed from
Step 3, and evaluate the discrepancies.

\begin{figure}[ht!]
\hspace{-0.1in}
\includegraphics[width=3.5in]{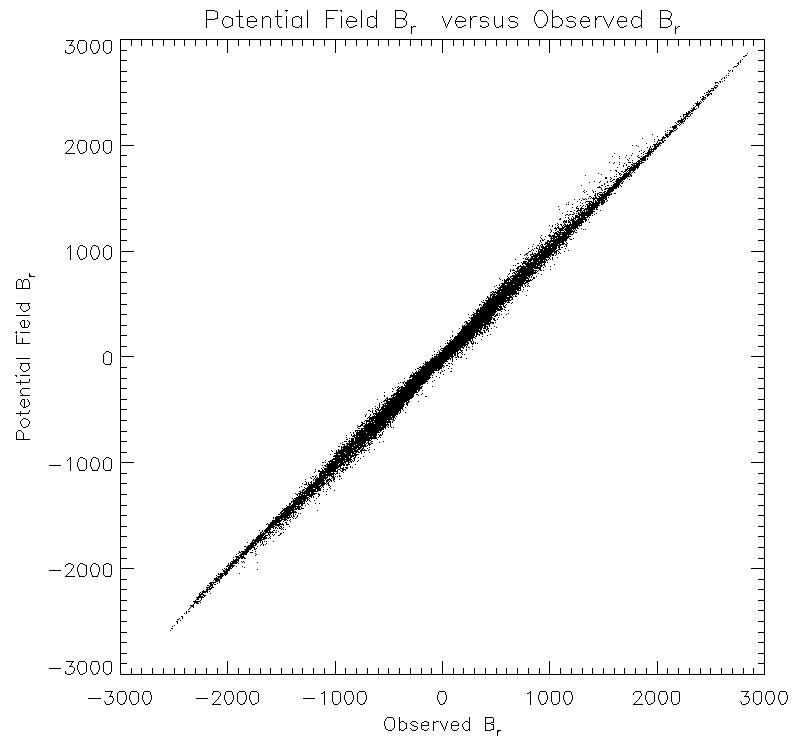}
\caption{Scatterplot of photospheric values of $B_r$ computed from potential 
field solution derived from $\vecB_h$, versus the observed values of $B_r$.  
The components of
$\vecB_h$ used as input were computed from the potential field solution that
was based on the observed values of $B_r$.  The scatter away from a straight
line measures the error introduced by the interpolation/extrapolation
procedures needed to get the magnetic field components located in their
correct positions on the grid.  RMS errors in $B_r$ are $\sim$ 18G,
smaller than quoted HMI errors for $B_r$.}
\label{fig:scatterbrpot}
\end{figure}

Figure \ref{fig:scatterbrpot} shows the recovered values of $B_r$ versus
the original values of $B_r$ from the data, showing an RMS difference of
$\sim$ 18G.  For comparison, the scatter-plots of the recovered values of
$B_{\theta}$ and $B_{\phi}$ (not shown) look like straight lines, and have
much smaller errors.  Similarly, the original potential field model based
on $B_r$ shows errors of $\sim 10^{-6}\,$G, in recovered versus observed values
of $B_r$, close to roundoff error.  Thus
the largest source of error seems to be the interpolation and extrapolation
procedures in subroutine
\texttt{bpot\_psi\_ss} that were needed to compute $B_r$ at radial face centers.
While these errors are visible here, they are smaller than the quoted HMI
errors in $B_r$, so we feel the solutions are accurate enough for many 
scientific studies.

\subsubsection{Applications of Potential Field Solutions from $\vecB_h$}
\label{sec:applicationsvecBh}

\citet{Welsch2016} proposed the idea that potential field models derived
from vector magnetograms can include observed data from $\vecB_h$ as well
as $B_r$, and proposed composite models where both solutions can be used,
with weights for each based on measurement errors for the different
components of $\vecB$ considered separately.  Because our potential field
software in PDFI\_SS includes the ability to compute both solutions, this can
be done in a straightforward way.  

\begin{figure}[ht!]
\hspace{-0.1in}
\includegraphics[width=3.5in]{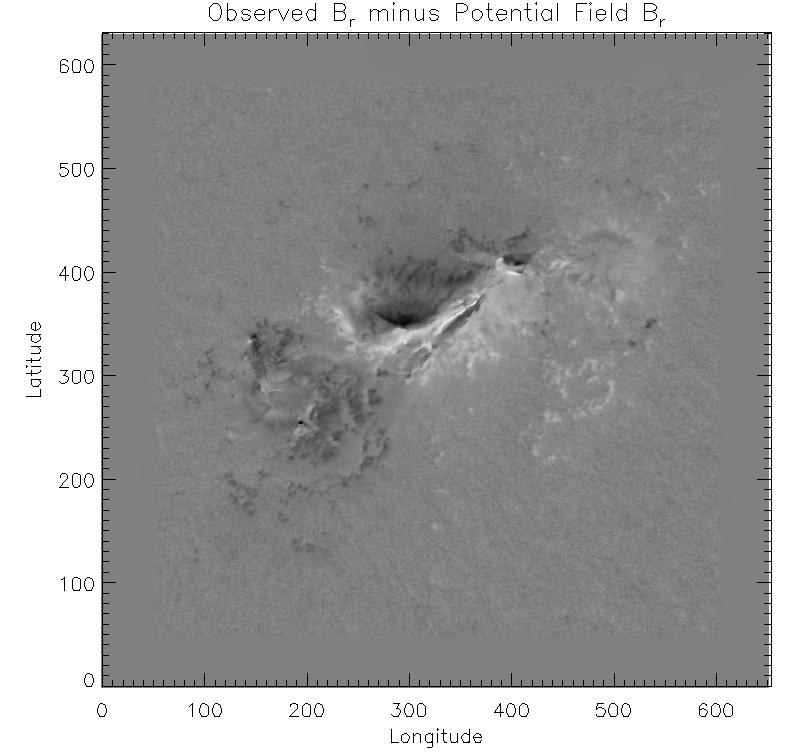}
\caption{Difference between $B_r$ computed from the
potential field that matches the observed values of $\vecB_h$ and the
observed values of $B_r$, from February 15, 2011, 22:47UT.  
\citet{Welsch2016} suggest that difference images
such as this result from horizontal currents flowing in the
solar atmosphere.  Using the solutions from subroutines \texttt{psipot\_ss} and
\texttt{bpot\_psi\_ss}, these difference images are straightforward to compute.
The linear grey-scale range used to display this image is -1500G to 1500G.}
\label{fig:brpotbhminusbr}
\end{figure}

We also note that \citet{Welsch2016} proposed that differences in $B_r$ between
the observations and the $\vecB_h$ potential-field solutions may provide
a diagnostic for the existence of horizontal currents.  For AR 11158, we show
such a difference image of $B_r$ in Figure \ref{fig:brpotbhminusbr}.  
It is interesting that in the sunspots there is only a slight difference in
$B_r$, but there are also large-scale patterns elsewhere
in the active region showing
a significant difference.  This potential-field software makes more detailed
studies a practical possibility.

\section{Using PDFI\_SS to Compute Electric Field Inversions in Cartesian 
Coordinates}
\label{sec:cartesian}

There are times when it makes more sense to compute electric field
solutions in Cartesian
coordinates rather than in the spherical coordinates assumed in PDFI\_SS.
How can we adapt the PDFI\_SS library for Cartesian coordinates without creating
a completely separate version?  Our approach to answering this question is
to note that to an excellent approximation, a very small patch near the equator
of a very large sphere will be, for all intents and purposes, a Cartesian
coordinate system.

Suppose that we want to perform electric field inversions in a Cartesian
coordinate system with $N_x$ cells in the $x$ direction, and $N_y$ cells
in the $y$ direction, and that each cell in $x$ has a width of $\Delta x$
and each cell in $y$ has a width of $\Delta y$.  The total extent of the
domain in the $x$ and $y$ directions is thus $L_x = N_x \Delta x$,
and $L_y = N_y \Delta y$.  We want to map this domain
onto the surface of a large sphere with radius $R$, with the $y$ range
bisected by the equator, at latitude zero 
(or colatitude of $\half \pi$.)
The important point is to make the value of the
sine of the colatitude $\theta$ for all the cells close to unity, meaning that 
variations due to spherical geometry are negligible.  Specifically,
we want the colatitude range $b-a$ to subtend a small angle, $\Delta \Theta$,
such that $\sin a$ and $\sin b$ are close to unity.  Given $\Delta \Theta$,
we then have 
\be
a = \half \pi - \half \Delta \Theta ,
\label{eqn:acart}
\ee
and 
\be
b = \half \pi + \half \Delta \Theta .  
\label{eqn:bcart}
\ee
Setting $R \Delta \Theta = L_y$, 
we obtain
\be
R = L_y / \Delta \Theta  = N_y \Delta y / \Delta \Theta,
\label{eqn:Rcart}
\ee
where $R$ is the desired radius of the sphere.  Once $R$ is determined, the
longitude range $d-c$ can be determined as 
\be
d-c = L_x / R = N_x \Delta x / R .
\label{eqn:dmccart}
\ee
If one knows the longitude of the left boundary from solar disk observations,
and wishes to preserve it, then
that value can be assigned to $c$, and then $d$ can be assigned by adding
to $c$ the results of equation (\ref{eqn:dmccart}).  If the value of $c$
is unimportant, we can assign
\be
c=0
\label{eqn:ccart}
\ee
and
\be
d = L_x / R = N_x \Delta x / R .
\label{eqn:dcart}
\ee
These values of $a$, $b$, $c$, and $d$, along with $R$, define the spherical
geometry parameters for a Cartesian coordinate system on the surface of a
large sphere.  The only question is what value to assign for $\Delta \Theta$.
Our experience has been that setting $\Delta \Theta = 1 \times 10^{-4}$ has
worked well for most of the cases we have tried.  It is small enough that
the sine of colatitude is essentially unity, but large enough that roundoff
errors in equations (\ref{eqn:acart}) and (\ref{eqn:bcart}) are not important.

The subroutine\\
\texttt{car2sph\_ss}\\
will compute the resulting values of $R$, $a$, $b$, $c$, and $d$, given an
input value of $\Delta \Theta$ and input values of $\Delta x$ and $\Delta y$,
along with the number of cells in the colatitude and longitude directions,
$m$ and $n$.  This subroutine will assign a value of $0$ to $c$.
It is important to remember that $n$ and $m$ are the same as 
$N_x$ and $N_y$ in
the above discussion of the Cartesian grid.  If $\Delta \Theta$ is set to $0$,
then internal to the subroutine, a value of $1 \times 10^{-4}$ will be used.
After output from \texttt{car2sph\_ss}, if one wishes to keep an original
value of the left-most longitude to solar disk coordinates, then that value
should be added to the values of $c$ and $d$ on output from the subroutine.
The variable \texttt{rsun} used in many of the PDFI\_SS subroutines should
then be assigned to the output value of the variable \texttt{rsph} from
the \texttt{car2sph\_ss}.

One complication with going from Cartesian to spherical
coordinates using this scheme is that the input Cartesian data will be
arranged in longitude-latitude index order, whereas most of the mathematical
operations in PDFI\_SS are performed using colatitude-longitude index
order.  

If one is performing the entire PDFI electric field solution,
and the input data are not yet on the staggered grid, the
subroutine \texttt{pdfi\_wrapper4jsoc\_ss} can be used on the Cartesian
input data, since this subroutine expects the input data to be
in longitude-latitude order, and performs the needed interpolations to the
staggered grid locations.
If one is performing a more customized calculation, or some other
operation using the PDFI\_SS software such as potential magnetic field 
solutions, the user will need to use the transpose and interpolation
subroutines described in \S \ref{sec:interpol} to get the data into
colatitude-longitude index order on the staggered grid locations.

\section{Compiling the PDFI\_SS Library, and linking it to Other Software}
\label{sec:compiling}
In this section of the article, we will first describe the history
of the PDFI\_SS library development.  Then we will discuss some choices
made in writing the Fortran source code for PDFI\_SS. We will then describe 
how to compile
the library, followed by discussions of how to link the library to other 
software written in Fortran, C/C++, and Python.  We will end by describing 
the use of the legacy PDFI\_SS software written in IDL.

\subsection{The History of the PDFI\_SS Software}
\label{sec:history}

The first published study in which time dependent vector magnetic fields were
used to derive electric fields was that of \citet{Fisher2010} (earlier,
\citet{Mikic1999}
described electric field solutions determined from time derivatives
of the radial component of $\vecB$).  In this case,
ANMHD magnetoconvection simulation
data, \citep{Welsch2007} which had known electric field
solutions, were used to create a vector magnetogram data sequence, which could
then be analyzed by computing
PTD solutions for $\vecE$.  While there was a broad resemblance between the
inverted PTD solutions and the actual electric fields, there were also a number
of artifacts.  The authors developed
an ``iterative'' technique to compute an additional scalar potential, 
whose gradient could
be added to the PTD solutions to make $\vecE$ and $\vecB$ perpendicular to each
other, resulting in a moderately better agreement between the inverted
and actual electric fields.  A further investigation in that article
used a variational approach, to impose a
``smallness'' constraint to the electric field solutions, which resulted in
a poor match with the actual ANMHD electric fields.  We subsequently gave 
up on using the variational approach.

The PTD Poisson equations were
solved in \citet{Fisher2010} using a Fortran
version of the Newton-Krylov technique, originally developed
for the first version of RADMHD \citep{Abbett2007}, since the required
boundary conditions were inconsistent with the use of FFTs.
Solutions obtained with the iterative method, which used repeated solutions 
of a Poisson equation, were performed in IDL using FFTs.

A great improvement in the accuracy of the electric field inversions of the
ANMHD simulations was made in \citet{Fisher2012b}, in which it was realized that
adding information about Doppler shifts, which can be measured, resulted in
dramatically better solutions for the electric field.  They derived Poisson
equations for contributions from both Doppler shifts and from horizontal flows
derived from Local Correlation Tracking, which were then solved in 
IDL using FFTs, with the solutions added to the PTD
solutions obtained with the Newton Krylov software.

In 2011-2012, co-authors Maria Kazachenko, Brian Welsch, and 
George Fisher realized
they needed more efficient software for solving the Poisson equations, in
which many more types of boundary conditions could be applied, and which would
be faster than their existing Newton-Krylov code.
They tested several numerical techniques, and concluded that the elliptic
equation package FISHPACK was ideally suited to these tasks.  They proceeded
to write a very general executable program in Fortran, which could be 
spawned from IDL, which would read input data, compute the solutions
using FISHPACK, and then write
the solutions to a file which could then be read back into the IDL session.  
This software model for PDFI existed from roughly 2012 through 2015, 
during which
the centered, Cartesian version of the PDFI software was developed.  We now
refer to this version as PDFI\_CC, where ``CC'' refers to 
``Cartesian-Centered''.  This is the version of the software that was used to
perform the research described in \citet{Kazachenko2014,Kazachenko2015}.

Starting in 2013, the above co-authors received funding for the CGEM 
project, \citep{Fisher2015}
in which they proposed to take the existing PDFI software and (1) convert it to
spherical coordinates, and (2) re-write it in an efficient computer language 
that could be run automatically from the SDO JSOC.  This process happened in
several stages.  First, the Cartesian IDL source code had to be 
converted from 
Cartesian to spherical coordinates.  This process took roughly six months,
and maintained the use of
a centered grid. (This version was called PDFI\_SC, where ``SC'' 
denotes ``spherical
centered'').  In the meantime, by studying MuRAM MHD simulation
results obtained from Matthias Rempel at HAO/NCAR, which had turbulent 
structures at the scale of the grid, the authors realized
that the centered grid finite difference formulation was simply
unable to obey Faraday's law accurately when the solutions
were so highly structured.  They realized they needed to convert their
finite difference equations into a conservative,
staggered grid coordinate system.  After investigating
several different formulations of staggered grid systems, they 
finally arrived at the system described in \S \ref{sec:stagger}.

Once the PDFI\_SC version was written and working in IDL, the next
step was to convert the IDL code from the spherical
centered grid to the spherical staggered grid
scheme.  This process occurred during the first half of 2015.  By July of 2015,
an IDL version of PDFI\_SS was operational, and had successfully
undergone a number of tests.

To deliver the software in a form which could be run automatically at the
SDO JSOC,
the co-authors knew that the IDL code would have to be converted to Fortran, 
since it relies so heavily on the FISHPACK Fortran library.  The conversion of 
the code from IDL to Fortran was done during the last half of 2015 and early
2016.  Since that time, nearly all development effort has been on the Fortran
version of the software.  The fortran version of PDFI\_SS
now contains a much broader spectrum of capabilities than the original IDL
version did.  While we continue to keep the IDL legacy
version within the PDFI\_SS
developer site, we no longer actively maintain the IDL branch of the software.
We do find that the existing IDL code, particularly the procedures for
performing vector calculus operations,  can still be useful when analyzing 
output from PDFI\_SS.

\subsection{Comments on PDFI\_SS Fortran Source Code Choices}
\label{sec:commentsfortran}

There were a number of choices made in how the PDFI\_SS Fortran code
was written.  Here
we briefly comment on these, and discuss the motivations for
these choices.

First, the calling arguments for all the subroutines include input and
output arrays, as well as other important information provided as single real
scalar values or as integers.  All quantities defined as arrays have their
dimensions defined in terms of integer values passed into the 
subroutine by
the user.  Modern Fortran allows one to determine array sizes and shapes
by querying the attributes of these arrays, potentially reducing the number
of necessary calling arguments.  However, we found that these advanced features
did not work when the PDFI\_SS subroutines were invoked from
other C and Python software.   Thus we define all array dimensions from
other calling arguments in the subroutines.

Second, we have avoided any use of ``common-block'' variables, 
or other global parameters or
variables, which can obscure the dependencies of output variables on input
arguments.  All input data are passed explicitly as calling arguments into
the Fortran subroutines.  This constraint eases the ability to use the library
from languages other than Fortran.

Third, all floating point
operations are performed using 64-bit reals.  All reals, either
scalars or arrays, are declared as
\texttt{real*8} variables in the source code, a choice which seems to work 
correctly with
all Fortran compilers attempted thus far.  All integer arguments to PDFI\_SS
subroutines are assumed to be default integers in Fortran, which are
32-bit integers.

Fourth, the source code assumes that all input and output arrays are dimensioned
or allocated (and deallocated) by the user in the calling programs.  This is
essential for the software to be used from languages other than Fortran.  Thus
very few of the arrays in PDFI\_SS are dynamically allocated within the
source code.  The one exception to this rule are the work arrays needed by
FISHPACK subroutines.  In this case, for each PDFI\_SS subroutine that 
calls a FISHPACK
subroutine, the work array is both allocated and then 
de-allocated within that same subroutine.  

Fifth, to facilitate
ease of interoperability with C code, character string arguments have been
completely avoided in the PDFI\_SS software.  Character strings have a
different representation in memory between Fortran and C.

Finally, the Fortran
syntax is implemented using the older .f suffix for the source-code file names, 
rather than the more modern
.f90 suffix.  While the latter choice results in more flexible syntax for
$e.g.$ line continuation,
the former choice helps enforce 80 character line limits, which makes viewing
the source code much easier from the default 80-character width of a 
terminal window.

\subsection{How to Compile the PDFI\_SS Fortran Library}
\label{sec:howtocompile}

The first step in compiling PDFI\_SS is to download, compile, and 
install the FISHPACK fortran library.  Links for the FISHPACK version 4.1
source code are given in the introduction to \S \ref{sec:pdfinum}.

After unpacking the tarball, we recommend that you replace the contents of
file \texttt{make.inc}
in the top folder of the FISHPACK distribution, with the contents of the file 
\texttt{fishpackmake.inc} located in the 
\texttt{doc} folder in the PDFI\_SS distribution, then replace the file
\texttt{Makefile} in the top folder of the FISHPACK distribution
with the contents of the file 
\texttt{fishpackmake} in the \texttt{doc}
folder in PDFI\_SS, and finally replace the file \texttt{Makefile} 
in the \texttt{src} folder of the
Fishpack distribution with the contents of the file
\texttt{fishpackmakesrc} in the
\texttt{doc} folder in PDFI\_SS.  You may
need to edit the file \texttt{make.inc} to: (1) make sure that the name of
the fortran compiler coincides with the name of the fortran compiler you have.
We have specifically included lines for the gfortran and intel compilers in
the \texttt{Linux} part of the \texttt{make.inc} file, and the gfortran
compiler for the Mac (Darwin) portion of \texttt{make.inc}.  If you are using
another compiler, you will need to edit the compiler definition \texttt{F90}
so that it reflects your compiler.
(2) Make sure that the options included in defining the
fortran compiler also ensure that all reals
are set to 64-bit reals.
(3) Check that compiler options for compiling
position-independent code, needed if FISHPACK will be used for languages
other than Fortran, are invoked. (4) Edit definitions for \texttt{make} and 
\texttt{ar}, if they are different from what is defined in this file.

We cannot overemphasize how important it is to invoke the 
compiler option that all reals are treated as 64-bit reals.
If this is not done, the
attempted use of FISHPACK with PDFI\_SS is doomed to fail, in ways that are
not always easy to diagnose.

To compile the FISHPACK library, type ``make''.
Once the FISHPACK library is compiled, you can install it into a location of
your choosing by typing ``make install'' (or ``sudo make install'' if this
requires root priviledge).  Alternatively, you will need to remember the
exact path to the location of the library file \texttt{libfishpack.a}.

Once FISHPACK has been compiled and installed, we are ready to 
compile PDFI\_SS.
As noted earlier, the PDFI\_SS software developer site is
\url{http://cgem.ssl.berkeley.edu/cgi-bin/cgem/PDFI_SS/index}.  By clicking
on ``Login'', one can log in as anonymous, and then by clicking on 
``Files'' one should find a blue hexidecimal link, the ID for the latest
software release.  By clicking on that link, one should then be led to links
for Tarball or Zip archives for the software.

Once the tarball has been downloaded and unpacked, you should see three
sub-folders:  \texttt{IDL}, \texttt{doc}, and \texttt{fortran}.  
Descend into the \texttt{fortran} folder with a terminal window.

The next step is to open the file ``Makefile'' with an ascii text editor,
such as \texttt{vi} or \texttt{emacs}.  You will most likely need to edit 
this file before you can compile the library.  Currently, the Makefile is 
set up assuming you will
be running on either a Mac or on a Linux machine of some kind; and that you
have access to a Fortran compiler.  The file assumes you will have access to
either the gnu/gfortran compiler or the Intel compiler, ifort.  If you plan to
use a different fortran compiler, you will need to edit Makefile to add the
name of that compiler and to add compiler options for it that coincide with
the meanings of the compiler options for gfortran or ifort.

To compile the PDFI\_SS library file, \texttt{libpdfi\_ss.a}, type ``make''.
To install the library into a specified location, edit the definition of
\texttt{INSTALLDIR}, and then type ``make install'' (or possibly
``sudo make install'', if you need root privilege for the specified location).
The default value of \texttt{INSTALLDIR} is \texttt{/usr/local/lib}.

\subsection{Linking PDFI\_SS to other Fortran programs}
\label{sec:linkingfortran}

Once the PDFI\_SS library has been installed, linking to other Fortran programs
is straightforward.  For the \texttt{gfortran} and \texttt{ifort} compilers,
linking to the library is invoked with the \texttt{-lpdfi\_ss -lfishpack} 
(in that
order) linking commands.  If the libraries are not stored in ``standard''
locations, you may need to specify the location of each library with the 
\texttt{-L<dir>} directive.   Specific examples can be found in the test
programs, described in further detail
\S \ref{sec:testing}.

\subsection{Linking to PDFI\_SS subroutines from C/C++}
\label{sec:linkingC}

If there are no character string arguments, calling a Fortran subroutine from
a C function is very straightforward, if one just remembers some basic rules:
(1) From C, a Fortran subroutine is a function of type void
($i.e.$ the function returns nothing).  All input and output is handled 
through the calling arguments. (2) The name of a Fortran subroutine is changed 
by the Fortran compiler (``Fortran name mangling''); typically this is done by
adding a trailing underscore.  This practice is observed by both the gfortran
and ifort compilers.  In other words, in C, if one wants to call 
\texttt{ahpot\_ss}, the corresponding function name in 
C is \texttt{ahpot\_ss\_};
(3) In Fortran, all arguments are called by reference, not by value.  This
means that when calling a Fortran subroutine from C, all arguments {\it must} 
be passed by reference, $i.e.$ as pointers.  For example, if in a C function
calling a Fortran subroutine, the variables \texttt{m} and \texttt{n} 
are declared as 
integers, their pointers \texttt{\&m} and \texttt{\&n} would be used in the 
call to the subroutine.
(4) Fortran is a column-major language.  For multi-dimensional arrays in
Fortran, the first index always varies in memory the fastest.  For example
a two-dimensional array \texttt{brll}, dimensioned $(n+1,m+1)$ in Fortran, 
assumed
to be in longitude-latitude orientation, is ordered such that we start with
the smallest latitude value, increase the longitude index from the smallest 
to the maximum value,
then repeat the process with the next lowest latitude index value, etc.  C
is considered to be a row-major language, so that given a 
two-dimensional array in C, the second index varies the fastest in memory.

From our experience, the easiest way to deal with this possible source of
confusion is first, to stick with using
one-dimensional arrays in C of length $(n+1)*(m+1)$ using the above example, 
and 
second, to make sure that all input one-dimensional arrays are arranged in 
column-major order before calling the Fortran subroutine.
On output, we also recommend defining one-dimensional arrays in C, 
keeping in
mind that the output data will be ordered by the Fortran subroutine into 
column-major order.  If you need
the data arranged in a different order, you will need to do that re-arrangement
after the subroutine call.
Fortunately, one-dimensional arrays in C map neatly onto multi-dimensional
arrays in Fortran, provided one keeps in mind the assumed column major order.
For Fortran arrays of three or more dimensions, the same principle works:
Define an array in C of length equal to the product of the Fortran dimensions,
and make sure that the first index varies the fastest, followed by the second
index, followed by the next index.

The size of default integers in Fortran is 32-bits, so the C calling program
should be sure to not use 16-bit or 64-bit integers when calling the
subroutines.  For nearly all systems,
a declaration of \texttt{int} in C should be compatible with Fortran integer
arguments to PDFI\_SS subroutines.  Similarly, all real variables in PDFI\_SS
are 64-bit reals, compatible with the double precision (\texttt{double}) 
declaration in C.  The PDFI\_SS subroutines assume that the calling program
has already allocated memory for both input and output arrays.  All memory
management for the calling arguments to PDFI\_SS subroutines is assumed to
be handled by the calling program, and is not done within PDFI\_SS itself.

We have written as one of our test programs (see \S \ref{sec:testing}) a simple
C program that calls the \texttt{brll2tp\_ss} subroutine.  The program shows
explicitly how the input array is constructed and
ordered into column-major order before
calling the subroutine, and when the output array is printed, one sees that
the output array is also arranged in column major order, using the transposed
dimensions.

We strongly recommend defining function prototype statements for any PDFI\_SS
subroutines you call from a C program (in C99, these statements are 
required).  This reduces the chance of making errors in calling the subroutine
from C, and can be helpful in debugging the code by warning the user when
calling arguments disagree with those of
the function prototype.  We have written an include file (\texttt{pdfi\_ss.h})
which contains the function prototypes in C for all of the user-callable 
subroutines
in PDFI\_SS.  This file can be included in any C-code that calls PDFI\_SS
Fortran suboutines.
In our test
C program (\S \ref{sec:testing}), our test program C source code
includes this file.

There is little difference in calling PDFI\_SS subroutines from a C++ program
compared to a C program.  The same rules about passing arguments (all
arguments are passed by reference, $i.e.$ as pointers) applies.  The main
difference is that (1) in the C++ program, you'll need to set the 
\texttt{lang=C} option, and (2) the compiler options for position-independent
code must be invoked when compiling PDFI\_SS.  In our Makefile, we have
endeavored to make sure this option is chosen for the gfortran and ifort
compilers.

\subsection{Linking The PDFI\_SS library into Python} 
\label{sec:python}
 
Linking the PDFI\_SS library into the Python programming language allows 
effective use 
of the software with solar physics related Open-Source Python packages, such 
as \texttt{SunPy} \citep{Mumford2015} and \texttt{astropy} 
\citep{Astropy2013}, as well as easy manipulation, analysis and plotting of 
the input and output data using the basic Python modules \texttt{NumPy}, 
\texttt{SciPy} \citep{Jones2001} and \texttt{matplotlib} 
\citep{Hunter2007}. The PDFI\_SS-Python linking has also been used to 
implement the PDFI\_SS electric field inversion into ELECTRIC field Inversion 
Toolkit, ELECTRICIT \citep{Lumme2017,Lumme2019}. 
ELECTRICIT is an easy-to-use Python software toolkit for downloading and 
processing of SDO/HMI data, and inverting the photospheric electric field 
from the data using a range of state-of-the-art methods.

We have successfully created a working Python interface for several 
PDFI\_SS functions using the F2PY Fortran to Python interface generator 
\url{https://docs.scipy.org/doc/numpy/f2py/}, which is a part of the 
\texttt{NumPy} package. F2PY is compatible with Fortran 77/90/95 
languages and allows partly automated creation and compilation of Python 
interfaces for Fortran routines and functions. 
The generator includes several methods of creating the interface, from which 
we have chosen to use the method based on signature files. The process has 
the following steps: (1) The F2PY package is used to automatically create a 
\emph{signature file} (e.g. \texttt{pdfi\_ss.pyf}) from the Fortran 
source code. The signature file specifies the Python wrapping of the 
PDFI\_SS routines of interest (e.g. \texttt{pdfi\_wrapper4jsoc\_ss}). 
(2) The automatically created signature file is then modified to ensure 
working wrapping of the Fortran routines (usually only modest changes are 
required). (3) Finally F2PY is used to compile an \emph{extension module} 
(e.g. \texttt{pdfi\_ss.so}) from the modified signature file and 
Fortran source code and/or compiled libraries. The extension module and its 
functions are then importable and callable in Python 
(\texttt{import pdfi\_ss}, \texttt{output = 
pdfi\_ss.pdfi\_wrapper4jsoc\_ss(arg1,arg2,...)}).
 
\subsection{Using the Legacy IDL code for PDFI\_SS}
\label{sec:idl}

We have retained the original IDL procedures that we used in the early
phases of the development of the PDFI\_SS library, although this software
is no longer maintained.  We find the software is sometimes useful in
the analysis of magnetic and electric field data generated by the library.

In this version of the software, there is still the need for a fortran
executable to solve the PDFI Poisson equations, but this executable is
spawned from the IDL code when needed, and nearly all of the computational
results apart from the solutions themselves are performed in IDL.  The
source code for the fortran executable \texttt{xpoisson} 
is contained within the file
\texttt{poisson\_arguments\_stag.f}, and is compiled and installed
with the Makefile that
is in the IDL folder.  The \texttt{xpoisson} executable is a very general
wrapper for the FISHPACK subroutines \texttt{HWSCRT}, \texttt{HSTCRT},
\texttt{HWSSSP}, and \texttt{HSTSSP}, and allows one to select either
Cartesian or spherical coordinates, and either centered or staggered grid
solutions.  The \texttt{xpoisson} executable does extensive error checking
on all the input parameters for the FISHPACK subroutines before solving
the Helmholtz or Poisson equation.  To communicate the input data to
\texttt{xpoisson}, and to read the output solutions from \texttt{xpoisson},
the ``Simple Data Format'' or \texttt{sdf} binary data format is used, and
this library must be compiled and installed before \texttt{xpoisson} can
be compiled and run.  

The \texttt{sdf} library is written in C, but is designed to
be used from either C or Fortran.  The objective of the sdf library is to
read and write binary files containing both simple variables and large
arrays, by calling simple subroutines or functions from Fortran or C.
It was developed to aid in the debugging of numerical codes, making it easy
to output and examine the contents of large arrays.
The \texttt{sdf} library also has a set
of IDL procedures to read and write \texttt{sdf} files, making this a
convenient way of communicating between an IDL session and the Fortran
executable.  Co-author Fisher developed and maintains the \texttt{sdf} library.
The source code for \texttt{sdf} can be downloaded from
\url{http://solarmuri.ssl.berkeley.edu/~fisher/public/software/SDF/}.  Use the
latest version.  An archive of the latest version of this
software at the time this article was
published can also be downloaded from Zenodo \citep{Fisher2020sdf}.

The PDFI\_SS IDL source code contains the core abilities to compute
the PTD and FLCT terms, the relaxation procedure (needed for the
Doppler and Ideal contributions) to the PDFI electric field,
but lacks much of the additional capabilities of the Fortran library.  
Computing solutions with the IDL code is also
much more time consuming than using
the Fortran library software.  Nevertheless, we sometimes
find the vector calculus procedures, which have nearly the same names as
the corresponding Fortran subroutines, can be useful in analyzing the results
from PDFI\_SS solutions.

\section{Test Programs Using the PDFI\_SS Software}
\label{sec:testing}

In the course of writing the PDFI\_SS library, it was necessary to 
develop a series of test programs to detect bugs accidentally
introduced into PDFI\_SS subroutines from code revisions, and
to test new capabilities as they are being developed.  Output from
the test programs can then be examined to see whether the results make sense.
The test programs are contained within the
\texttt{test-programs} folder of the \texttt{fortran} folder of the PDFI\_SS
distribution.

Most of the test programs need to read in binary data from input files, and
write out the binary results.  All of this input/output data (except for
the Python test) are assumed to be written
using the Simple Data Format (\texttt{sdf}) format that was introduced 
in \S \ref{sec:idl}.  The names of the needed input files are provided in
the document \texttt{README.txt} contained in the \texttt{test-programs}
sub-folder.  Copies of the needed input files can be obtained from
\url{http://cgem.ssl.berkeley.edu/~fisher/public/data/test_data_pdfi_ss/},
also available as a dataset on Zenodo \citep{Fisher2020}.
The test programs (aside from the python test) assume that the input files
are located in the \texttt{test-programs} folder 
mentioned above.  For the python
linking test, the input files are assumed to be placed into the
\texttt{python-linking} folder within \texttt{test-programs}.
We have typically analyzed the output from the test programs using an IDL
session, in which we read in all the contents of the output file written by
the test program.  If all of the IDL procedures from the \texttt{sdf}
distribution are in your IDL path, this command is very simple:\\
\texttt{sdf\_read\_varlist, 'outputfile'}, where \texttt{outputfile} is
the filename created by the test program.  Then typing ``help'' in the IDL
session will display all of the variables and arrays that were written out.
These results can then be studied and analyzed in IDL.

Next, we provide the names of the test programs and their
purpose, and then will describe how to compile the test programs.  A detailed
description of each test program will then be provided in subsections of this
section.

The names of the test program source code files,
and the purpose of the
test program, are given in Table \ref{tab:testnames}.

\begin{table}[h!]
\renewcommand{\thetable}{\arabic{table}}
\centering
\caption{Test Program File Name and Purpose:} \label{tab:testnames}
\begin{tabular}{c c }
\tablewidth{0pt}
\hline
\hline
Source Code & Purpose \\
\hline
\texttt{test\_wrapper.f} & Test \texttt{pdfi\_wrapper4jsoc\_ss} \\
\texttt{test\_anmhd.f} & Test \texttt{ANMHD} Electric Field Inversions \\
\texttt{test\_bpot.f} & Test Potential Field (from $B_r$) \\
\texttt{test\_bptrans.f} & Test 3D Transpose of Potential Field \\
\texttt{test\_psipot.f} & Test Potential Field (from $\vecB_h$) \\
\texttt{test\_global.f} & Test Global PTD Solution for $\vecE$ \\
\texttt{test\_interp.f} & Test 9th Order B-spline Interpolation \\
\texttt{test\_pdfi\_c.c} & Test PDFI\_SS library function from C \\
\texttt{python-linking} & Test PDFI\_SS linking from Python \\
\hline
\end{tabular}
\end{table}

To compile the suite of Fortran and C test programs, there is a Makefile in the
\texttt{test-programs} folder.  Edit the Makefile to make sure that the
definitions of the Fortran and C compilers are consistent with your system.
Make sure that the library locations for the
\texttt{sdf} and \texttt{fishpack} libraries are correct in the Makefile, 
and that the PDFI\_SS library
has been compiled in the overlying \texttt{fortran} folder.  
Then typing ``make'' in a terminal window should compile all the test programs.
The test programs can be removed by typing ``make clean''.

To run the Python test script, first make sure that the \texttt{NumPy}
and \texttt{SciPy} packages are installed for the version of python you plan
to use.  Edit the Makefile in the \texttt{python-linking} folder to set
the version of the python executable.  Then typing ``make'' should compile
the shared object file \texttt{pdfi\_ss.so}.  This enables one to then run the
script \texttt{pdfi\_wrapper4jsoc\_script.py}, allowing the Fortran subroutines
to be called from Python.  The names of the needed
input data files are given in the \texttt{README.txt} file in this folder.  The
files themselves are available at the URL referenced above.

We must caution that the test programs \texttt{test\_bpot.f}, 
\texttt{test\_bptrans.f}, and \texttt{test\_psipot.f}, when compiled
as the executables \texttt{xbpot}, \texttt{xbptrans}, and \texttt{xpsipot},
respectively, use huge amounts of memory and create huge output
files, and can take a long time to run, particularly if you have insufficient
memory.  We recommend running these test programs only on systems with at least
16GB of memory free, and with a Solid-State Disk.

\subsection{\texttt{test\_wrapper.f} (executable \texttt{xwrapper})}
\label{sec:xwrapper}
The source-code in \texttt{test\_wrapper.f} is designed to
mimic the JSOC's call to subroutine \texttt{pdfi\_wrapper4jsoc\_ss}.  In
a nutshell, it reads in test magnetogram and Doppler data from HMI, 
along with stored FLCT estimates of
horizontal flows, then adds padding to the data in a manner consistent with how
this process is done upstream of the PDFI\_SS call by the JSOC, and then
calls the subroutine \texttt{pdfi\_wrapper4jsoc\_ss}.  The output from the
subroutine is written to an output file.  The test program also independently
computes each of the four electric field contributions, and also writes these
to the output file, so that a detailed and independent comparison can be made.
The program computes and writes to the output file many other diagnostic
quantities.  

We must caution that the input data used here are taken from
a preliminary test data series for NOAA AR 11158
generated several years ago, and do not reflect
a number of improvements to the data analysis that have occurred since that
time.  The FLCT flow velocities, in particular, were generated with
non-optimal parameter choices.  Nevertheless, since these data are fixed here
for the purpose of testing the PDFI\_SS code, not the data analysis procedures,
we have retained their use in this test program.

The documentation at the front of \texttt{test\_wrapper.f} includes a list
of the variables from the output file, if read into an IDL session with
the procedure \texttt{sdf\_read\_varlist}.  Here, we will discuss only
a summary of the overall PDFI\_SS electric field results for this particular
test case.  Particularly important diagnostics one can examine with
the output data include
comparison of the curl of $\vecE$ with the temporal difference in magnetic field
components between the two adjacent vector magnetic field measurements.

One can use quantities in the output file to examine a detailed breakdown of the
PDFI solutions into their four contributions, which are computed independently
within \texttt{test\_wrapper.f}.  For further details, see the documentation
at the head of the \texttt{test\_wrapper.f} source code file.

\begin{figure}[ht!]
\hspace{-0.1in}
\includegraphics[width=3.5in]{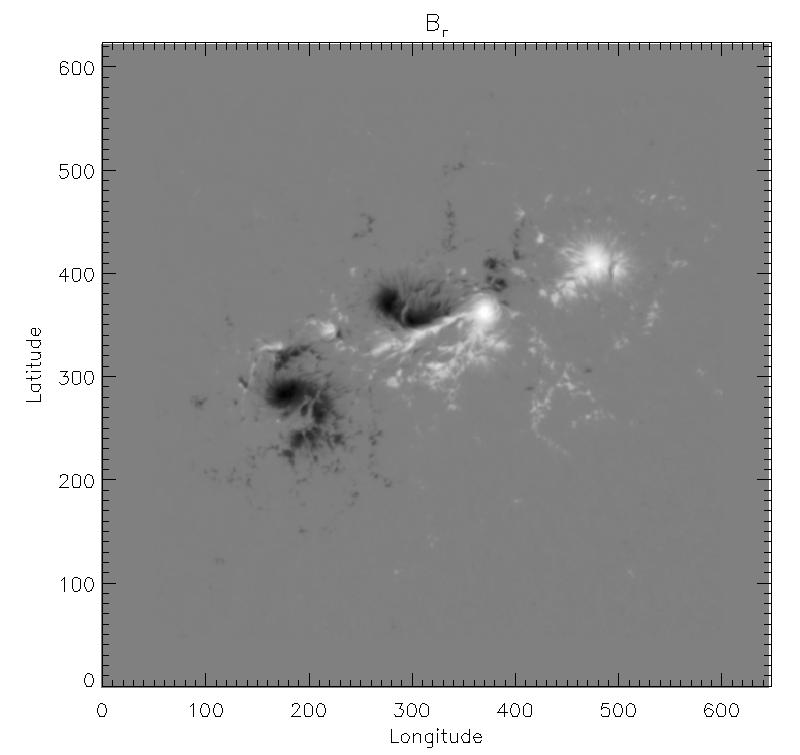}
\caption{Average $B_r$ taken from test data series, from 
February 14, 2011, 23:35-23:47.  The linear grey-scale is from -2500G to 2500G.
The image of $B_r$ shown here is from the average of the two magnetograms.}
\label{fig:brlltest}
\end{figure}

\begin{figure}[ht!]
\hspace{-0.1in}
\includegraphics[width=3.5in]{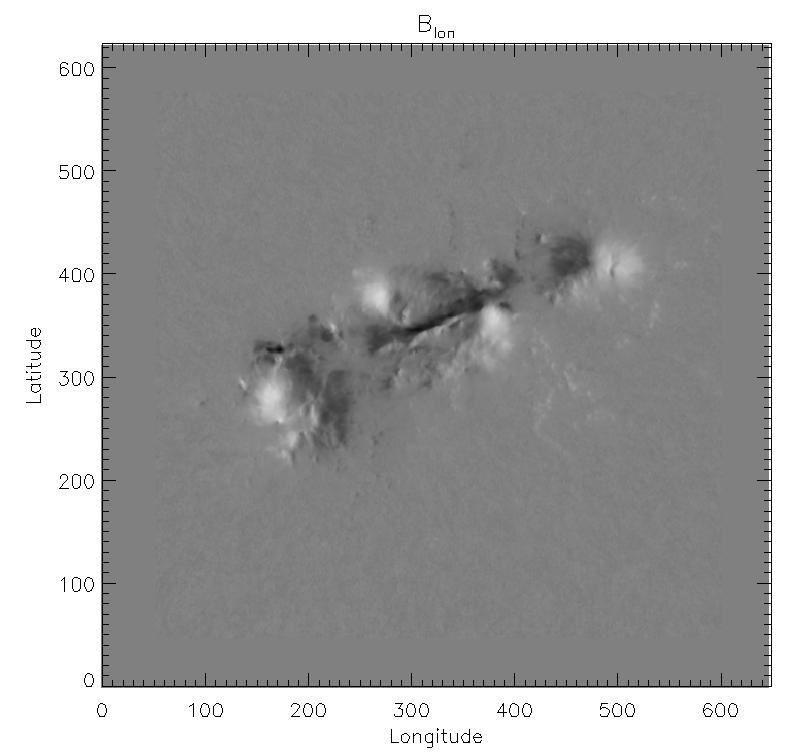}
\caption{Average $B_{lon}$ taken from test data series, from 
February 14, 2011, 23:35-23:47.  The linear grey-scale is from -2000G to 2000G.
The image of $B_{lon}$ shown here is from the average of the two magnetograms.}
\label{fig:blontest}
\end{figure}

\begin{figure}[ht!]
\hspace{-0.1in}
\includegraphics[width=3.5in]{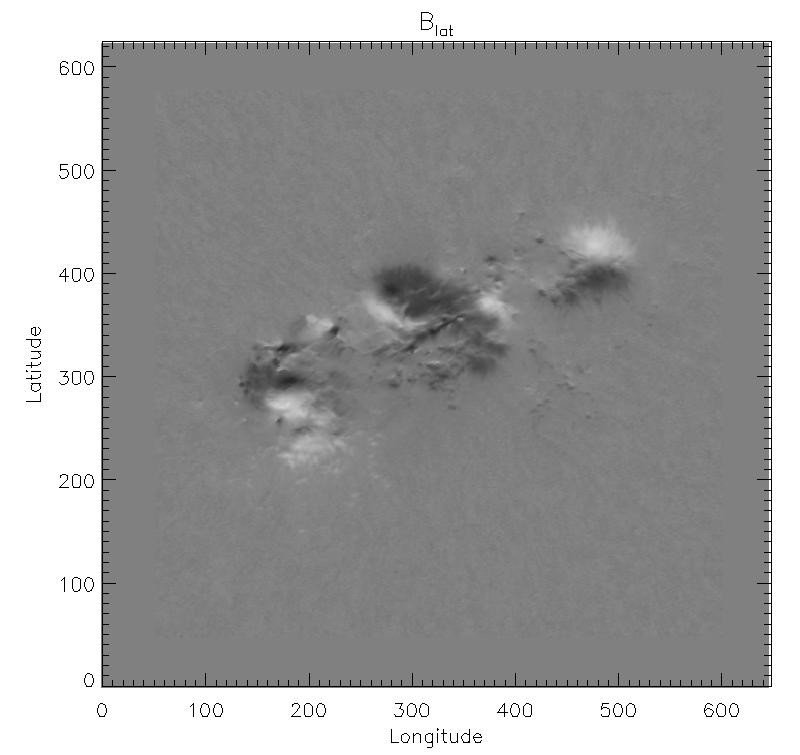}
\caption{Average $B_{lat}$ taken from test data series, from 
February 14, 2011, 23:35-23:47.  The linear grey-scale is from -2000G to 2000G.
The image of $B_{lat}$ shown here is from the average of the two magnetograms.}
\label{fig:blattest}
\end{figure}

\begin{figure}[ht!]
\hspace{-0.1in}
\includegraphics[width=3.5in]{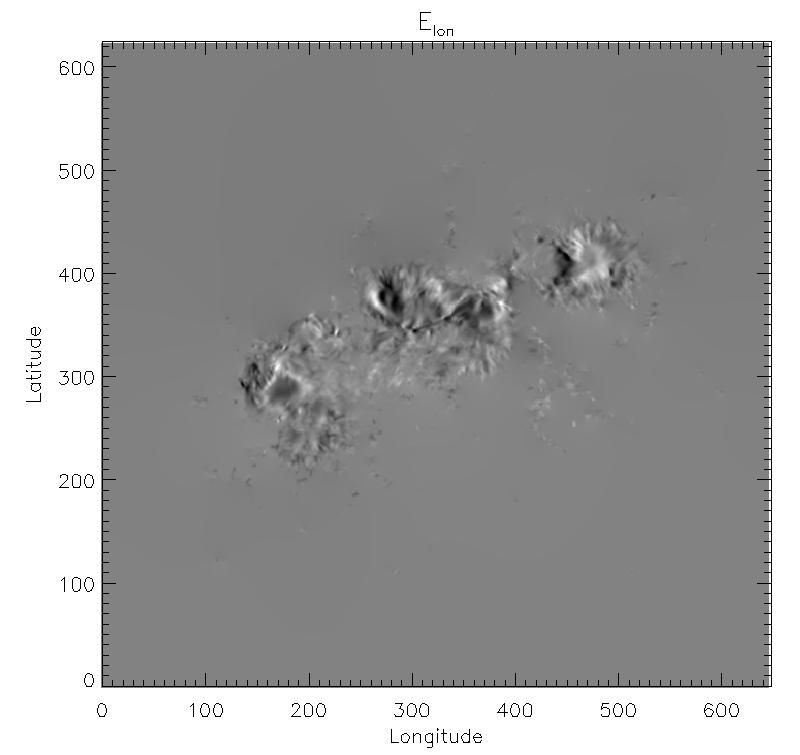}
\caption{Average $E_{lon}$ taken from test data series, from 
February 14, 2011, 23:35-23:47.  The linear grey-scale is from -0.5 to 0.5 
V\ cm$^{-1}$.  The image of $E_{lon}$ shown here is the solution 
evaluated half-way between the times of the two magnetograms.}
\label{fig:elontest}
\end{figure}

\begin{figure}[ht!]
\hspace{-0.1in}
\includegraphics[width=3.5in]{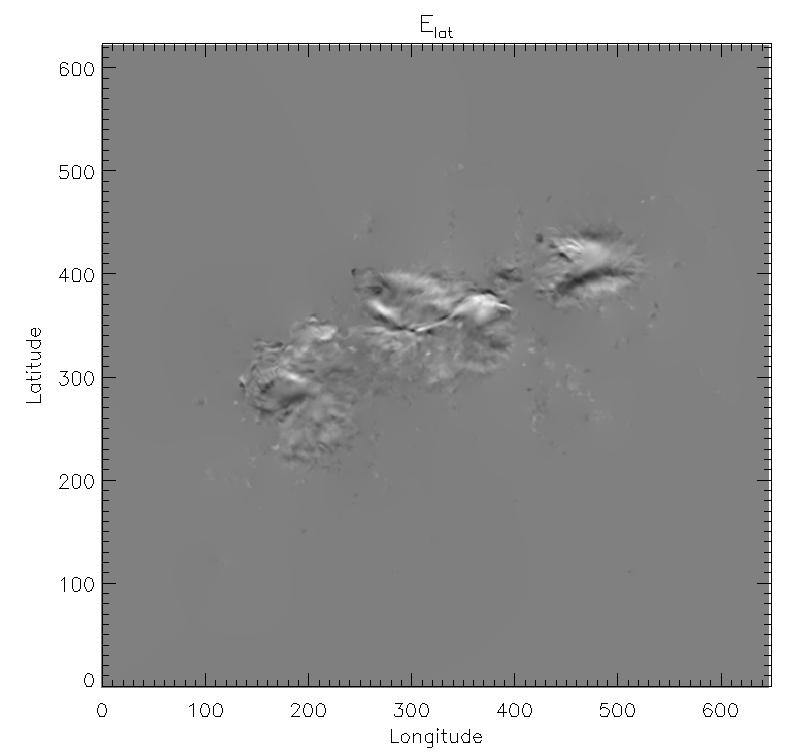}
\caption{Average $E_{lat}$ taken from test data series, from 
February 14, 2011, 23:35-23:47.  The linear grey-scale is from -0.75 to 0.75 
V\ cm$^{-1}$.  The image 
of $E_{lat}$ shown here is the solution evaluated half-way 
between the times of the two magnetograms.}
\label{fig:elattest}
\end{figure}

\begin{figure}[ht!]
\hspace{-0.1in}
\includegraphics[width=3.5in]{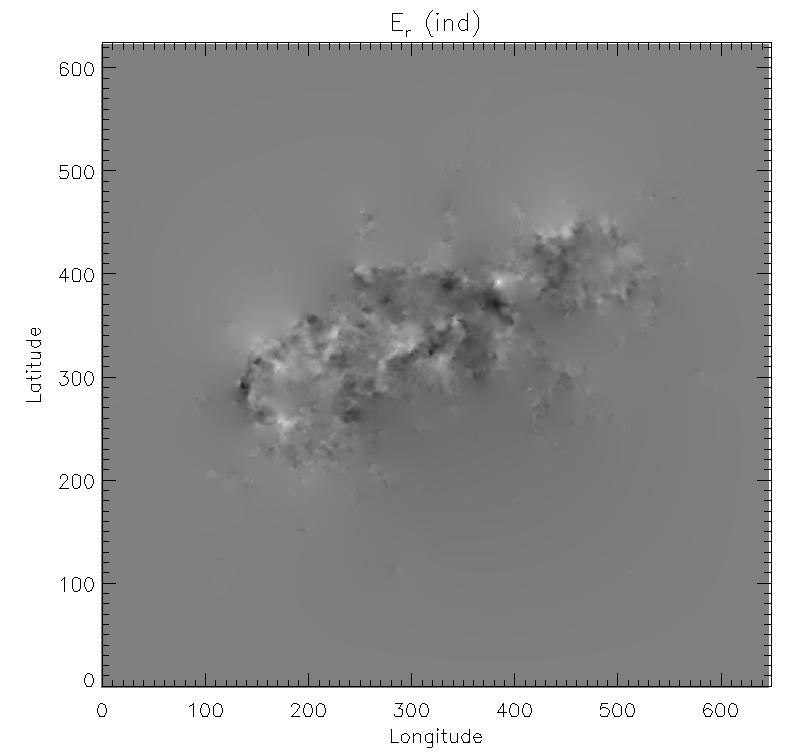}
\caption{Average $E^{ind}_{r}$ taken from test data series, from 
February 14, 2011, 23:35-23:47.  This figure shows only the inductive
contribution to $E_r$.  The linear grey-scale is from -0.5 to 0.5 
V\ cm$^{-1}$.  The 
image of $E^{ind}_{r}$ shown here is the solution evaluated half-way 
between the times of the two magnetograms.}
\label{fig:erllindtest}
\end{figure}

\begin{figure}[ht!]
\hspace{-0.1in}
\includegraphics[width=3.5in]{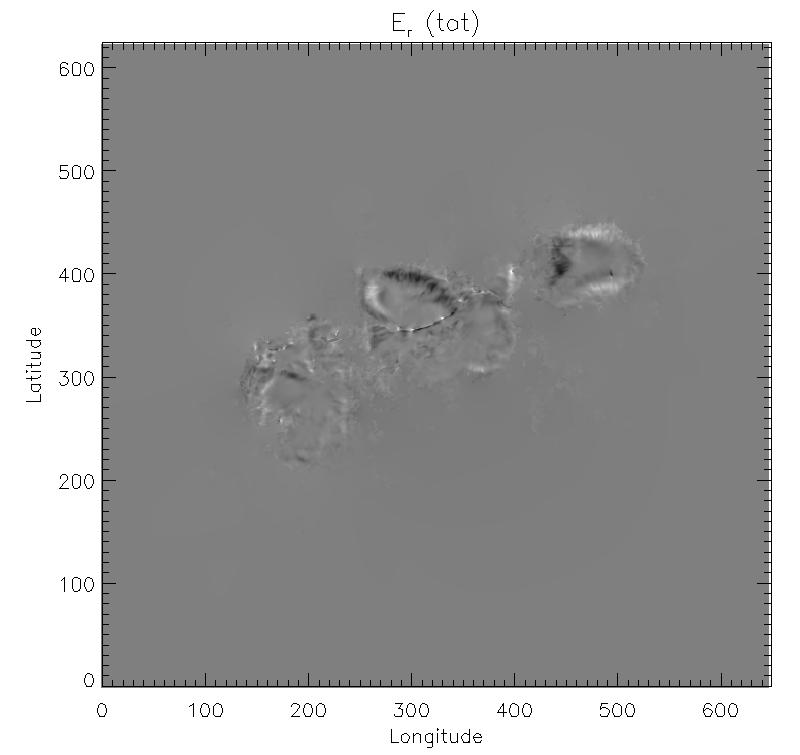}
\caption{Average $E_{r}$ taken from test data series, from 
February 14, 2011, 23:35-23:47.  This figure shows the total electric field
contribution $E_r$.  The linear grey-scale is from -2 to 2 
V\ cm$^{-1}$.  
The image of $E_{r}$ shown here is the solution evaluated half-way 
between the times of the two magnetograms.}
\label{fig:erlltest}
\end{figure}

\begin{figure}[ht!]
\hspace{-0.1in}
\includegraphics[width=3.5in]{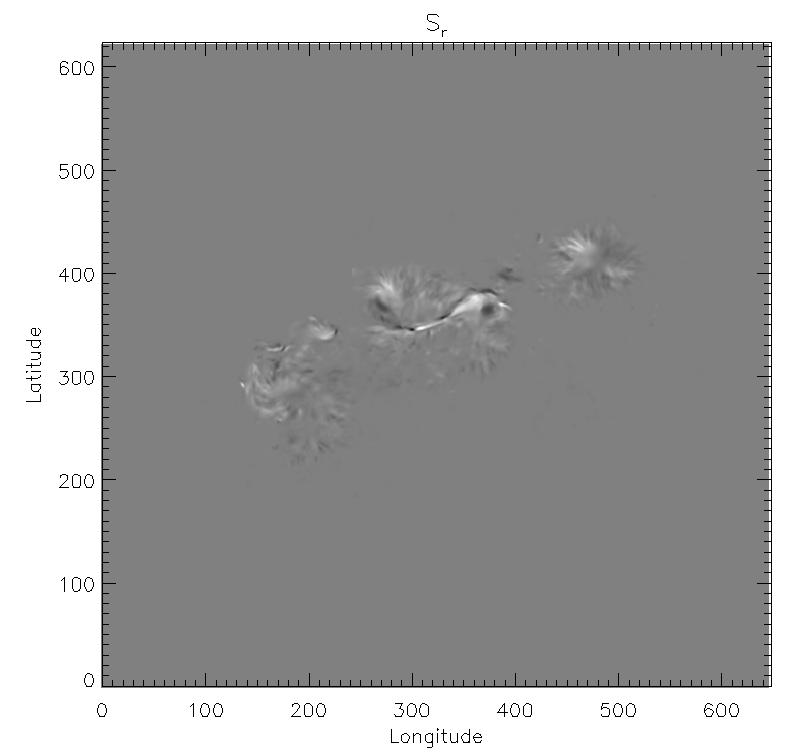}
\caption{Average radial component of Poynting flux $S_r$
taken from test data series, from 
February 14, 2011, 23:35-23:47.  The linear grey-scale range is from
-8 to 8 $\times 10^{9} \rm{\ erg\ } \rm{cm}^{-2} \rm{\ s}^{-1}$.
The image of $S_{r}$ shown here is the solution evaluated half-way 
between the times of the two magnetograms.}
\label{fig:srlltest}
\end{figure}

We end our discussion of \texttt{test\_wrapper.f} by displaying in a 
series of grey-scale figures (Figures \ref{fig:brlltest}-\ref{fig:erlltest})
the three magnetic field 
components for the test data, along with computed PDFI electric field 
inversions for the three components of $\vecE$.  Both the inductive and
total contributions to $E_r$ are shown.  We also show in Figure 
\ref{fig:srlltest} the radial
component of the Poynting flux computed for the pair of magneograms.
These figures provide an overall picture for the electric field and Poynting
flux morphology which can be compared to the magnetic field components for
context.

\subsection{\texttt{test\_anmhd.f}  (executable \texttt{xanmhd})}
\label{sec:xanmhd}

The purpose of the \texttt{test\_anmhd.f} program is to use the PDFI\_SS
software to compute the electric fields for the ANMHD test case using 
vector magnetic field and Doppler data from a horizontal slice of
an ANMHD simulation of magnetoconvection
\citep{Welsch2007,Kazachenko2014}, and then
compare that solution with the $-\vecV \times \vecB$ electric field computed
from the simulation itself.  This provides a good independent test of the
PDFI\_SS solution technique, and it can also be compared with the results
obtained by \citet{Kazachenko2014} in \S 4 of that article using 
PDFI solutions that assume a centered grid formalism in Cartesian coordinates.

Because the ANMHD
simulation was performed in Cartesian coordinates, the first task
is to use the formalism described in \S \ref{sec:cartesian} to map the
Cartesian domain onto a small surface patch bisected by the equator on a very
large sphere.  The resulting radius of the sphere in this case is
$9.998 \times 10^8\rm{\ km}$, well over a thousand times larger than 
$R_{\odot}$.
The colatitude range $b-a = 10^{-4} \rm{\ radians}$, and is also equal to the
longitude range $d-c$, since the domain is a square in Cartesian coordinates.

There is a subroutine in the PDFI\_SS library,\\
\texttt{pdfi\_wrapper4anmhd\_ss},\\ 
which closely mimics the functionality of subroutine
\texttt{pdfi\_wrapper4jsoc\_ss}, but with some differences needed
to accommodate this special case (for example, no ``zero padding'' is done
by the latter subroutine, since padding was also not done in 
\citet{Kazachenko2014}).  \texttt{test\_anmhd.f} calls this
subroutine, and then writes the results to an output file.  When 
the resulting electric field solutions
are compared with those from the ANMHD simulation itself, our use of the
staggered grid means the comparison is a little more complicated than it was
in \citet{Kazachenko2014}.  First, so that we can compare quantities directly,
we must interpolate the simulation magnetic and electric fields from a 
centered grid (which
the simulation used) to our staggered grid locations; this step was not
necessary for the comparison in \S 4 of \citet{Kazachenko2014}.  Second, because
the magnetic field time derivatives were not masked in \citet{Kazachenko2014},
we also do not mask them here (in contrast to what is done with HMI data in
\texttt{pdfi\_wrapper4jsoc\_ss}).  Similarly, we also did not mask the FLCT
electric field or the Ideal electric field.  However, we found that if we
did not mask the Doppler contribution to the electric field, noise in the
definition of the $\unitv{q}$ unit vector in the weak field regions would
wreak havoc on the solutions; therefore, we did use the strong field mask
on the Doppler electric field solutions.  The threshold for the strong
magnetic field mask is 370G, chosen to be consistent with the threshold 
used in \citet{Kazachenko2014}.

\begin{figure}[ht!]
\hspace{-0.2in}
\includegraphics[width=4.0in]{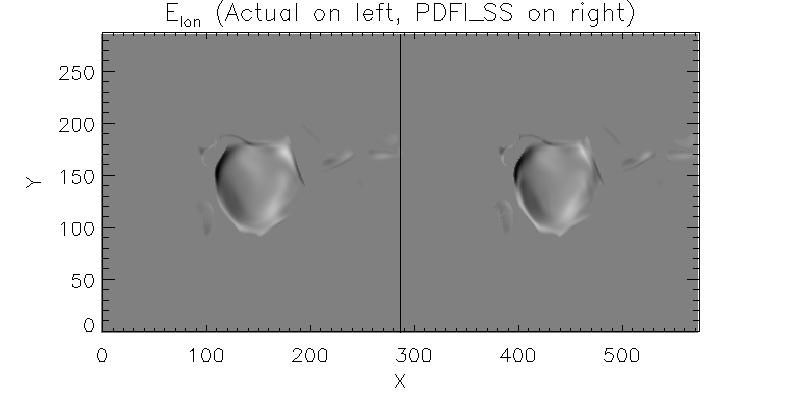}
\caption{Longitudinal component of the electric field from the ANMHD simulation
(left side) and from the PDFI\_SS inversion (right side).  The linear grey-scale
range is from -1 to 1 V\ cm$^{-1}$.  
Both contributions have been multiplied by the
strong-field mask for the TE grid.}
\label{fig:elonactualpdfi}
\end{figure}

\begin{figure}[ht!]
\hspace{-0.2in}
\includegraphics[width=4.0in]{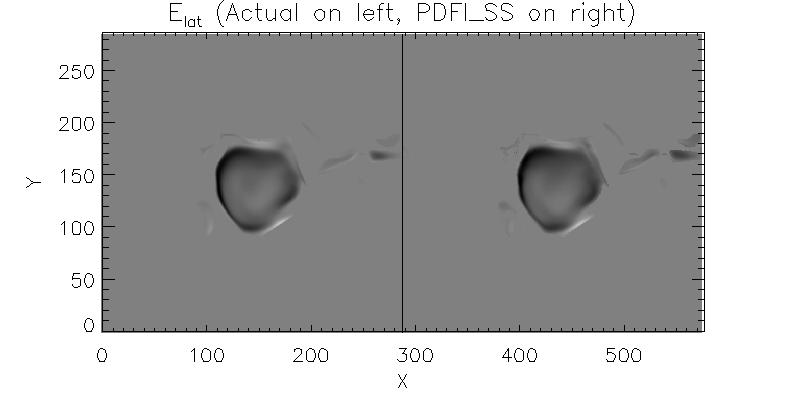}
\caption{Latitudinal component of the electric field from the ANMHD simulation
(left side) and from the PDFI\_SS inversion (right side).  The linear grey-scale
range is from -1 to 1 
V\ cm$^{-1}$.  Both contributions have been multiplied by the
strong-field mask for the PE grid.}
\label{fig:elatactualpdfi}
\end{figure}

\begin{figure}[ht!]
\hspace{-0.2in}
\includegraphics[width=4.0in]{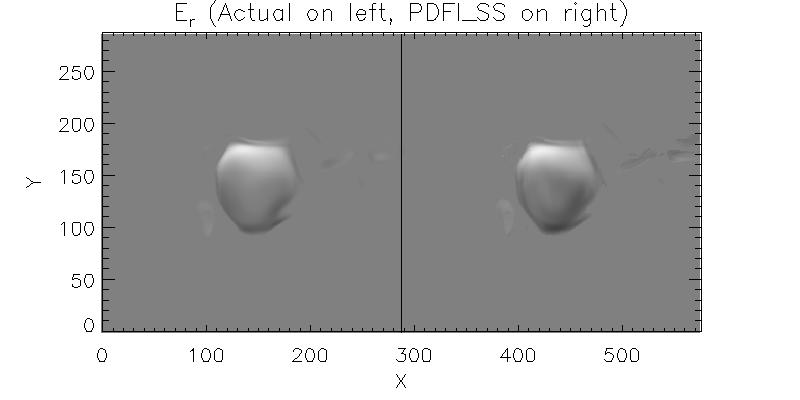}
\caption{Radial component of the electric field from the ANMHD simulation
(left side) and from the PDFI\_SS inversion (right side).  The linear grey-scale
range is from -3 to 3 
V\ cm$^{-1}$.  Both contributions have been multiplied by the
strong-field mask for the COE grid.}
\label{fig:eractualpdfi}
\end{figure}

\begin{figure}[ht!]
\hspace{-0.2in}
\includegraphics[width=4.0in]{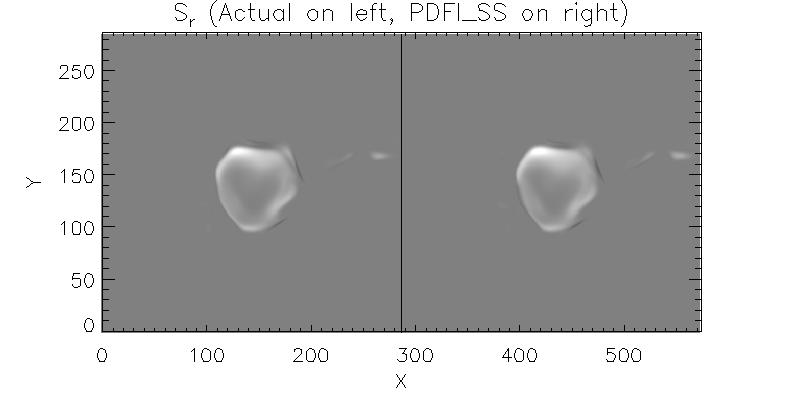}
\caption{Radial component of the Poynting flux from the ANMHD simulation
(left side) and from the PDFI\_SS inversion (right side).  The linear grey-scale
range is from
-5 to 5 $\times 10^{10} \rm{\ erg\ } \rm{cm}^{-2} \rm{\ s}^{-1}$.  Both
contributions have been multiplied by the strong-field mask for the CE grid.}
\label{fig:sractualpdfi}
\end{figure}

\begin{figure}[ht!]
\hspace{-0.3in}
\includegraphics[width=3.8in]{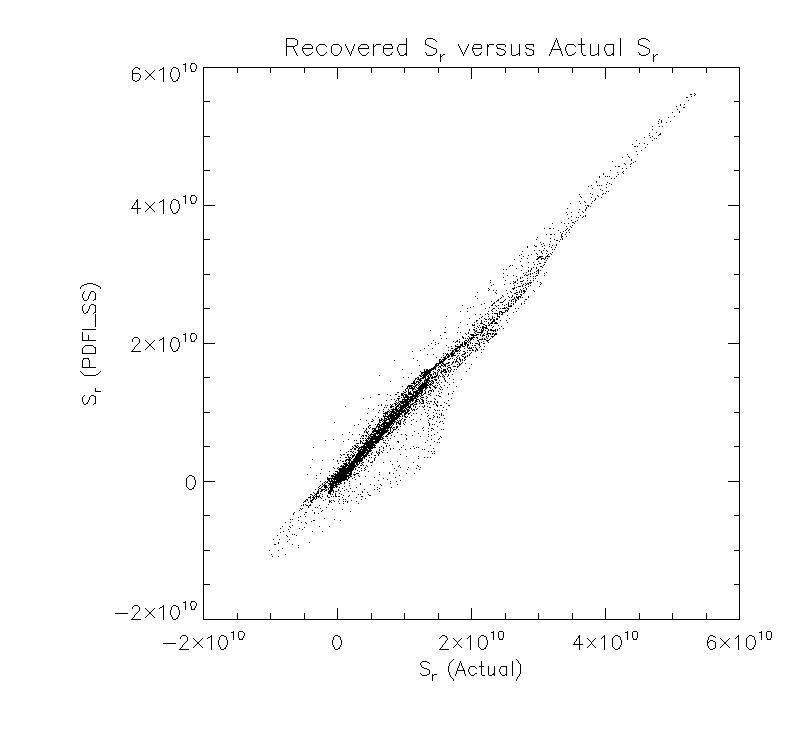}
\caption{Scatter-plot of the Radial component of the Poynting flux 
from the PDFI 
inversion (y-axis) versus that from the ANMHD simulation
(x-axis).  Both
contributions have been multiplied by the strong-field mask for the CE grid.}
\label{fig:sractualpdfiscatter}
\end{figure}

The output file \texttt{anmhd\_output\_file\_pdfi\_ss.sdf} from 
\texttt{test\_anmhd.f} can be read into an IDL session
with \texttt{sdf\_read\_varlist}.  There is an extensive amount of diagnostic
data that can be analyzed, as detailed in the documentation near the front
of the file \texttt{test\_anmhd.f}.  Here, we display just a few
aspects of this output data, where the results can be compared with those of
\citet{Kazachenko2014}.  Figures
\ref{fig:elonactualpdfi}-\ref{fig:eractualpdfi} show side-by-side comparisons
of the longitudinal, latitudinal, and radial electric field images, with
the ANMHD simulation results on the left side of the figures, while the PDFI\_SS
inversion results are shown on the right hand side of the figures.  Both the
ANMHD simulation results and the PDFI\_SS results have been multiplied by
the strong magnetic field masks, as was done for similar figures in
\citet{Kazachenko2014}.  Figure \ref{fig:sractualpdfi} shows the side-by-side
comparison of the radial Poynting flux.  Finally, 
Figure \ref{fig:sractualpdfiscatter} shows a scatter-plot of the radial
component of the Poynting flux from the inversion versus that from the ANMHD 
simulation, and provides a good indication of the resulting error levels from 
the inversion.

Overall, while we find that the ANMHD electric field results are recovered
well, we find that the quality of the inversion is not as good as it
was for the centered grid case used in \citet{Kazachenko2014}.  There are a
number of possible reasons for this, including the fact that the simulation
data must be interpolated to the staggered grid locations, and the fact that in 
the PDFI\_SS inversions,
Faraday's law is obeyed to roundoff error, whereas in the simulation data
it is not, as one can see in the lower right panel of Figure 1 
of \citet{Welsch2007}.  The latter 
is a consequence of the ANMHD
simulations being run with spectral
techniques used to compute spatial derivatives, whereas the curl of the 
simulation data was computed using
finite differences.  In spite of these differences, these figures show 
clearly that the PDFI\_SS
technique is able to reproduce the main morphological features and amplitudes
of the ANMHD electric fields.

\subsection{\texttt{test\_bpot.f} (executable \texttt{xbpot})}
\label{sec:xbpot}

The potential field software described in \S \ref{sec:potential} is tested
by \texttt{test\_bpot.f}, in which the radial magnetic field on the CE grid
at the photosphere is used to compute a potential magnetic field distribution
in the volume above the photosphere.  In the test program, the angular
resolution of the solution is the same as that for the photospheric data.  In
radius, the test program assumes 1000 voxels in the radial direction, 
with a source-surface
height of $2 R_{\odot}$ ($i.e.$ one solar radius above the photosphere).
The user can choose whether to assume periodic boundary conditions in $\phi$ by
setting the variable \texttt{bcn} to $0$, or homogenous Neumann
boundary conditions in $\phi$
on the poloidal potential $P$ by setting \texttt{bcn} to $3$.
The test program takes at least several minutes to run (about ten minutes
on a MacBook Pro with 16GB of memory and a solid-state disk).  With limited
amounts of memory, it will likely take considerably longer.

The executable \texttt{xbpot} produces an output file,
\texttt{test\_bpot\_output.sdf}, with a series
of 2D and 3D arrays.  The file can be read in using 
IDL procedure \texttt{sdf\_read\_varlist}.
Reading in the file can take a long time, as the output file is very large.
Once the file is read in, one can type the ``help'' command in IDL to see the
list of output arrays.
The solution for the poloidal potential $P$ is stored
in the 3D array \texttt{scrb3d}.  The three magnetic field components 
for the
solution are computed two different ways:  First, subroutines \texttt{brpot\_ss}
and \texttt{bhpot\_ss} are called to compute $B_r$ and $\vecB_h$ using
the poloidal potential $P$ as input.  The resulting 3D magnetic field arrays
are \texttt{brpot3d}, \texttt{btpot3d}, and \texttt{bppot3d}.  
Second, subroutine \texttt{ahpot\_ss}
is used to compute the vector potential $\vecA^P$ from $P$.  The theta and phi
components of of the vector potential are returned into the 3D arrays
\texttt{atpot} and \texttt{appot}.  Then the 3 magnetic field components can
be computed from the vector potential with subroutine \texttt{curlahpot\_ss}.
The resulting three 3D magnetic field arrays are \texttt{brpotvp},
\texttt{btpotvp}, and \texttt{bppotvp}.  We find that the magnetic field
components computed in the two different ways differ very little.

It is important to note that the potential field solution is computed on
the faces of the voxels, as described in \S \ref{sec:potential}, and that
the horizontal magnetic field components at the bottom layer in the 3D arrays
are located half a voxel above the photosphere.  The photospheric values of
the potential field components are computed with subroutine 
\texttt{bhpot\_phot\_ss}, and output into the 2D arrays \texttt{btpotphot}
and \texttt{bppotphot}.

Note that all of these output arrays are stored in colatitude-longitude-radius 
index
order.  The following test program will take the output file from this
test program and rotate
some of the 3D arrays into longitude-latitude-radius index order.

\subsection{\texttt{test\_bptrans.f} (executable \texttt{xbptrans})}
\label{sec:xbptrans}

The test program \texttt{test\_bptrans.f} uses the 3D transpose subroutines
\texttt{ahpottp2ll\_ss}, \texttt{bhpottp2ll\_ss}, and \texttt{brpottp2ll\_ss}
to transpose the 3D arrays computed by \texttt{test\_bpot.f} from
colatitude-longitude index order to longitude-latitude index order.
The program uses the ouput file from \texttt{test\_bpot.f} as an input file,
and writes transposed output arrays into the file 
\texttt{test-inplace-transpose.sdf}.  The contents of the file can be
read into an IDL session using \texttt{sdf\_read\_varlist}.  The arrays
in the file include the 3D arrays \texttt{alon} and \texttt{alat}, 
the longitudinal and
latitudinal components of the vector potential respectively, and 
\texttt{blonpot},
\texttt{blatpot}, and \texttt{brllpot}, the 3D arrays of the longitudinal,
latitudinal, and radial components of $\vecB^P$.  The transposed 2D
photospheric magnetic field components are in the arrays
\texttt{blonphot} and \texttt{blatphot}.

When working with large 3D arrays such as these, we used a memory saving
technique in Fortran which is worth describing.  Instead of having to create two
separate arrays of the same size, but with different shapes, it is possible,
using a combination of C and Fortran pointers,
to create the two separate arrays such that they occupy the same locations
in memory, but still have different shapes.  The two different array names
can then be successfully passed as arguments into our PDFI\_SS 3D transpose
subroutines.  We will now illustrate how we do this using the radial
magnetic field component of the potential field solution as an example.

The ability to combine the usage of C and Fortran pointers can be invoked
by the declaration\\
\texttt{use, intrinsic :: iso\_c\_binding}\\
within a Fortran program.  The original array \texttt{brpot} in
colatitude-longitude index order is declared as a 3rd rank allocatable array:\\
\texttt{real*8, allocatable,target :: brpot(:,:,:)}
The transposed array, \texttt{brllpot}, is declared as a 3rd rank Fortran
pointer:\\
\texttt{real*8, pointer :: brllpot(:,:,:)}\\
We also define the two shape arrays\\
\texttt{integer :: shapebrpot(3), shll(3)}.\\
Then one defines and reads in the integers \texttt{m,n,p}.  The
array \texttt{brpot} is then allocated as\\
\texttt{allocate brpot(m,n,p+1)}.\\
Once the array is allocated, its shape can be determined with the Fortran
\texttt{shape} function:\\
\texttt{shapebrpot=shape(brpot)}.\\
The shape of the transpose array can then be determined by the following
three statements:\\
\texttt{shll(1)=shapebrpot(2)}\\
\texttt{shll(2)=shapebrpot(1)}\\
\texttt{shll(3)=shapebrpot(3)}.\\
The next step is to read in the \texttt{brpot} array from the input file.
Once that is done, we can assign the Fortran pointer for the transposed array:\\
\texttt{call c\_f\_pointer(c\_loc(brpot),brllpot,shll)}.\\
What this statement does is define the \texttt{brllpot} array to occupy the same
location in memory as the \texttt{brpot} array, but with the shape of the
transpose array.  One can then call the PDFI\_SS subroutine 
\texttt{brpottp2ll\_ss}, with the \texttt{brpot} array as input, and 
\texttt{brllpot} as the output array:\\
\texttt{call brpottp2ll\_ss(m,n,p,brpot,brllpot)}.\\
The source code for \texttt{brpottp2ll\_ss} is written in such a way that
the transpose is done sequentially for each horizontal layer of the 3D arrays,
with a copy of the 2D array made before any transpose operation has been done.
Thus the 3D transpose can be done ``in place'', without destroying any data.
It is very important to note that once the 3D transpose subroutine has been 
called,
the original un-transposed array is essentially destroyed by scrambling it
into the new shape and is hence useless.  Thus the 3D transpose operation
should be the last operation done using the original array.

This concept is used for all the 3D transpose operations in
\texttt{test\_bptrans.f}.  The contents of the output file were used to
generate the potential field figures shown 
in \S \ref{sec:magneticroutines}.

\subsection{\texttt{test\_psipot.f} (executable \texttt{xpsipot})}
\label{sec:xpsipot}

The purpose of the \texttt{test\_psipot.f} test program is to test the ability
to compute the potential magnetic field by using the observed horizontal
components of the magnetic field  at the photosphere as input.  
These solutions are computed by
subroutines \texttt{psipot\_ss} and \texttt{bpot\_psi\_ss}, as described in
\S \ref{sec:psipotbh}.  On input, the test program reads in arrays of
the observed horizontal magnetic fields at the photosphere at the staggered grid
locations from file \texttt{test-fortran-input.sdf}, and on output computes 
the scalar potential $\Psi$
and the 3D magnetic field components, all in colatitude-longitude-radius index
order.  The results of the potential field calculation are written to
the output file \texttt{test-psipot-output.sdf}.

The solution for the scalar potential $\Psi$ is returned as two
separate 3D arrays, \texttt{psi3d}, which assumes zero net radial flux, and
\texttt{psi3df}, in which a net radial flux is imposed with subroutine
\texttt{psi\_fix\_ss} that coincides with
the net radial flux from the observed radial components of the magnetic field.
The potential magnetic field components are computed with 
\texttt{bpot\_psi\_ss}, and are given in the arrays 
\texttt{btpot3d}, \texttt{bppot3d}, and \texttt{brpot3d}.
Photospheric values of the horizontal potential magnetic field components
are given in the 2D arrays \texttt{btpot} and \texttt{bppot}.  The horizontal
divergence of the observed photospheric horizontal magnetic fields are given
in the 2D array \texttt{divbh}, and the divergence of the potential photospheric
horizontal magnetic fields are given in the 2D array \texttt{divbhpot}.

As an alternative to the observed horizontal magnetic field components in
\texttt{test-fortran-input.sdf}, one can instead change the input file
name to\\ 
\texttt{test-fortran-inputpotphot.sdf}, where in this case the
horizontal components of $\vecB$ are computed from the potential magnetic
field solution chosen to match $B_r$.  In this case, the radial magnetic
field component computed by \texttt{test\_psipot.f} should be close to the
observed radial magnetic field component.  The error between the two was
shown earlier in Figure \ref{fig:scatterbrpot}.

One can choose to use periodic boundary conditions in $\phi$
by setting the variable
\texttt{bcn} to 0 in \texttt{test\_psipot.f}, or homogenous Neumann boundary
conditions by setting \texttt{bcn} to 3.  For this test case, homogenous
Neumann boundary conditions require more compute time.

\subsection{\texttt{test\_global.f} (executable \texttt{xglobal})}
\label{sec:xglobal}

The test program \texttt{test\_global.f} is designed to test the ability of
subroutine \texttt{enudge3d\_gl\_ll} to compute a global ($4 \pi$ steradians)
PTD solution covering the entire surface of the Sun.  On input, arrays 
of $\dot B_{lon}$,
$\dot B_{lat}$, and $\dot B_r$ defined on the PE, TE, and CE grids, 
respectively, spanning
the global Sun, are computed from two sets of vector magnetogram data, using
a temporal finite difference to define the time derivative.  In this  
case, we've used AR 11158 vector magnetogram data, and mapped it onto 
the global
Sun geometry, setting $a=0$, $b=\pi$, $c=0$, and $d=2 \pi$, $i.e$ we've 
allowed AR 11158 to take over the entire surface of the Sun.  Because in 
a global geometry
we can't have a non-zero net radial magnetic flux, or its time derivative,
we use subroutine \texttt{fluxbal\_ll} to zero out the average before calling
\texttt{enudge3d\_gl\_ll}, which computes $\vecE$ on all the rails of the
spherical voxels, which are bisected in radius by the photosphere.

To test whether this global solution for $\vecE$ obeys all three components of
Faraday's law, we compute the curl of $\vecE$ with subroutine
\texttt{curle3d\_ll}, which can accommodate both spherical wedge and global
geometries.  The resulting components of the curl of $\vecE$ can then be
compared against the input time derivatives $\dot B_{lon}$, $\dot B_{lat}$,
and $\dot B_r$.  The documentation near the top of \texttt{test\_global.f}
provides the names of the appropriate arrays for the time derivatives and
the three components of the curl of $\vecE$, as well as the arrays for
$\vecE$ itself.  Scatter-plots of the curl of $\vecE$ versus the magnetic
field time derivatives should show straight lines for each component.  The 
output file
for \texttt{test\_global.f} is \texttt{global\_test\_out\_ar11158.sdf}, and
the contents of the file can be read in with IDL procedure
\texttt{sdf\_read\_varlist}.

\subsection{\texttt{test\_interp.f} (executable \texttt{xinterp})}
\label{sec:xinterp}

In \S \ref{sec:changeresolution}, we described several subroutines designed
to use a B-spline to interpolate the input data for computing PDFI electric
field inversions
to a different resolution.  Here, the test program \texttt{test\_interp.f}
performs a test of this B-spline interpolation, by first interpolating an
observed HMI image of $B_r$, which includes regions of zero-padding, to a new 
resolution that 
is about 20\% higher than
the original resolution, and then re-interpolates that image back to the
original resolution.  The original $B_r$ data array can then be compared 
directly
to the twice-interpolated array of the same size, and the quality of the
twice-interpolated image can be evaluated.  The output file,
\texttt{brint.sdf}, contains the original image as the array
\texttt{brdat}, the interpolation to the higher resolution as the array
\texttt{brint}, and the twice-interpolated array as \texttt{brback}.  The
test program uses the subroutine \texttt{interp\_hmidata\_ll} to perform
the interpolations.  The default value of the degree of the spline,
\texttt{deg} is set to 9, but the user can experiment with other values.
Allowed values are 3,5,7, and 9.  Thus far we've found setting
\texttt{deg=9} seems to produce the most accurate results.  The output file
can be read in with the IDL procedure \texttt{sdf\_read\_varlist} to study
the results.

\subsection{\texttt{test\_pdfi\_c.c} (executable \texttt{xctest})}
\label{sec:xctest}

In \S \ref{sec:linkingC}, we described the principles of calling PDFI\_SS
subroutines from C/C++ programs.  In the test program \texttt{test\_pdfi\_c.c},
we illustrate many of the important points made in that section with this
very simple test program, which calls a single PDFI\_SS subroutine,
\texttt{brll2tp\_ss}.  This test program illustrates (1) how to define a
one dimensional array in C that maps onto two-dimensional arrays in Fortran,
(2) how to define the input array in column-major order, consistent with its use
in Fortran, (3) shows that calling arguments are all called by reference, 
and (4) uses the
output array to demonstrate the column-major nature of the
output generated by the Fortran subroutine.

In the C test program, pointers for the arrays \texttt{brll} (the input
array), and \texttt{br} (the output array) are defined and initialized to
\texttt{NULL}.  Next, integer variables \texttt{np1} and \texttt{mp1}
(representing $n+1$ and $m+1$, respectively) are defined and set to 12
and 10, respectively.  The integers \texttt{m} and \texttt{n} are then
defined by decrementing \texttt{np1} and \texttt{mp1} by one.  Next, the
input and output arrays \texttt{brll} and \texttt{br} are each allocated to
have the size $(n+1)*(m+1)$, $i.e.$ the product of the Fortran dimensions.
Next, the array \texttt{brll} is defined so that regarded as a Fortran
array, the value of \texttt{brll(j,i)=(i-1)+(j-1)}.  This is done by using
an outer loop over the latitude index $i$ that goes from $0$ to $m$, 
and an inner loop over the longitude index $j$ that goes from $0$ to $n$.
The calculation of the array value is\\
\texttt{brll[i*np1+j]=(double) i + j;} 
This is the essence of constructing the array in C to have column major
order, before the Fortran subroutine is called.  Note that the $j$ index
is the most rapidly varying.
(The values of $i$ and $j$
differ between C and Fortran because in Fortran, the default index values
start from $1$ whereas in C they start from $0$.)

Next, the Fortran subroutine \texttt{brll2tp\_ss} is called:\\
\texttt{brll2tp\_ss\_ (\&m, \&n, brll, br);}
Note the trailing underscore in the subroutine name, and the fact
that the pointers to \texttt{m} and \texttt{n} are used in the call
(the arrays \texttt{brll} and \texttt{br} are already defined as pointers).

Following the subroutine call, both the \texttt{brll} and \texttt{br} arrays
are printed to the screen.  In the loop that prints the values
of \texttt{brll}, the outer loop is over the latitude index and the inner
loop is over the longitude index, consistent with column major order in Fortran.
When printing out the output array \texttt{br} to the screen, column major
ordering results in the colatitude index $i$ varying most rapidly, and
the longitude index $j$ varies more slowly: the outer loop is over $j$,
and the inner loop is over $i$.  The fortran array \texttt{br(i,j)} is
indexed in C as \texttt{br[j*mp1+i]}.

These same principles in setting up one dimensional arrays in C that
map onto multi-dimensional Fortran arrays should
work for any of the subroutines in PDFI\_SS, as long as one pays careful
attention to the dimensions of the arrays defined in the Fortran subroutines,
and remembers that Fortran assumes the arrays are defined in column major
index order.

When running the executable \texttt{xctest}, the array \texttt{brll} is
printed to the screen, followed by its colatitude-longitude 
transpose, \texttt{br}.
\newline

\subsection{Python linking Test Program}
\label{sec:pythonlinking}

We have prepared an example of the PDFI\_SS-Python linking described in 
\S \ref{sec:python}
in the subfolder \texttt{fortran/test-programs/python-linking}. 
The package contains a working signature file \texttt{pdfi\_ss.pyf} created 
for routines \texttt{add\_padding\_as\_ss}, 
\texttt{pad\_abcd\_as\_ss}, \texttt{pad\_int\_gen\_ss}, 
and \texttt{pdfi\_wrapper4jsoc\_ss} in the PDFI\_SS library. 
The F2PY extension module \texttt{pdfi\_ss.so} can be compiled by 
typing ``make'' in the terminal while within the subfolder 
\texttt{python-linking}. Typing ``make clean'' removes the extension module. 
The \texttt{fishpack} library location must be correctly specified in the 
\texttt{Makefile} and the FISHPACK library must be compiled with 
the \texttt{-fPIC} flag (see discussion in \S \ref{sec:howtocompile}). 

The Python script \texttt{pdfi\_wrapper4jsoc\_script.py} imports the Python 
interfaces of the PDFI\_SS subroutines from the compiled extension module, 
and calls them to process the input data and to estimate the electric field 
by calling the \texttt{pdfi\_wrapper4jsoc\_ss} subroutine 
(\S \ref{sec:pdfiwrapper4jsoc}). Thus, the script reproduces the 
first part of the test program \texttt{test\_wrapper.f} (\S  
\ref{sec:xwrapper}). The input data required by the 
\texttt{pdfi\_wrapper4jsoc\_ss} can be downloaded in 
Python-compatible \texttt{.sav} format from the URL described in the
introduction to \S \ref{sec:testing}.
In \texttt{pdfi\_wrapper4jsoc\_script.py}, the output arrays of 
\texttt{pdfi\_wrapper4jsoc\_ss} are returned as \texttt{NumPy} arrays, 
and the script saves them to a \texttt{NumPy} \texttt{.npz} file. 
This output can be then compared to the output of the pure Fortran version 
of \texttt{pdfi\_wrapper4jsoc\_ss} executed within \texttt{test\_wrapper.f}. 
Example output files created by executing \texttt{test\_wrapper.f} can 
be downloaded from the above URL in
Python-compatible \texttt{.sav} format, and 
their content can be compared to the output of the 
\texttt{pdfi\_wrapper4jsoc\_script.py} using the \\ 
\texttt{compare\_wrapper\_outputs.py} script. The differences printed by 
the script should be small and close to floating point precision.

If the user wishes to do the comparison from scratch on his/her local system, 
the package includes also IDL helper scripts for transforming the 
\texttt{.sdf} input and output files of \texttt{test\_wrapper.f} 
program into Python-compatible \texttt{.sav} format. 
The \texttt{README.txt} file in this sub-folder contains also step-by-step 
instructions for creating the F2PY interface from scratch.

\section{List of Subroutines and Common Arguments used in PDFI\_SS}
\label{sec:lists}

This article discusses the most important Fortran subroutines within PDFI\_SS.
We first make a brief note on some conventions used in the software 
regarding suffixes of the subroutine names.
Most of the subroutine names in PDFI\_SS end with the suffix
\texttt{\_ss}, to denote the ``spherical-staggered'' grid assumptions, but
there are some important exceptions.  For those subroutine names that end
in \texttt{\_ll}, it is assumed that the input and output array arguments
are arranged in ``longitude-latitude'' index order, in contrast to the
``colatitude-longitude'' array index order used by most of the subroutines.  
See \S \ref{sec:transpose} for a general
discussion of the distinction between these
two array indexing schemes.  A few
subroutines end with the suffix \texttt{\_sc}, denoting ``spherical centered'',
meaning that for these cases, a centered rather than a staggered grid
description is assumed.

We provide a list in alphabetical order of the most important
subroutines in the PDFI\_SS library (in Table \ref{tab:subroutinelist}),
as well as a list of commonly used arguments in the subroutines, along with
brief descriptions.  The list of subroutines also includes a brief statement
of purpose, and links to the section in the article where more detailed 
discussion of the subroutine occurs.

\startlongtable
\begin{deluxetable*}{l l r}
\renewcommand{\thetable}{\arabic{table}}
\centering
\tablecaption{Subroutine Name, Purpose, and Section: \label{tab:subroutinelist}}
\tablewidth{0pt}
\tablehead{\colhead{Subroutine Name} & \colhead{Purpose} & \colhead{Section}}
\startdata
\texttt{abcd2wcs\_ss} & Compute WCS/FITS keywords from a,b,c,d & 
\S \ref{sec:interpol}\\
\texttt{add\_padding\_as\_ss} & Insert unpadded data array into padded data 
array & \S \ref{sec:padding}\\
\texttt{ahpot\_ss} & Compute Vector Potential in 3D for Potential Magnetic 
Field from Poloidal Potential $P$ &  \S \ref{sec:magneticroutines}\\
\texttt{ahpottp2ll\_ss} & Transpose 3D Vector Potential for Potential Field 
from Colat-Lon to Lon-Lat Index Order & \S \ref{sec:transpose3d}\\
\texttt{angle\_be\_ss} & Compute Angle Between $\vecE$ and $\vecB$ & 
\S \ref{sec:relaxation}\\
\texttt{berciktest\_ss} & Test Accuracy of Solution for $P$ to Bercik's 
Equation & \S \ref{sec:berciktest}\\
\texttt{bhll2tp\_ss} & Transpose COE Arrays of $\vecB_h$ from Lon-Lat to 
Colat-Lon Order & \S \ref{sec:interpol}\\
\texttt{bhpot\_phot\_ss} & Compute Horizontal Potential Magnetic Field at 
Photosphere & \S \ref{sec:magneticroutines}\\
\texttt{bhpot\_ss} & Compute $\vecB_h$ for Potential Magnetic Field in 3D 
Volume from Poloidal Potential $P$ &  \S \ref{sec:magneticroutines}\\
\texttt{bhpottp2ll\_ss} & Transpose 3D arrays of Potential Field $\vecB_h$ 
from Colat-Lon to Lon-Lat Order & \S \ref{sec:transpose3d}\\
\texttt{bhtp2ll\_ss} & Transpose COE Arrays of $\vecB_h$ from Colat-Lon to 
Lon-Lat Order & \S \ref{sec:interpol}\\
\texttt{bhyeell2tp\_ss} & Transpose Staggered Grid Arrays of $\vecB_h$ from 
Lon-Lat to Colat-Lon Order & \S \ref{sec:interpol}\\
\texttt{bhyeetp2ll\_ss} & Transpose Staggered Grid Arrays of $\vecB_h$ from 
Colat-Lon to Lon-Lat Order & \S \ref{sec:interpol}\\
\texttt{bpot\_psi\_ss} & Compute Potential Magnetic Field from $\Psi$ in 
3D & \S \ref{sec:bpotpsiss}\\
\texttt{br\_voxels3d\_ss} & Compute $B_r$ on Top and Bottom Faces of Voxels 
from $\vecB_h$, $B_r$ at Radial Mid-Point & \S \ref{sec:bpotpsiss}\\
\texttt{brll2tp\_ss} & Transpose $B_r$ on COE grid from Lon-Lat to Colat-Lon 
Order & \S \ref{sec:interpol}\\
\texttt{brpot\_ss} & Compute $B_r$ for Potential Magnetic Field in 3D Volume 
from Poloidal Potential $P$ & \S \ref{sec:magneticroutines}\\
\texttt{brpottp2ll\_ss} & Transpose 3D Potential Field Array $B_r$ from 
Colat-Lon to Lon-Lat Order & \S \ref{sec:transpose3d}\\
\texttt{brtp2ll\_ss} & Transpose $B_r$ on COE Grid from Colat-Lon to 
Lon-Lat Order & \S \ref{sec:interpol}\\
\texttt{bryeell2tp\_ss} & Transpose $B_r$ on Staggered Grid from Lon-Lat 
to Colat-Lon Order & \S \ref{sec:interpol}\\
\texttt{bryeetp2ll\_ss} & Transpose $B_r$ on Staggered Grid from Colat-Lon 
to Lon-Lat Order &  \S \ref{sec:interpol}\\
\texttt{bspline\_ss} & Low-Level Routine for B-spline Interpolation & 
\S \ref{sec:changeresolution}\\
\texttt{car2sph\_ss} & Compute Radius of Sphere To Use for Cartesian 
Solutions & \S \ref{sec:cartesian}\\
\texttt{curl\_psi\_rhat\_ce\_ss} & Compute $\grad \times \unitv{r} \psi$ 
for $\psi$ on CEG grid & \S \ref{sec:veccalc}\\
\texttt{curl\_psi\_rhat\_co\_ss} & Compute $\grad \times \unitv{r} \psi$ 
for $\psi$ on COE grid &  \S \ref{sec:veccalc}\\
\texttt{curlahpot\_ss} & Compute $\grad \times \vecA^P$ for Potential Magnetic
Field in 3D Volume & \S \ref{sec:magneticroutines}\\
\texttt{curle3d\_ll} & Compute $\grad \times \vecE$ for $\vecE$ Voxel Arrays 
in Lon-Lat Order & \S \ref{sec:curlofe}\\
\texttt{curle3d\_ss} & Compute $\grad \times \vecE$ for $\vecE$ Voxel Arrays 
in Colat-Lon Order & \S \ref{sec:curlofe}\\
\texttt{curle3dphot\_ss} & Compute $\grad \times \vecE$ for $\vecE$ evaluated
at Photosphere & \S \ref{sec:curlofe}\\
\texttt{curlehr\_ss} & Compute $\unitv{r} \cdot \grad \times \vecE$ 
for $\vecE_h$ arrays in Colat-Lon Order & \S \ref{sec:curlofe}\\
\texttt{curlh\_ce\_ss} & Compute $\unitv{r} \cdot \grad \times \vecU$ 
evaluated on CE grid & \S \ref{sec:veccalc}\\
\texttt{curlh\_co\_ss} & Compute $\unitv{r} \cdot \grad \times \vecU$ 
evaluated on CO grid & \S \ref{sec:veccalc}\\
\texttt{dehdr\_ss} & Compute Radial Derivatives of Horizontal Electric Fields 
Evaluated at Photosphere & \S \ref{sec:curlofe}\\
\texttt{delh2\_ce\_ss} & Compute Horizontal Laplacian of $\psi$ at CE Grid 
Locations for $\psi$ on CEG grid & \S \ref{sec:veccalc}\\
\texttt{delh2\_co\_ss} & Compute Horizontal Laplacian of $\psi$ at CO Grid 
Locations for $\psi$ on COE grid & \S \ref{sec:veccalc}\\
\texttt{divh\_ce\_ss} & Compute Divergence of Horizontal Components of a 
Vector at CE Grid Locations & \S \ref{sec:veccalc}\\
\texttt{divh\_co\_ss} & Compute Divergence of Horizontal Components of 
a Vector at CO Grid Locations & \S \ref{sec:veccalc}\\
\texttt{divh\_sc} & Compute Divergence of Horizontal Components of a Vector
using Centered Grid Formalism & \S \ref{sec:relaxation}\\
\texttt{downsample3d\_ll} & Flux Preserving Downsampling of 3 Component 
Electric Field in Lon-Lat Order & \S \ref{sec:laplace}\\
\texttt{downsample3d\_ss} & Flux Preserving Downsampling of 3 Component 
Electric Field in Colat-Lon Order & \S \ref{sec:laplace}\\
\texttt{downsample\_ll} & Flux Preserving Downsampling of 2 Component 
Electric Field in Lon-Lat Order & \S \ref{sec:laplace}\\
\texttt{downsample\_ss} & Flux Preserving Downsampling of 2 Component 
Electric Field in Colat-Lon Order & \S \ref{sec:laplace}\\
\texttt{e\_doppler\_rpils\_ss} & Experimental Technique to Compute Doppler 
Electric Field Using Radial and LOS PILs & \S \ref{sec:dopplerrelax}\\
\texttt{e\_doppler\_ss} & Default Technique to Compute Non-inductive Doppler 
Electric Field & \S \ref{sec:dopplerrelax}\\
\texttt{e\_flct\_ss} & Compute Non-inductive Electric Field from FLCT 
Velocities & \S \ref{sec:flctsubroutine}\\
\texttt{e\_ideal\_ss} & Compute Non-inductive Ideal Electric Field, 
Minimizing $\vecE \cdot \vecB$& \S \ref{sec:idealsubroutine}\\
\texttt{e\_laplace\_ll} & Compute Curl-Free Electric Field Using $E_t$
assuming Lat-Lon Order & \S \ref{sec:laplace}\\
\texttt{e\_laplace\_ss} & Compute Curl-Free Electric Field Using $E_t$
assuming Colat-Lon Order & \S \ref{sec:laplace}\\
\texttt{e\_ptd\_ss} & Compute Inductive (PTD) Electric Field Components 
Given $\dot P$ and $\dot T$ & \S \ref{sec:ptdsolve}\\
\texttt{ehyeell2tp\_ss} & Transpose Horizontal Electric Field on Staggered Grid
from Lon-Lat to Colat-Lon Order & \S \ref{sec:interpol}\\
\texttt{ehyeetp2ll\_ss} & Transpose Horizontal Electric Field on Staggered Grid
from Colat-Lon to Lon-Lat Order & \S \ref{sec:interpol}\\
\texttt{emagpot\_psi\_ss} & Compute Potential Field Magnetic Energy 
from $B_r$ and $\psi$ at Photosphere & \S \ref{sec:magenergies}\\
\texttt{emagpot\_srf\_ss} & Compute Potential Field Magnetic Energy 
from $\vecB_h^P$ and $\vecA^P$ at Photosphere & \S \ref{sec:magenergies}\\
\texttt{emagpot\_ss} & Compute Potential Field Magnetic Energy by Integrating
$B^2/(8 \pi)$ over Volume & \S \ref{sec:magenergies}\\
\texttt{enudge3d\_gl\_ll} & Compute 3 Component Global Nudging Electric Field 
in Lon-Lat Order & \S \ref{sec:global}\\
\texttt{enudge3d\_gl\_ss} & Compute 3 Component Global Nudging Electric Field 
in Colat-Lon Order & \S \ref{sec:global}\\
\texttt{enudge3d\_ss} & Compute 3 Component Nudging Electric Field in 
Colat-Lon Order& \S \ref{sec:nudging}\\
\texttt{enudge\_gl\_ll} & Compute 2 Component Global Nudging Electric Field 
in Lon-Lat Order & \S \ref{sec:global}\\
\texttt{enudge\_gl\_ss} & Compute 2 Component Global Nudging Electric Field 
in Colat-Lon Order & \S \ref{sec:global}\\
\texttt{enudge\_ll} & Compute 2 Component Nudging Electric Field in 
Lon-Lat Order & \S \ref{sec:nudging}\\
\texttt{enudge\_ss} & Compute 2 Component Nudging Electric Field in 
Colat-Lon Order& \S \ref{sec:nudging}\\
\texttt{eryeell2tp\_ss} & Transpose Radial Electric Field from Lon-Lat to 
Colat-Lon Order & \S \ref{sec:interpol}\\
\texttt{eryeetp2ll\_ss} & Transpose Radial Electric Field from Colat-Lon to 
Lon-Lat Order & \S \ref{sec:interpol}\\
\texttt{find\_mask\_ss} & Compute Strong Field Mask on COE grid & 
\S \ref{sec:interpol}\\
\texttt{fix\_mask\_ss} & Convert Intermediate Interpolated Mask Values to 
0 or 1 & \S \ref{sec:interpol}\\
\texttt{fluxbal\_ll} & Remove Net Radial Magnetic Flux from $B_r$ assuming 
Lon-Lat Order & \S \ref{sec:global}\\
\texttt{fluxbal\_ss} & Remove Net Radial Magnetic Flux from $B_r$ assuming 
Colat-Lon Order & \S \ref{sec:global}\\
\texttt{gradh\_ce\_ss} & Compute Horizontal Gradient of $\psi$ for $\psi$ on CEG
Grid & \S \ref{sec:veccalc}\\
\texttt{gradh\_co\_ss} & Compute Horizontal Gradient of $\psi$ for $\psi$ on COE
Grid & \S \ref{sec:veccalc}\\
\texttt{gradh\_sc} & Compute Horizontal Gradient of $\psi$ Using Centered Grid
Formalism & \S \ref{sec:relaxation}\\
\texttt{hm\_ss} & Compute Helicity Injection Rate Contribution 
Function & \S \ref{sec:efieldproducts}\\
\texttt{hmtot\_ss} & Compute Helicity Injection Rate over Photospheric 
Domain & \S \ref{sec:efieldproducts}\\
\texttt{interp\_data\_ss} & Interpolate Several Arrays from COE Grid to 
Staggered Grid Locations & \S \ref{sec:interpol}\\
\texttt{interp\_eh\_ss} & Interpolate Horizontal Electric Fields From 
Staggered Grid Locations to CO Grid & \S \ref{sec:relaxation}\\
\texttt{interp\_hmidata\_3d\_ll} & Interpolate 18 COE Input Data Arrays 
to a Different Resolution Using B-Spline & \S \ref{sec:changeresolution}\\
\texttt{interp\_hmidata\_ll} & Interpolate a Single COE Input Data Array
to a Different Resolution Using B-Spline & \S \ref{sec:changeresolution}\\
\texttt{interp\_var\_ss} & Interpolate 3 Components of a Vector from COE to
Staggered Grid Locations & \S \ref{sec:interpol}\\
\texttt{kcost\_ss} & Compute Wavenumbers Assuming 
Homogenous Neumann Boundary Conditions in $\phi$ 
& \S \ref{sec:azimuthboundary}\\
\texttt{kfft\_ss}  & Compute Wavenumbers Assuming Periodic
Boundary Conditions in $\phi$ & \S \ref{sec:azimuthboundary}\\
\texttt{laplacetest\_ss} & Test the Accuracy of the solution for $\psi$ to the
3D Laplace Equation & \S \ref{sec:psipot}\\
\texttt{mflux\_ss} & Compute the Net Magnetic Flux over the Photospheric 
Spherical Wedge Domain & \S \ref{sec:magneticroutines}\\
\texttt{pad\_abcd\_as\_ss} & Compute New Values of $a$, $b$, $c$, and $d$ 
Given the Old Values and the amounts of padding & \S \ref{sec:padding}\\
\texttt{pad\_int\_gen\_ss} & Compute Amounts of Padding on all 4 Boundaries & 
\S \ref{sec:padding}\\
\texttt{pdfi\_wrapper4anmhd\_ss} & Compute PDFI Solution for $\vecE$ for the
ANMHD Test Case & \S \ref{sec:xanmhd}\\
\texttt{pdfi\_wrapper4jsoc\_ss} & Compute PDFI Solution for $\vecE$ for a
Cadence of Vector Magnetogram and Doppler Data & \S \ref{sec:pdfiwrapper4jsoc}\\
\texttt{psi\_fix\_ss} & Remove $1/r$ artifact from $\Psi$ and Impose Observed
Net Magnetic Flux (if Desired) & \S \ref{sec:psipot}\\
\texttt{psipot\_phot\_ss} & Compute $\Psi$ at Photosphere from $B_r$ and 
Poloidal Potential $P$ & \S \ref{sec:magenergies}\\
\texttt{psipot\_ss} & Compute $\Psi$ for a Potential Field using $\vecB_h$ at 
Photosphere, solving the Laplace Equation & \S \ref{sec:psipot}\\
\texttt{ptdsolve\_eb0\_ss} & Compute $\dot P$, $\partial \dot P / \partial r$,
and $\dot T$ by Solving Poisson Equations assuming $E_t=0$ & 
\S \ref{sec:ptdsolve}\\
\texttt{ptdsolve\_ss} & Compute $\dot P$, $\partial \dot P / \partial r$, and
$\dot T$ by Solving Poisson Equations using non-zero $E_t$ & 
\S \ref{sec:ptdsolve}\\
\texttt{relax\_psi\_3d\_ss} & Solve for Scalar Potential $\psi$ using the 
``Iterative'' Method formulated by Brian Welsch & \S \ref{sec:relaxation}\\
\texttt{scrbpot\_ss} & Solve Bercik's Equation for the Poloidal Potential $P$
For a Potential Magnetic Field in 3D & \S \ref{sec:scrbpotss}\\
\texttt{sinthta\_ss} & Compute $\sin \theta$ at Colatitude Cell Edges and 
Cell Centers & \S \ref{sec:veccalc}\\
\texttt{sr\_ss} & Compute Radial Component of the Poynting Flux at the
Photosphere & \S \ref{sec:efieldproducts}\\
\texttt{srtot\_ss} & Integrate the Radial Poynting Flux over Area to Derive
Magnetic Energy Input Rate & \S \ref{sec:efieldproducts}\\
\texttt{wcs2abcd\_ss} & Compute $a$, $b$, $c$, and $d$ from WCS/FITS keywords 
for COE grid & \S \ref{sec:interpol}\\
\texttt{wcs2mn\_ss} & Compute $m$ and $n$ from WCS/FITS keywords & 
\S \ref{sec:interpol}\\
\enddata
\end{deluxetable*}

Now we list and describe some of the most commonly used calling argument 
variables used in the subroutines within PDFI\_SS, as well as important 
information about these variables:
\subsection{Common Input Variables Defining Domain Geometry}
\label{sec:geometryvariables}
\begin{itemize}

\item \texttt{rsun} [km]: - This real*8 scalar variable defines the radius
of the photospheric surface upon which the PDFI calculations will be done.
The expected units are km.  For most solar applications, this can be set to
$6.96d5$.  But if you are computing solutions in Cartesian geometries, this
variable is typically set to a much larger number (see \S \ref{sec:cartesian}).

\item \texttt{a},\texttt{b},\texttt{c},\texttt{d} [radians]:  These four
real*8 scalar variables define the two-dimensional ``spherical wedge''
subdomain used in PDFI\_SS (see \S \ref{sec:fishconv}).  The quantity
\texttt{a} defines the colatitude of the northern domain edge, \texttt{b}
defines the colatitude of the southern domain edge, and \texttt{c} and
\texttt{d} define the longitude values of the left and right domain edges, 
respectively.  The quantity \texttt{a} is less than \texttt{b}, and both
are bounded below by $0$ and above by $\pi$.  The quantity \texttt{c} is
less than \texttt{d} and both are nominally in the range from $0$ to $2 \pi$.

\item \texttt{m},\texttt{n} :  These 32-bit integer values set the number
of cell interiors in colatitude and longitude, respectively.  Nearly all
array dimensions in PDFI\_SS subroutines are defined in terms of these
integers.

\item \texttt{p} :  This 32-bit integer sets the number of radial voxels used
in calculations of potential magnetic field solutions in three dimensions.
The array sizes for the 3D arrays in the potential magnetic field software
are defined in terms of \texttt{m}, \texttt{n}, and \texttt{p}.

\item \texttt{dtheta}, \texttt{dphi} [radians] :  These two real*8 scalar
variables describe the size of the colatitude and longitude cells in a 
Plate Carr\'ee Grid, and are assumed to be constant within the spherical wedge
domain.  Their values are defined by
equations (\ref{eqn:dtheta}-\ref{eqn:dphi}) in \S \ref{sec:fishconv}.
The PDFI\_SS software does not make any assumptions about the relative size
of \texttt{dtheta} and \texttt{dphi}, but in the HMI magnetic pipeline
software, we attempt to keep \texttt{dtheta} and \texttt{dphi} nearly
equal.

\item \texttt{dr} [km]:  This real*8 scalar variable describes the radial
depth of spherical voxels used in subroutines \texttt{e\_voxels3d\_ss},
\texttt{br\_voxels3d\_ss}, \texttt{enudge3d\_ss}, \texttt{enudge3d\_gl\_ss},
\texttt{enudge3d\_gl\_ll}, \texttt{curle3d\_ss}, and \texttt{curle3d\_ll}.
In these subroutines, the bottom faces and edges of the voxels lie
\texttt{0.5*dr} below the photosphere, and the top faces and edges of the
voxels lie \texttt{0.5*dr} above the photosphere.

\item \texttt{sinth(m+1)}, \texttt{sinth\_hlf(m)}:  These two real*8 arrays
contain values of $\sin\,\theta$ (where $\theta$ is colatitude), 
evaluated at colatitude cell edges
(\texttt{sinth}), and colatitude cell centers (\texttt{sinth\_hlf}).  These
two arrays can be computed with subroutine \texttt{sinthta\_ss}.

\item \texttt{rssmrs} [km]:  This real*8 scalar variable is used by the
potential magnetic field software, and denotes the distance between the
radius of the Sun (\texttt{rsun}), and the source-surface outer boundary.

\end{itemize}

\subsection{Input Magnetic Field and Velocity Variables passed into
\texttt{pdfi\_wrapper4jsoc\_ss}}
\label{sec:pdfiwrapper4jsocinputs}

All the Input Magnetic Field, LOS unit vector, and Velocity Variables that
are passed into subroutine \texttt{pdfi\_wrapper4jsoc\_ss} are defined on the
COE grid, and are given in longitude-latitude index order.  All 18 of these 
real*8 arrays
are dimensioned $(n+1,m+1)$.  Variable names ending in \texttt{0} refer to
the first of the two timesteps, and names ending in \texttt{1} refer to
the second of the two timesteps.  See discussion in \S \ref{sec:interpol}.
We also include in this list the scalar \texttt{bmin}, which sets the
threshold for the strong-field mask.

\begin{itemize}

\item\texttt{bmin} : [G] real*8 scalar which determines the threshold for
the strong field mask array on the COE grid.  The quantity $|\vecB|$ must
be larger than \texttt{bmin} at both input timesteps for the mask value to
be set to $1$.

\item \texttt{bloncoe0}, \texttt{bloncoe1} [G]: arrays of the longitudinal
component of the magnetic field at the first and second timesteps, respectively.

\item \texttt{blatcoe0}, \texttt{blatcoe1} [G]: arrays of the latitudinal
component of the magnetic field at the first and second timesteps, respectively.

\item \texttt{brllcoe0}, \texttt{brllcoe1} [G]: arrays of radial component of
the magnetic field at the first and second timesteps, respectively.

\item \texttt{lloncoe0}, \texttt{lloncoe1} : arrays of longitudinal component of
the unit vector pointing toward the observer at the first and second timesteps,
respectively.

\item \texttt{llatcoe0}, \texttt{llatcoe1} : arrays of latitudinal component
of the unit vector pointing toward the observer at the first and second 
timesteps, respectively.

\item \texttt{lrllcoe0}, \texttt{lrllcoe1} : arrays of the radial component
of the unit vector pointing toward the observer at the first and second
timesteps, respectively.

\item \texttt{vloncoe0}, \texttt{vloncoe1} [km\ sec$^{-1}$] : 
arrays of the longitudinal
component of the optical flow velocity computed by FLCT at the first and second
timesteps, respectively (see discussion in \S \ref{sec:horizvel}).

\item \texttt{vlatcoe0}, \texttt{vlatcoe1} [km\ sec$^{-1}$] : 
arrays of the latitudinal
component of the optical flow velocity computed by FLCT at the first and second
timesteps, respectively (see discussion in \S \ref{sec:horizvel}).

\item \texttt{vlosllcoe0}, \texttt{vlosllcoe1} [m\ sec$^{-1}$] : 
arrays of the
line-of-sight component of the velocity, with positive values denoting
redshifts, at the first and second timesteps, respectively.  Note units
difference when compared to FLCT velocities.

\end{itemize}

\subsection{Output Magnetic Field and Electric Field Variables from
\texttt{pdfi\_wrapper4jsoc\_ss}}
\label{sec:pdfiwrapper4jsocoutputs}

On output from \texttt{pdfi\_wrapper4jsoc\_ss}, magnetic field variables
at both timesteps are returned on their staggered grid locations, as well
as the electric field solution variables, computed midway between the two
timesteps, also on their staggered grid
locations.  The radial Poynting flux array is returned, as well as its spatial
integral.  The Helicity injection rate contribution function is also returned,
along with its spatial integral, the Relative Helicity injection rate.  Strong
field masks for the COE, CO, CE, TE, and PE grids are also returned.  All 
output arrays are in longitude-latitude index order.

\begin{itemize}

\item \texttt{blon0(n+1,m)}, \texttt{blon1(n+1,m)} [G] : 
These real*8 arrays of the
longitudinal magnetic field component are 
defined on the PE grid, for the first and second timesteps.

\item \texttt{blat0(n,m+1)}, \texttt{blat1(n,m+1)} [G] : 
These real*8 arrays of the
latitudinal magnetic field component are 
defined on the TE grid, for the first and second timesteps.

\item \texttt{brll0(n,m)}, \texttt{brll1(n,m)} [G] : These 
real*8 arrays of the radial magnetic field component are 
defined on the CE grid, for the first and second timesteps.

\item \texttt{elonpdfi(n,m+1)} [V\ cm$^{-1}$] : 
This real*8 array of the longitudinal
component of the PDFI electric field is
defined on the TE grid, evaluated midway between the two timesteps.
To convert to units of [G km\ sec$^{-1}$], multiply by $1000$.

\item \texttt{elatpdfi(n+1,m)} [V\ cm$^{-1}$] : 
This real*8 array of the latitudinal
component of the PDFI electric field is
defined on the PE grid, evaluated midway between the two timesteps.
To convert to units of [G km\ sec$^{-1}$], multiply by $1000$.

\item \texttt{delondr(n,m+1)} [V\ cm$^{-2}$] : This real*8 array of the radial
derivative of the longitudinal component of the PTD electric field is
defined on the TE grid, evaluated midway between
the two timesteps.  To convert to units of [G\ sec$^{-1}$], multiply by
$10^8$.

\item \texttt{delatdr(n+1,m)} [V\ cm$^{-2}$] : This real*8 array of the radial
derivative of the latitudinal component of the PTD electric field is
defined on the PE grid, evaluated midway between
the two timesteps.  To convert to units of [G\ sec$^{-1}$], multiply by $10^8$.

\item \texttt{erllpdfi(n+1,m+1)} [V\ cm$^{-1}$] : 
This real*8 array of the radial component
of the PDFI electric field is defined on the COE grid,
evaluated midway between the two timesteps.  To convert to units of 
[G km\ sec$^{-1}$],
multiply by $1000$.  Do not use this array when evaluating 
the horizontal components of $\grad \times \vecE$.

\item \texttt{erllind(n+1,m+1)} [V\ cm$^{-1}$] : This real*8 array of the 
inductive (PTD)
contribution to the radial electric field is defined
on the COE grid, evaluated midway between the two timesteps.  To convert to
units of [G km\ sec$^{-1}$], multiply by $1000$.  This is the array to use when
evaluating the horizontal components of $\grad \times \vecE$.

\item \texttt{srll(n,m)} [erg\ cm$^{-2}$\ s$^{-1}$] : 
This real*8 array of the radial
component of the Poynting flux is defined on the CE grid,
evaluated midway between the two timesteps.

\item \texttt{srtot} [erg\ s$^{-1}$] : This real*8 scalar is the area 
integral of the
radial component of the Poynting flux, evaluated midway between the two
timesteps.

\item \texttt{hmll(n,m)} [Mx$^2$\ cm$^{-2}$\ s$^{-1}$] : This real*8 
array of the contribution
function for the Helicity injection rate is defined on
the CE grid, evaluated midway between the two timesteps.

\item \texttt{hmtot} [Mx$^2$\ s$^{-1}$] : This real*8 scalar is the area 
integral of
the contribution function for the Helicity injection rate, evaluated midway
between the two timesteps.

\item \texttt{mcoell(n+1,m+1)} : This real*8 array is the strong-field mask
for the COE grid, evaluated midway between the two timesteps.

\item \texttt{mcoll(n-1,m-1)} : This real*8 array is the strong-field mask
for the CO grid, evaluated midway between the two timesteps.

\item \texttt{mcell(n,m)} : This real*8 array is the strong-field mask for the
CE grid, evaluated midway between the two timesteps.

\item \texttt{mtell(n,m+1)} : This real*8 array is the strong-field mask for
the TE grid, evaluated midway between the two timesteps.

\item \texttt{mpell(n+1,m)} : This real*8 array is the strong-field mask for
the PE grid, evaluated midway between the two timesteps.

\end{itemize}

\subsection{The Poloidal and Toroidal Potentials for Electric Field Solutions}
\label{sec:ptdvariables}

The subroutine \texttt{ptdsolve\_ss} returns the poloidal and toroidal
potentials (or their time derivatives), as well as the radial derivative 
of the poloidal potential (or its time derivative),
at the photospheric surface within our staggered spherical wedge domain.  
Here we
summarize the output variables returned from \texttt{ptdsolve\_ss}.  Note that
when solving the Poisson equations for the PTD potentials, all input and
output variables are arranged in colatitude-longitude index order:

\begin{itemize}

\item \texttt{scrb(m+2,n+2)} [G km$^2$ or G km$^2$\ s$^{-1}$] :  This real*8
array contains the poloidal potential $P$ (or $\dot P$)
evaluated on the CEG grid.  The name \texttt{scrb} originates from our original
notation in KFW14, in which we called the poloidal potential $\scrB$ (or
``script B'').  While we now use the notation $P$, the array name in
the software refers to its original variable name.

\item \texttt{dscrbdr(m+2,n+2)} [G km or G km s$^{-1}$] : This real*8 array
contains the radial derivative of the poloidal potential
$\partial P / \partial r$ (or $\partial {\dot P} / \partial r$) 
evaluated on the CEG grid.

\item \texttt{scrj(m+1,n+1)} [G km or G km s$^{-1}$] : This real*8 array
contains the toroidal potential $T$ (or $\dot T$) evaluated on the
COE grid.  The name \texttt{scrj} originates from our original notation in
KFW14, in which we called the toroidal potential $\scrJ$ (or ``script J'').
While we now use the notation $T$, the array name in the software refers
to its original variable name.

\end{itemize}

\subsection{Staggered Grid Variable Names of Magnetic and Electric Field 
Components Commonly Used In PDFI\_SS Calculations}
\label{sec:bevariables}

As described in \S \ref{sec:pdfinum}, most of the calculations involving
magnetic field and electric field components are performed in
colatitude-longitude index order, using the staggered grid locations described
in \S \ref{sec:stagger}.  When magnetic field components are needed (rather
than their time derivatives) we use values averaged between the two input
timesteps, effectively evaluated midway between the two timesteps.
The electric fields are also evaluated midway between the two timesteps.
Here we describe magnetic field variable names used in vector
calculus subroutines \texttt{divh\_ce\_ss} and \texttt{curlh\_co\_ss}, and
the electric field variable names used on output by \texttt{e\_ptd\_ss},
as well as on input to the vector calculus subroutines \texttt{divh\_co\_ss} and
\texttt{curlh\_ce\_ss}.  These magnetic and electric field
variable names are also frequently used as
input or output arguments to many of the transpose subroutines discussed in
\S \ref{sec:interpol}.  In the subroutines \texttt{e\_flct\_ss},
\texttt{e\_doppler\_ss}, and \texttt{e\_ideal\_ss}, variations of
these electric field variable names are used on output from these subroutines.
Note that when used in the calculation of electric
fields, electric field components are computed in units of $c \vecE$, $i.e.$
[G km\ s$^{-1}$], rather than in the [V\ cm$^{-1}$] units passed 
on output from
\texttt{pdfi\_wrapper4jsoc\_ss}.

\begin{itemize}

\item \texttt{bt(m+1,n)} [G] : This real*8 array is the colatitude component
($B_{\theta}$) of the magnetic field, computed on the TE grid, evaluated
midway between the two timesteps.

\item \texttt{bp(m,n+1)} [G] : This real*8 array is the longitude component
($B_{\phi}$) of the magnetic field, computed on the PE grid, evaluated midway
between the two timesteps.

\item \texttt{br(m,n)} [G] : This real*8 array is the radial component of the
magnetic field, computed on the CE grid, evaluated midway between the two 
timesteps.

\item \texttt{et(m,n+1)} [G km\ s$^{-1}$] : 
This real*8 array is the colatitude
component of $c \vecE$ ($c E_{\theta}$), computed on the PE grid, evaluated
midway between the two timesteps.

\item \texttt{ep(m+1,n)} [G km\ s$^{-1}$] : 
This real*8 array is the longitudinal
component of $c \vecE$ ($c E_{\phi}$), computed on the TE grid, evaluated 
midway between the two timesteps.

\item \texttt{er(m+1,n+1)} [G km\ s$^{-1}$] : This real*8 array is the radial
component of $c \vecE$ ($c E_{r}$), computed on the COE grid, evaluated midway
between the two timesteps.

\end{itemize}

\subsection{Variables Returned from Potential Magnetic Field Subroutines}
\label{sec:potvariables}

The subroutine \texttt{scrbpot\_ss} returns the 3d array representing the
poloidal potential $P$, subroutine \texttt{ahpot\_ss} returns the two 3D arrays
representing the vector potential $\vecA^P$, subroutine
\texttt{bhpot\_ss} returns the two 3D arrays representing the horizontal
components of $\vecB^P$, and subroutine \texttt{brpot\_ss} returns the
3D array of the radial component of $\vecB^P$.  If desired, all three components
of $\vecB^P$ can be computed from the vector potential $\vecA^P$ using
subroutine \texttt{curlahpot\_ss}.  Here is a list of these
returned variables:

\begin{itemize}

\item \texttt{scrb3d(m,n,p+1)} [G km$^2$] : This 3D real*8 array is the poloidal
potential $P$ for the potential magnetic field solution returned from 
subroutine \texttt{scrbpot\_ss}.  This array can be used to generate the
vector potential and magnetic field components for the Potential Magnetic Field
solution.  This variable is evaluated on the CE grid (horizontal directions),
on radial shells.

\item \texttt{mflux} [G km$^2$] : This real*8 scalar is the net signed 
photospheric radial magnetic flux, returned from subroutine \texttt{mflux\_ss}.
This variable is used on input to subroutines \texttt{ahpot\_ss} and
\texttt{brpot\_ss} to compute potential magnetic field solutions with non-zero
net radial photospheric magnetic flux.  To compute solutions with zero net
radial magnetic flux, one can set \texttt{mflux} to zero.

\item \texttt{atpot(m,n+1,p+1)} [G km] : This 3d real*8 array represents
$\vecA^P_{\theta}$, the colatitude component of the vector potential for the
Potential Magnetic Field solution, computed by subroutine \texttt{ahpot\_ss}.
This variable is evaluated on the PE grid (horizontal directions), on radial
shells.

\item \texttt{appot(m+1,n,p+1)} [G km] : This 3d real*8 array represents
$\vecA^P_{\phi}$, the longitudinal component of the vector potential for the
Potential Magnetic Field solution, computed by subroutine \texttt{ahpot\_ss}.
This variable is evaluated on the TE grid (horizontal directions), on radial
shells.

\item \texttt{btpot(m+1,n,p)} [G] : This 3d real*8 array represents
$\vecB^P_{\theta}$, the colatitude component of the Potential Magnetic Field
solution, computed by subroutine \texttt{bhpot\_ss} or
\texttt{curlahpot\_ss}.
This variable is evaluated on the TE grid (horizontal directions), midway
between radial shells.

\item \texttt{bppot(m,n+1,p)} [G] : This 3d real*8 array represents
$\vecB^P_{\phi}$, the longitudinal component of the Potential Magnetic field
solution, computed by subroutine \texttt{bhpot\_ss} or
\texttt{curlahpot\_ss}.
This variable is evaluated on the PE grid (horizontal directions), midway
between radial shells.

\item \texttt{brpot(m,n,p+1)} [G] : This 3d real*8 array represents
$\vecB^P_{r}$, the radial component of the Potential Magnetic Field solution,
computed by subroutine \texttt{brpot\_ss} or \texttt{curlahpot\_ss}.
This variable is evaluated on the CE grid (horizontal directions), on radial
shells.

\end{itemize}

\subsubsection*{Acknowledgements}
This work was supported by NASA and NSF through their funding of the ``CGEM''
project through NSF award AGS1321474 to UC Berkeley, NASA award 80NSSC18K0024
to Lockheed Martin, and NASA award NNX13AK39G to Stanford University.  This
work was also supported by NASA through the one-year
extension to the CGEM project, ``ECGEM'', through award 80NSSC19K0622 to
UC Berkeley.  
E. Lumme acknowledges the doctoral program in particle physics and 
universe sciences (PAPU) of the University of Helsinki, and Emil Aaltonen 
Foundation for financial support.

We wish to thank the reviewer of this article for many valuable comments which
greatly improved its clarity and accuracy.

We wish to thank the US Taxpayers for their generous support for this
project.

We which to thank Todd Hoeksema for valuable discussions regarding regarding
orbital artifacts in the HMI data.  We wish to thank Sandy McClymont for
his insights into the plots in \S \ref{sec:advstag} and for his valuable
comments.
We wish to thank Vemareddy Panditi for pointing out some bugs in
the legacy IDL software in the PDFI\_SS distribution.

\clearpage
\bibliography{apj-jour,master}



\end{document}